\documentclass[12pt]{article}

\usepackage{color}
\usepackage[dvips]{graphicx}
\usepackage{graphicx}
\usepackage{amsmath,amssymb}
\usepackage{feynmp}
\usepackage{epsfig}

\usepackage{mathrsfs}

\def\N{\mathbb{N}}
\def\Z{\mathbb{Z}}

\def\R{\mathbb{R}}

\input{colordvi.tex}

\newcommand{\sfrac}[2]{\left(\frac{#1}{#2}\right)}

\newcommand{\bra}[1]{ \langle {#1} | }
\newcommand{\ket}[1]{ | {#1} \rangle }

\def\diag{\mathop{\rm diag}\nolimits}
\def\tr{\mathop{\rm tr}}

\def\SO{\mathop{\rm SO}}

\def\SU{\mathop{\rm SU}}
\def\U{\mathop{\rm U}}

\begin{document}

\begin{titlepage}
    
\begin{flushright}
UT-14-38 \\
IPMU14-0291
\end{flushright}
     
\vskip 1cm
\begin{center}
   
{\large \bf Skewness Dependence of GPD / DVCS, \\
    Conformal OPE and AdS/CFT Correspondence II: \\
  --- a holographic model of GPD --- }
 
 \vskip 1.2cm
     
  Ryoichi Nishio\footnote{current affiliation: SmartNews, Inc.}$^{,2,3}$
  and Taizan Watari$^3$
      
\vskip 0.4cm
  
 {\it   $^2$Department of Physics, University of Tokyo, Tokyo 113-0033, Japan  
    \\[2mm]
     
    $^3$Kavli Institute for the Physics and Mathematics of the Universe, University of Tokyo, Kashiwa-no-ha 5-1-5, 277-8583, Japan
     }
\vskip 1.5cm
     
\abstract{
Traditional idea of Pomeron/Reggeon description for hadron scattering 
is now being given theoretical foundation in gravity dual descriptions, 
where Pomeron corresponds to exchange of spin-$j\in 2\Z$ states in the 
graviton trajectory. Deeply virtual Compton scattering (DVCS) 
is essentially a 2 to 2 scattering process of a hadron and a photon, and 
hence one should be able to study non-perturbative aspects (GPD) of 
this process by the Pomeron/Reggeon process in gravity dual.
We find, however, that even one of the most developed formulations of gravity 
dual Pomeron (Brower--Polchinski--Strassler--Tan (BPST) 2006) is not able to
capture skewness dependence of GPD properly. 
In Part I (arXiv:1212.3322), therefore, we computed Reggeon wavefunctions 
on AdS$_5$ so that the formalism of BPST can be generalized. In this 
article, Part II, we use the wavefunctions to determine the DVCS amplitude, 
bring it to the form of conformal OPE/collinear factorization, 
and extract a holographic model of GPD, which naturally fits into the 
framework known as ``dual parametrization'' or ``(conformal) collinear 
factorization approach''.
} 
    
\end{center}
\end{titlepage}


\tableofcontents

\section*{Introduction to Part II}
\label{sec:intro-II}

This preprint is a continuation of the study in another 
preprint \cite{NW-skew1}; those two preprints share the same title 
and are regarded as part II and part I, respectively. 
They will be combined to be a single article when 
submitted to a journal. 
Since Part II has to refer to equations in Part I many times, 
the full text of Part I \cite{NW-skew1} (except Introduction) is 
included as a part of this preprint for convenience of the readers, after  
minimum corrections are made.
Sections \ref{ssec:review-conf-OPE-GPD}--\ref{ssec:repr-dilatation} and 
the appendices \ref{ssec:appendix-eigen-q=0}--\ref{ssec:vector} have 
appeared already in Part I \cite{NW-skew1}; the new material in part II is 
found in sections \ref{ssec:confine}--\ref{sec:model} and the 
appendices \ref{ssec:appendix-projector}--\ref{sec:conformal-coeff-AdS}.

We found that interesting preprints \cite{Costa-1209, Costa-1404} cover 
a subject that is closely related to our study in 
sections \ref{sec:mode-fcn}--\ref{sec:organize} and the 
appendix \ref{sec:appendix-mode-fcn}. References \cite{Costa-1209, Costa-1404} 
mainly deal with correlation functions of CFT's as functions of spacetime 
coordinates, whereas we deal with them in this article and in \cite{NW-skew1} 
as functions of incoming/outgoing momenta, and confinement effects are 
also implemented, so that we can study hadron scattering processes.

\section{Introduction: journal-article version}
\label{sec:intro-full}

Scattering processes of hadrons involve non-perturbative information 
of QCD. When it comes to scattering with the center of mass energy 
higher than the QCD scale, lattice computation will not have enough 
computation power in a near future, yet perturbative QCD is able to say 
something only about hard components involved in the scattering. This is where 
holographic descriptions of strongly coupled gauge theories may find 
a role to play. Although we cannot expect gravitational ``dual'' 
descriptions to be both calculable and perfectly equivalent to the QCD 
of the real world at the same time, we still hope to be able to learn 
non-perturbative aspects of hadrons at qualitative level, using calculable 
holographic dual descriptions of nearly conformal strongly coupled gauge 
theories.  

String theory started out as the dual resonance model describing scattering 
amplitudes of hadrons. One of its major problems as a theory of hadrons 
was a ``prediction'' that the amplitude of the elastic scattering of 
two hadrons falls off exponentially $e^{Bt}$ in the momentum transfer 
squared $t$ for some $B > 0$, although the amplitude is known in reality 
to fall off in a power-law in $|-t|$ for hard scattering. The ``prediction,'' 
however, is now understood as that of string theory with a flat background 
metric; the amplitude of elastic scattering turns into such a power-law indeed, 
when the target space of string theory has a warped metric. At the 
qualitative level, string theory on a warped spacetime---holographic 
(gravitational dual) descriptions---can be a viable theory of hadron 
scattering \cite{PS-01-PRL, PS-02-DIS, BPST-06}.

Holographic technique can be used to study not just amplitudes 
of hadron scattering as a whole, but also to extract information 
of partons within hadrons \cite{PS-02-DIS}. 
Parton distribution functions (PDFs) are defined by the inverse Mellin 
transformation of hadron matrix elements of gauge singlet parton-bilinear 
operators in QCD, and gravity dual descriptions can be used to determine 
matrix elements of the gauge singlet operators. The PDF extracted in this 
way satisfies DGLAP ($q^2$-evolution) and BFKL ($\ln (1/x)$-evolution) 
equations (e.g., \cite{Hatta-07, CC-DIS, BDST-10, NW-first}); just like 
in perturbative QCD \cite{saddle}, those two evolution equations follow 
from how the saddle point $j^*$ moves in the complex angular momentum 
$j$-plane integral (inverse Mellin transform).
The holographic description for the PDF and the generalized parton 
distribution (GPD) also shows crossover transition 
between this DGLAP/BFKL behavior and the Regge behavior \cite{BPST-06} 
(see also \cite{NW-first}). Thus, the parton information studied in this 
way may be used to understand non-perturbative issues associated with 
partons in a hadron at qualitative level. 

In this article, 
we study 2-body--2-body scattering between a hadron and 
a photon (that is possibly virtual) in gravitational dual descriptions; 
$\gamma^* (q_1) + h (p_1) \rightarrow \gamma^{(*)} (q_2) + h(p_2)$. 
A special case of this scattering---the forward scattering with 
$q_1 = q_2$ and $p_1 = p_2$---has been studied extensively in the
literature (e.g., \cite{PS-02-DIS, Hatta-07, CC-DIS, 
BDST-10, NW-first}) for study of DIS and PDF, and some references 
also cover the case of non-forward elastic scattering 
($(q_1)^2 = (q_2)^2$, $(q_1-q_2)^2 \neq 0$). This article extends the 
analysis so that all kinds of skewed ($q_1^2 \neq q_2^2$) cases are covered. 
In hadron physics, therefore, the kinematics needed for deeply virtual 
Compton scattering, hard exclusive vector meson production and time-like 
Compton scattering processes \cite{GPD-review} is covered in this analysis. 
With the full skewness dependence included in this analysis, it is also 
possible to use the result of this study to bridge a gap between data 
in such scattering processes at non-zero skewness \cite{DVCS-GPD-dispersion} 
and the transverse profile of partons in a hadron, which is encoded by 
GPD at zero skewness \cite{GPD-transv-profile}. 

From theoretical perspective, the task of this article is to generalize 
the formalism of \cite{PS-02-DIS, BPST-06}  (see also \cite{Hatta-07, 
BDST-10, NW-first}), so that it can be used for 2-body-to-2-body 
scattering that is not necessarily elastic. 
Pomeron/Reggeon propagators have been treated as if it were for a 
scalar field in \cite{PS-02-DIS, BPST-06, NW-first}, but 
they correspond to exchange of stringy states with non-zero (arbitrarily 
high) spins; for the study of scattering with non-zero skewness, the 
polarization of higher spin state propagator should also be treated
with care (see also the approach in \cite{Costa-1209, Costa-1404}). 

It is notoriously a difficult problem to compute scattering of strings 
on a curved background geometry. We do not pretend that the generalization 
of the formalism in this article is something derived from string theory 
without a flaw. This is rather an attempt at capturing an approximately 
right picture of non-perturbative aspects in hadron scattering that 
string theory would predict in a distant future.  We are forced to rely 
sometimes on physics intuition, and to ignore subtleties or corrections 
that are not under control, when we face situations where not enough 
techniques have been developed in string theory at the moment. 

This article is organized as follows. 
We begin in section \ref{ssec:review-conf-OPE-GPD} with a review of 
parameterization of GPD in terms of conformal OPE, because the expansion 
in a series of conformal primary operators becomes the key concept 
in using AdS/CFT correspondence (cf \cite{CC-DIS}).  
After plainly stating what needs to be done in the gravity dual approach  
in section \ref{ssec:idea-simply}, we proceed to explain our basic
gravity dual setting and idea of how to construct a scattering
amplitude of our interest by using string field theory in sections 
\ref{sec:set-up} and \ref{sec:cubic-sft}. 
Section \ref{sec:mode-fcn} shows the results of computing 
wavefunctions of spin-$j$ fields on AdS$_5$, while more detailed 
account of derivation of the wavefunctions is given in the 
appendix \ref{sec:appendix-mode-fcn}. 
Classification of eigenmodes that turn out to be relevant for 
the ``twist-2'' operators in later sections is given in 
section \ref{ssec:eigen-q=0}, and wavefunctions are presented as 
analytic functions of the complex spin (angular momentum) variable $j$ 
in section \ref{ssec:eigen-q=not0}.
Those wavefunctions are organized into irreducible representations 
of conformal algebra in section \ref{ssec:repr-dilatation}; 
the representation for spin-$j$ primary operators contain more 
eigenmode components than those treated by the Pomeron exchange 
amplitude in the formalism of \cite{BPST-06}, indicating that more 
contributions are needed in the scattering amplitude with non-vanishing 
skewness than in the formalism of \cite{BPST-06}. 
These wavefunctions (and propagators) are used in section \ref{sec:organize}
in organizing scattering amplitude on AdS$_5$. The amplitude obtained in 
this way can be cast into the form of conformal OPE, from which 
one can also extract GPD as a function of kinematical variables.
We are not committed to a particular form of implementing confining 
effects in the holographic description, as discussed in 
section \ref{ssec:confine}. Some qualitative aspects of the GPD profile 
are examined in section \ref{sec:model}.

Not surprisingly, holographic models of GPD so obtained 
provide a special subclass of GPD models that have been called 
``dual parametrization'' or ``(conformal) collinear factorization approach''
in QCD/hadron community \cite{dual-para, dual-para-2, Mueller-S-05, 
KK-Mueller-07}. After all, it is the combination of the dual resonance 
model and the AdS/CFT correspondence that are being used. 

\section{Our Approach: Conformal OPE and Gravity Dual}

\subsection{Review: Conformal OPE of DVCS Amplitude}
\label{ssec:review-conf-OPE-GPD}

\subsubsection{Notation and Conventions}

\begin{figure}[htbp]
\begin{center}
\begin{tabular}{cc}
  \includegraphics[scale=0.4]{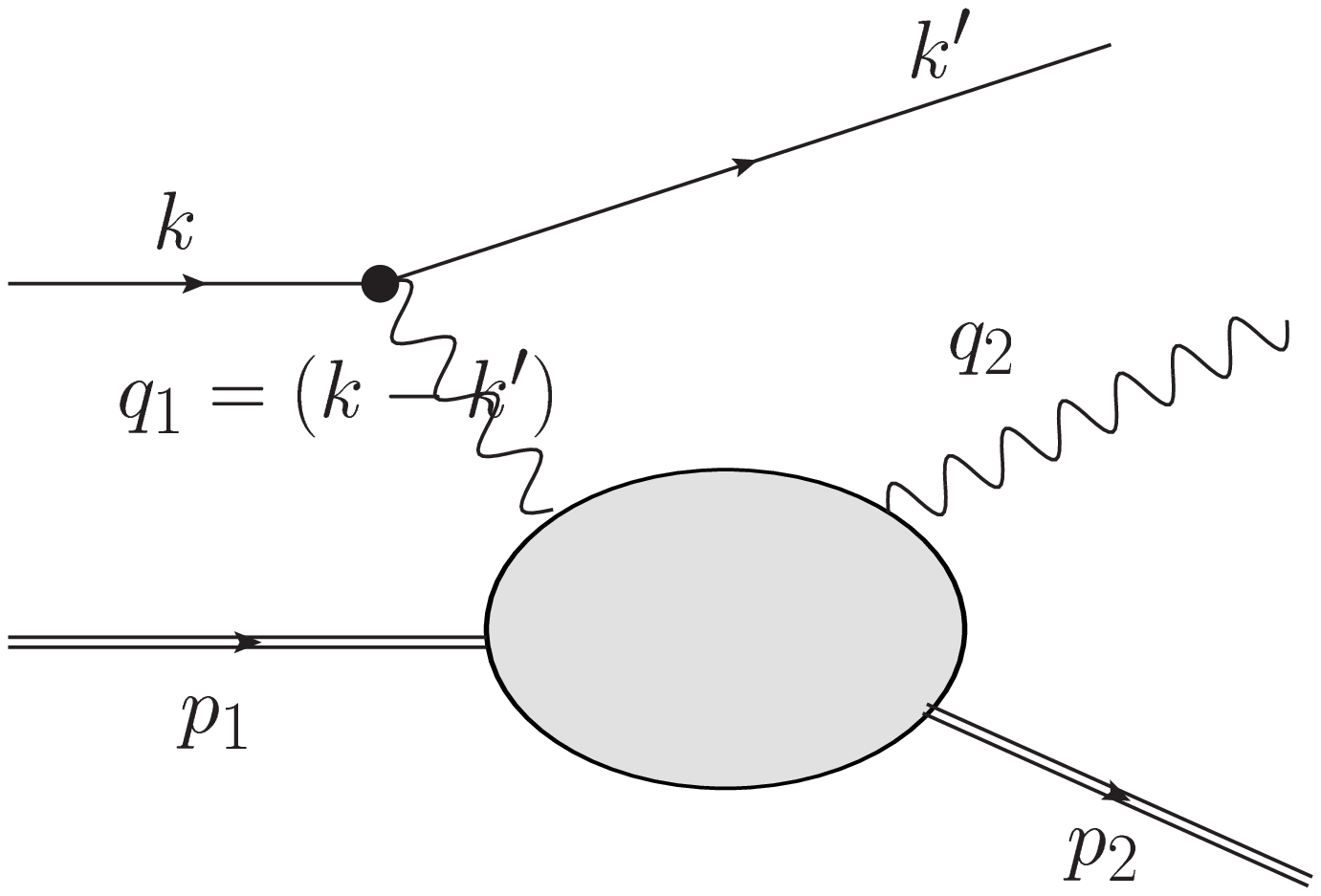} &
  \includegraphics[scale=0.4]{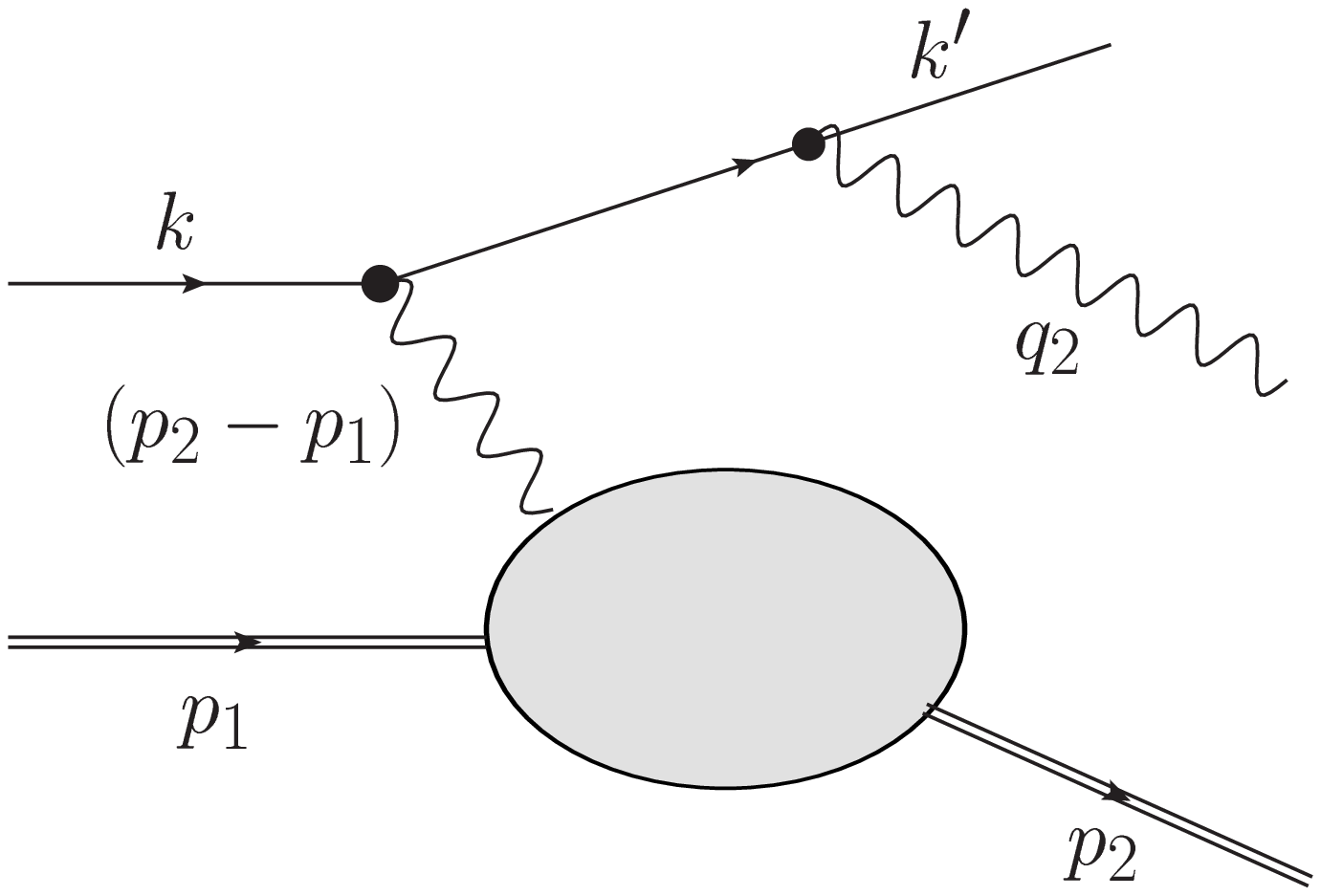} \\
 (a) DVCS & (b) Bethe--Heitler \\ 
\hline
  \includegraphics[scale=0.4]{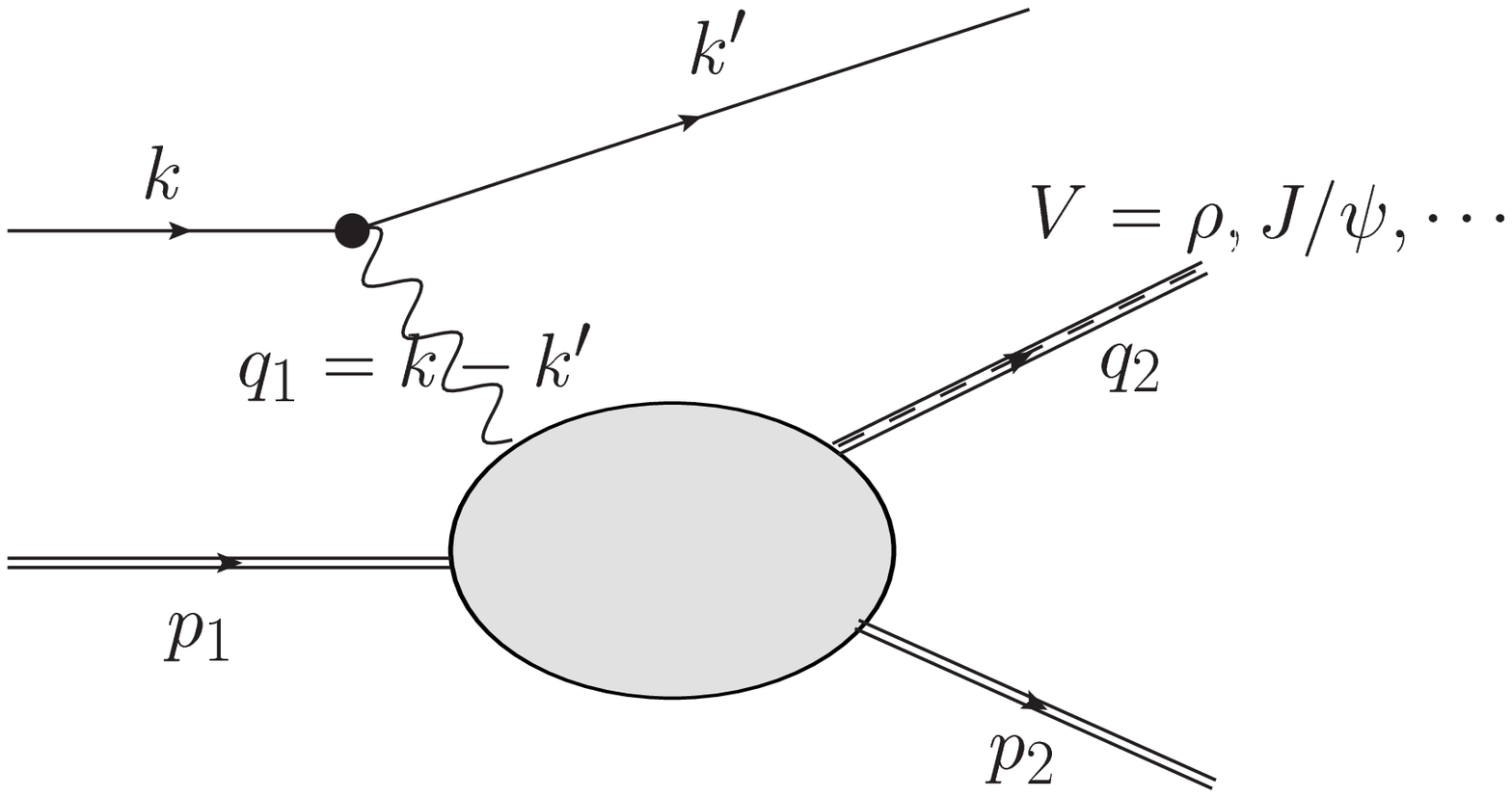} &
  \includegraphics[scale=0.4]{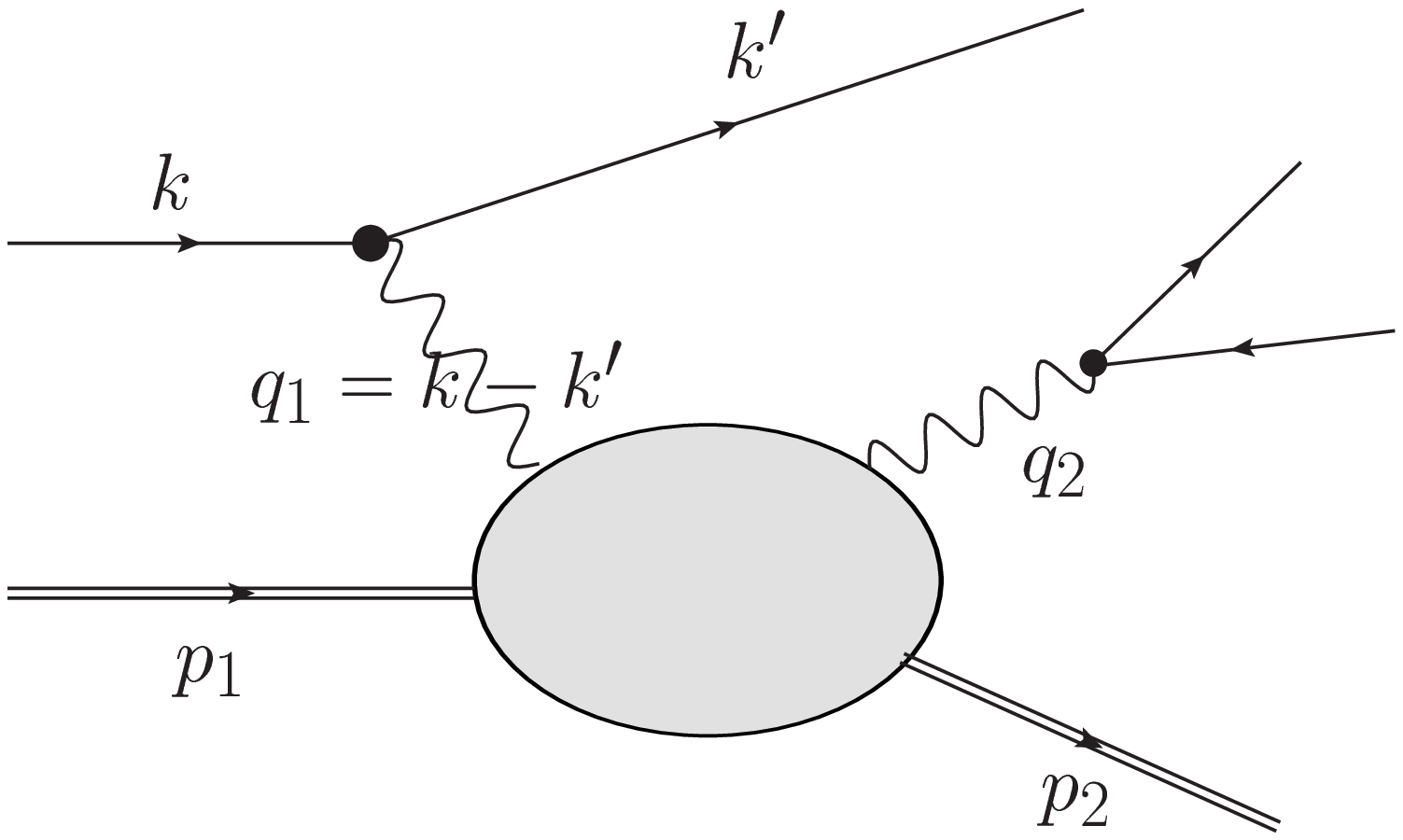}  \\
(c) VMP & (d) TCS 
\end{tabular}
  \caption{\label{fig:DVCS-TCS} 
(a, b) are diagrams contributing to the leptoproduction process of photon on 
a hadron, $\ell + h \longrightarrow \ell + \gamma + h$, 
(c) is the exclusive vector meson production process, and 
finally (d) the time-like Compton scattering process. 
}
\end{center}
\end{figure}

\begin{figure}[tbp]
 \begin{center}
  \includegraphics[scale=0.8]{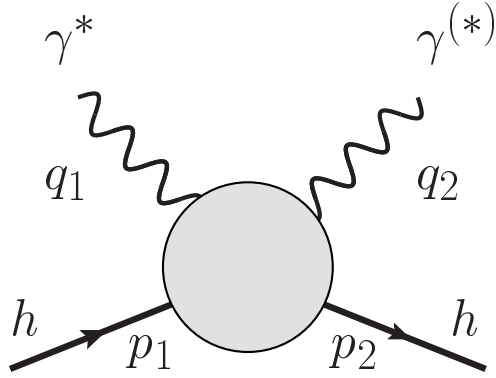} 
  \caption{\label{fig:DVCS-DDVCS} photon--hadron 2-body to 2-body
  scattering amplitude}
 \end{center}
\end{figure}

Deeply virtual Compton scattering $\gamma^{*}+h \rightarrow h+\gamma$ (DVCS), 
hard exclusive vector meson production $e+h \rightarrow e+h+V$ (VMP) and 
time-like Compton scattering processes $e+h \rightarrow e+h+e^+e^-$ (TCS)
are shown in Figure \ref{fig:DVCS-TCS} (a), (c) and (d), respectively. 
As a part of all these processes, the photon--hadron 2-body to 2-body
scattering amplitude,  
\begin{equation}
 {\cal M}(\gamma^\ast h \rightarrow \gamma^{(\ast)}h) = (\epsilon_1^\mu) T_{\mu\nu} (\epsilon_2^\nu)^\ast, 
\end{equation}
is involved.\footnote{There are two contributions from (a) the 
$\gamma^* +h \rightarrow\gamma+h$ virtual Compton scattering and (b) 
Bethe--Heitler process in the leptoproduction process of a photon 
on a target hadron $h$: $\ell + h \longrightarrow \ell + \gamma + h$, 
and they interfere. They can be separated experimentally, however, by 
exploiting kinematical dependence and polarization \cite{DVCS-vs-BH}. 
It thus makes sense to focus only on the amplitude (a).} 
This 2-body to 2-body scattering amplitude with this exclusive choice of the 
final states (Figure~\ref{fig:DVCS-DDVCS}) is truly non-perturbative
information, and this is the subject of this article. 
Because the ``final state'' photon is required to be on-shell 
$q_2^2 = 0$ in DVCS and time-like\footnote{We use the $(-,+++)$ metric 
throughout this paper. } $q_2^2 < 0$ in VMP and TCS, 
we are interested in developing a theoretical framework to calculate 
this non-perturbative amplitude in the case $q_2^2$ is different from 
space-like $q_1^2 > 0$.

Just like in QCD / hadron literature, we use the following 
notation for Lorentz invariant kinematical variables:
\begin{align}
\label{kinematics1}
  p^\mu &= \frac{1}{2}(p_1^\mu+p_2^\mu),& q^\mu &= \frac{1}{2}(q_1^\mu+q_2^\mu),& \Delta^\mu &=p_2^\mu -p_1^\mu = q_1^\mu - q_2^\mu,
\end{align}
\begin{align}
\label{kinematics2}
  x &= \frac{-q^2}{2p\cdot q},& \eta &= \frac{-\Delta \cdot q}{2 p\cdot
 q}, &s&= W^2 = -(p+q)^2, & t&=-\Delta^2.
\end{align}
$\eta$ is called skewness; in the scattering process of our interest, 
$q_1^2 = q^2 + \Delta^2/4 + q \cdot \Delta$ and 
$q_2^2 = q^2 + \Delta^2/4 - q \cdot \Delta$ are not the same, and 
hence the skewness does not vanish.
We will focus on high-energy scattering; for typical energy scale 
of hadron masses / confinement scale $\Lambda$, we assume that 
\begin{equation}
 \Lambda^2 \ll q_1^2, \; \; W^2, \quad {\rm while} \quad 
 |t| \lesssim {\cal O}(\Lambda).
\label{eq:Bjorken-2}
\end{equation}

The photon--hadron scattering amplitude (Figure~\ref{fig:DVCS-DDVCS}) 
in the real-world QCD (where all charged partons are fermions), the 
Compton tensor is given by the hadron matrix element with insertion 
of two QED currents $J^\mu$,
\begin{equation}
T^{\mu\nu} = i
 \int d^4 x e^{-iq \cdot x} 
   \langle {h(p_2)}|T \{ J^\nu(x/2) J^\mu(-x/2) \} |{h(p_1)}\rangle.
\label{eq:Compton tensor = <p|JJ|p>}
\end{equation}
For simplicity, we assume that the target hadron is a scalar, and 
further pay attention only to the structure function $V_1$ appearing 
in the gauge-invariant decomposition\footnote{
Here, we introduced a convenient notation 
\begin{align}
  P[q]_{\mu\nu}=\left[\eta_{\mu\nu}-\frac{q_\mu q_\nu}{q^2}\right].
\end{align}
} of the Compton tensor:
\begin{align}
   T^{\mu\nu}=&V_1 P[q_1]^{\mu\rho} P[q_2]^{\nu}_{\rho}
           +V_2 (p\cdot P[q_1])^\mu (p\cdot P[q_2])^\nu
           +V_3 (q_2\cdot P[q_1])^\mu (q_1\cdot P[q_2])^\nu
\notag \\ \label{eq:structure functions of Compton tensor}
           &+V_4 (p\cdot P[q_1])^\mu (q_1\cdot P[q_2])^\nu
           +V_5 (q_2\cdot P[q_1])^\mu (p\cdot P[q_2])^\nu
           +A \epsilon^{\mu\nu\rho\sigma}q_{1\rho}q_{2\sigma}.
\end{align}
Those structure functions, $V_{1,2,3,4,5}(x,\eta,t,q^2)$, should be 
expressed in terms of the kinematical variables $x, \eta$ and $t$, and 
one of our primary purposes of this article is to study how the structure 
functions depend on the skewness $\eta$.

\subsubsection{Light-cone Operator Product Expansion}

The light-cone operator product expansion (OPE)
can be applied to the product of currents 
$T\left\{ J^\nu  J^\mu \right\}$, before evaluating it as a hadron 
matrix element. Let the expansion be 
\begin{align}
 i  \int d^4 x e^{-i q \cdot x} T \{ J^\nu(x/2) J^\mu(-x/2) \}
= \sum_I {\cal C}_{I\: \rho_1 \dots \rho_{j_I}}^{\mu\nu}(q) 
    {\cal O}_I^{\rho_1\dots \rho_{j_I}}(0;q^2)
\label{eq:OPE of two currents}
\end{align}
for some basis of local operators 
${\cal O}_{I}^{\rho_1 \cdots \rho_{j_I}}$ renormalized at $\mu^2 = q^2$.
$C^{\mu\nu}_{I\; \rho_1 \cdots \rho_{j_I}}$'s are the corresponding Wilson 
coefficients renormalized at $\mu^2 = q^2$. If we were to evaluate these 
local operators on the right-hand side with the same state for both bra and
ket, $\bra{h(p_2)}$ and $\ket{h(p_1)}$ with $p_2^\mu=p_1^\mu$, then 
the Compton tensor and its structure functions do not receive non-zero 
contributions from local operators that are given by total derivative of 
some other local operators. In the case of our interest, however, 
such operators do contribute. 

Let us take a series of operators in QCD that are called twist-2
operators in the weak coupling limit. The twist-2 operators in the 
flavor non-singlet sector are labeled by two integers, $j, l$,
\begin{equation}
 {\cal O}^\alpha_{j,l}:=
   \left[(-i)^{j+l-1} \partial^{\mu_{j+1}} \cdots \partial^{\mu_{j+l}} 
    \bar{\Psi}_a \gamma^{\mu_1}
           \left(\overleftrightarrow{D}\right)^{\mu_2} \cdots 
           \left(\overleftrightarrow{D}\right)^{\mu_j} \; \lambda^\alpha_{ab} 
    \Psi_b\right]_{{\rm t.s.t.l.}}(0;q^2),
\label{eq:fl-nonsgl-twist2-op}
\end{equation}
with an $N_F \times N_F$ flavor matrix $(\lambda^\alpha)_{ab}$. Similarly,  
in the flavor-singlet sector, 
there are two series of twist-2 operators with the label $j, l$, given by 
quark bilinear and gluon bilinear. Here, these operators are made 
totally symmetric and traceless (t.s.t.l) in the $j+l$ Lorentz indices 
so that they transform in irreducible representations of the Lorentz
group $\SO(3,1)$. 
$\overleftrightarrow{D} := \overrightarrow{D}-\overleftarrow{D}$. 

Suppose that the hadron matrix element of the operator ${\cal
O}^\alpha_{j,l}$ is given by 
\begin{equation}
 \bra{h(p_2)} {\cal O}^\alpha_{j,l} \ket{h(p_1)} = 
   \sum_{k=0}^j \left[ \Delta^{\mu_1} \cdots \Delta^{\mu_{k+l}}
       p^{\mu_{k+l+1}} \cdots p^{\mu_{j+l}}\right]_{{\rm t.s.t.l.}} 
  A_{j,k}(t; q^2) (-2)^{j-k};
\end{equation}
the reduced matrix element $A^\alpha_{j,k}(t)$ is non-perturbative
information and cannot be determined by perturbative QCD.
If we pay attention only to Wilson coefficients 
$C^{\mu\nu}_{j,l,\alpha; \mu_1 \cdots \mu_{j+l}}$ that are proportional
to $\eta^{\mu\nu}$, and are to write them as\footnote{
In the leading order of QCD perturbation, 
${\cal C}^\alpha_{j,0} = -[1+(-1)^j]$ for $j = 2, 4, \cdots$ and 
$(\lambda^\alpha)_{ab} = [\diag(4/9,1/9,1/9)]_{\rm t.l.}$.} 
\begin{equation}
 \eta^{\mu\nu} {\cal C}^\alpha_{j,l} 
   \frac{q_{\rho_1} \cdots q_{\rho_{j+l}} }{(q^2)^{j+l}}, 
\end{equation}
then the twist-2 flavor non-singlet contribution to the structure
function $V_1$ becomes
\begin{equation}
 V_1 \simeq \sum_{j,l} {\cal C}^\alpha_{j,l} \frac{1}{x^{j+l}} 
   \sum_{k=1}^{j}A^\alpha_{j,k}(t;q^2) \eta^{k+l} =: 
  \sum_j {\cal C}^\alpha_{j} \left( \vartheta \right) \frac{1}{x^j}  
    A^\alpha_{j}(\eta, t; q^2),
\end{equation}
where $\vartheta := (\eta/x)$, 
${\cal C}^\alpha_j(\vartheta) := \sum_{l=0}^{\infty}  
{\cal C}^\alpha_{j,l} \vartheta^l$, and 
$A^\alpha_j(\eta,t) := \sum_{k=0}^j \eta^k A^\alpha_{j,k}(t)$.
If the structure function $V_1$ receives contributions only from 
even $j \in \Z$, then this $j$-summation is rewritten as 
\begin{equation}
 V_1(x,\eta,t;q^2) \simeq - \int \frac{dj}{4i}
    \frac{1+e^{-\pi i j}}{\sin(\pi j)} {\cal C}^\alpha_j (\vartheta)
    \frac{1}{x^j} A^\alpha_j (\eta,t;q^2)
\end{equation}
in the form of inverse Mellin transformation; here, 
${\cal C}^\alpha_j(\vartheta; q^2)$ and $A^\alpha_j(\eta,t;q^2)$ are now 
meant to be holomorphic functions on $j$ (possibly with some poles and
cuts) that coincide with the original ones at $j \in 2 \Z$.
Precisely the same story holds true also for flavor-singlet sector.

Because the structure function is given by the inverse Mellin transform 
of a product of three factors, namely, (a) the signature factor 
$\mp [1\pm e^{-\pi i j}]/\sin (\pi j)$, (b) 
Wilson coefficients ${\cal C}^\alpha_j$ and (c) hadron matrix elements 
$A^\alpha_j$, it can be regarded as a convolution of inverse Mellin
transforms of those three factors. The inverse Mellin transform 
of the signature factor becomes 
\begin{equation}
 \int \frac{dj}{2\pi i} \frac{1}{x^j}
   \frac{\pi}{2} \frac{\mp[1 \pm e^{-\pi i j}]}{\sin (\pi j)} = \frac{-1}{2} 
   \left[ \frac{1}{1-x+i\epsilon} \pm \frac{1}{1+x } \right] ,
\end{equation} 
which corresponds to propagation of the parton in perturbative
calculation \cite{Ji}, and the inverse Mellin transform of the matrix
element is called the generalized parton distribution:
\begin{equation}
 H^\alpha(x,\eta,t;\mu^2 = q^2) = \int \frac{dj}{2\pi i} \frac{1}{x^j} 
  A^\alpha_j(\eta,t;\mu^2 = q^2).
\label{eq:GPD-def-natural}
\end{equation}

Generalized parton distribution (GPD) $H^\alpha(x,\eta,t;\mu^2)$ of a
hadron $h$ is a non-perturbative information, just like the ordinary
PDF, which is obtained by simply setting $\eta=0$ and $t=0$.
For phenomenological fit of experimental data of DVCS and VMP, 
some function form of the GPD needs to be assumed, because of the 
convolution involved in the scattering amplitude \cite{DVCS-GPD-dispersion}. 
Setting up a model (and assuming a function form) for the 
non-perturbative information in terms of $A_j(\eta, t;q^2)$ rather than 
the GPD itself $H(x,\eta,t;q^2)$ is called dual 
parameterization \cite{dual-para, dual-para-2, Mueller-S-05, KK-Mueller-07}, 
and some phenomenological ans\"{a}tze have been proposed.
In this article, we aim at deriving qualitative form of $A_{j}(\eta, t)$ 
by using gravitational dual (that is analytic in $j$), instead of 
assuming the form of $A_j(\eta,t)$ by hand.

\subsubsection{Renormalization and OPE in dilatation eigenbasis}

Remembering that the distinction between the $\gamma^*+h\rightarrow \gamma+h$ 
scattering amplitude and GPD originates from the factorization 
into the Wilson coefficients and local operators (and their matrix
elements), one will notice that the GPD defined in this way should 
depend on the choice of the basis of local operators.
Although the choice of operators ${\cal O}^\alpha_{j,l}$ with $j \geq 1$
and $l \geq 0$ in (\ref{eq:fl-nonsgl-twist2-op}) appears to be the most 
natural (and intuitive) one for the twist-2 operators in the flavor non-singlet
sector, there is nothing wrong to take a different linear combinations 
of these operators as a basis, when the corresponding Wilson
coefficients also become linear combinations of what they are for 
${\cal O}^\alpha_{j,l}$. Given the fact that the operators 
${\cal O}^\alpha_{j,l}$ mix with one another under renormalization,  
it should not be compulsory for us to stick to the basis ${\cal O}^\alpha_{j,l}$.

Under the perturbation of QCD, the flavor non-singlet twist-2 operators 
are renormalized under 
\begin{align}
  \mu\frac{\partial}{\partial \mu}\:
 [{\cal O}_{j-m,m}(0;\mu^2)] =
 -[\gamma^{(j)}]_{mm'} \:
  [{\cal O}_{j-m',m'}(0;\mu^2)];
\label{eq:q-evolution in nozero MT}
\end{align}
because operators can mix only with those with the same number of
Lorentz indices, the anomalous dimension matrix $[ \gamma ]$ is block diagonal 
in the basis of ${\cal O}^\alpha_{j,l}$; the $j\times j$ matrix 
for the operators ${\cal O}^\alpha_{j-m,m}$ ($m=0,\cdots,j-1$) is denoted
by $[\gamma^{(j)}]$. This matrix is upper triangular in this basis, 
and the diagonal entries are given by the anomalous dimensions  
of the twist-2 spin-$j$ operators without a total derivative:
\begin{equation}
 \left[\gamma^{(j)}\right]_{mm} = \gamma(j-m).
\end{equation}
Therefore, the eigenvalues of the anomalous dimension matrix is 
$\left\{ \gamma(j-m) \right\}_{m=0,\cdots,j-1}$ in this diagonal block, 
and the corresponding operator $\overline{{\cal O}}^\alpha_{j-m-1,m}$ 
is a linear combination of operators ${\cal O}_{j-m',m'}$ with 
$m'=m ,\cdots, j-1$ \cite{ER}. The corresponding Wilson coefficient 
$\overline{C}^\alpha_{j-m-1,m}$ for such an operator is a linear combination of 
${\cal C}^\alpha_{j-m',m'}$ with $m'=m,\cdots,0$.
In this operator basis, matrix elements and Wilson coefficients
renormalize multiplicatively, without mixing.\footnote{In reality, 
the anomalous dimension matrix depends on the coupling constant
$\alpha_s$, and $\alpha_s$ changes over the scale. Thus, the
eigenoperator of the renormalization / dilatation also changes over the 
scale. In scale invariant theories (and in theories only with slow running 
in $\alpha_s$), however, this multiplicative renormalization is exact or 
a good approximation. (c.f. \cite{BM-conf-viol})} 

In this new basis of local operators, the structure function becomes 
\begin{equation}
 V_1 \simeq \sum_{n,K} \overline{\cal C}^\alpha_{n,K} \frac{1}{x^{n+1+K}} 
   \sum_k \overline{A}^\alpha_{n+1,k}(t;\mu^2) \eta^{K+k} =:
   \sum_n \overline{C}^\alpha_n (\vartheta) \frac{1}{x^{n+1}}
          \overline{A}^\alpha_{n+1}(\eta, t;\mu^2), 
\end{equation}
where 
\begin{equation}
 \overline{C}^\alpha_n (\vartheta) = \sum_{K=0}^\infty \overline{C}^\alpha_{n,K}
  \vartheta^K, 
\end{equation}
and $\overline{A}^\alpha_{n+1,k}(t;\mu^2)$ is the reduced matrix element of 
the operator\footnote{Just like 
${\cal O}_{j,l} = (-i \partial)^l {\cal O}_{j,0}$, there is a relation 
$\overline{\cal O}_{n,K} = (-i \partial)^K \overline{\cal O}_{n,0}$ in the new
basis. This is why all the hadron matrix elements of 
$\overline{\cal O}_{n,K}$ can be parameterized by $\overline{A}_{n+1,k}$, just
like those of ${\cal O}_{j,l}$ are by $A_{j,k}$. Here, $n$ corresponds to 
the conformal spin, which is sometimes denoted by $j$ in the literature.
In this article, however, we maintain $j=n+1$. } 
$\overline{\cal O}^\alpha_{n,0}(0;\mu^2)$.
The structure function is therefore written as yet another inverse
Mellin transform
\begin{equation}
 V_1 \simeq - \int \frac{dj}{4i} \frac{1+e^{-\pi i j}}{\sin (\pi j)} 
    \overline{C}^\alpha_{j-1}(\vartheta) \frac{1}{x^j}
    \overline{A}^\alpha_{j}(\eta, t; \mu^2).
\label{eq:V1-inv-Mellin-conformal-QCD}
\end{equation}
Yet another GPD can also be defined by using
$\overline{A}^\alpha$, instead of $A^\alpha_j(\eta, t;q^2)$:
\begin{equation}
 \overline{H}^\alpha(x,\eta,t;\mu^2) = \int \frac{dj}{2\pi i} \frac{1}{x^j} 
    \overline{A}^\alpha_{j}(\eta, t; \mu^2).
\end{equation}
When it comes to the description of the 
$\gamma^* + h \rightarrow \gamma+h$ scattering amplitude as a whole, 
it does not matter which operator basis is used. Although we need GPD 
rather than the scattering amplitude in order to talk about the 
distribution of partons in the transverse directions in a hadron, yet 
we only need GPD at $\eta = 0$. Thus, the newly defined GPD 
$\overline{H}$ does just as good a job as $H$ defined 
in (\ref{eq:GPD-def-natural}); they are the same at $\eta = 0$.

Even within the dual parameterization approach, it has been advantageous 
to use this operator basis, because it becomes much easier to implement 
a phenomenological assumption (function form) of 
$\overline{A}^\alpha_j(\eta,, t; \mu^2)$ that is consistent with
renormalization group flow \cite{dual-para}.

\subsubsection{Conformal OPE}

Although the hadron matrix element is essentially non-perturbative, 
and is not calculable within perturbative QCD, more discussion has been 
made on the Wilson coefficients $\overline{C}^\alpha_{n,K}$. They still have 
to be calculated order by order in perturbation theory, if one is 
interested strictly in the QCD of the real world. If one is interested
in gauge theories that are more or less ``similar'' to QCD, however, 
stronger statements can be made for a system with higher symmetry: conformal
symmetry. One can think of ${\cal N}=4$ super Yang--Mills theory or 
${\cal N} = 1$ supersymmetric $\SU(N) \times \SU(N)$ gauge theory of 
\cite{Klebanov-W} as an example of theories with exact (super)
conformal symmetry. The QED probe in the real world QCD can be replaced 
by gauging global symmetries (such as (a part of) $\SU(4)$ R-symmetry 
of ${\cal N}=4$ super Yang--Mills theory and 
$\SU(2) \times \SU(2) \times \U(1)$ symmetry of \cite{Klebanov-W}).
By applying the conformal symmetry, one can derive stronger statements 
on the Wilson coefficients of primary operators appearing in the OPE.

Suppose that we are interested in the OPE of two primary operators, 
$A$ and $B$, that are both scalar under $\SO(3,1)$. If we take 
the basis of local operators for the expansion to be primary operators 
$\overline{\cal O}_n$ (with $j_n$ Lorentz indices and $l_n$ scaling dimension) 
and their descendants $\partial^K \overline{\cal O}_n$ (with $j_n + K$ 
Lorentz indices), then in the OPE, 
\begin{equation}
 T\left\{ A(x) B(0) \right\} = \sum_n 
   \left(\frac{1}{x^2}\right)^{\frac{1}{2}(l_A + l_B - l_n + j_n)} 
 \sum_{K=0}^{\infty} c_{n,K}
    \frac{ x^{\rho_1} \cdots x^{\rho_{j_n+K}} }{ (x^2)^{j_n+K} } 
    [\partial^K \overline{\cal O}_n(0)]_{\rho_1 \cdots \rho_{j_n+K}}, 
\end{equation}
the conformal symmetry determines all the coefficients of the 
descendants $c_{n,K}$ ($K \geq 1$) in terms of that of the primary
operator, $c_{n,0} =: c_n$.
Ignoring the mixture of non-traceless contributions, one finds that 
\cite{FGG-confl-HG}
\begin{equation}
 T \left\{ A(x) B(0) \right\} \simeq \sum_n
   \left( \frac{1}{x^2} \right)^{\frac{1}{2}(l_A + l_B - l_n + j_n)}
    \!\!\!\!\!\!\!\!\! \!\!\!\!\!\!\!\!\! \!\!\!\!\!\!\!\!\! 
    x^{\rho_1} \cdots x^{\rho_{j_n}} c_n \; 
    {}_1 F_1\left( \frac{l_A-l_B+l_n+j_n}{2} , l_n + j_n; 
                  x \cdot \partial \right) 
  [\overline{\cal O}_n(0)]_{\rho_1 \cdots \rho_{j_n}}.
\end{equation}

Questions of real interest to us is the OPE of conserved currents $J^\nu$ 
and $J^\mu$. They are not scalars of $\SO(3,1)$, 
but the same logic as in \cite{FGG-confl-HG} can be used also to show that, 
in the terms with Wilson coefficients proportional to $\eta^{\mu\nu}$, 
\begin{equation}
 T\left\{ J^\nu(x) J^\mu(0) \right\} \simeq \eta^{\mu\nu} \sum_n
   \left( \frac{1}{x^2} \right)^{3-\frac{\tau_n}{2}}
    \!\!\!\!\!\!\!\!\! \!\!
    x^{\rho_1} \cdots x^{\rho_{j_n}} c_n \; 
    {}_1 F_1\left( \frac{l_n+j_n}{2} , l_n + j_n; x \cdot \partial \right) 
  [\overline{\cal O}_n(0)]_{\rho_1 \cdots \rho_{j_n}} + \cdots,
\end{equation}
where $\tau_n := l_n - j_n$ is the twist, mixture of the non-traceless 
(and hence higher twist) contributions are ignored, and terms with 
Wilson coefficients without $\eta^{\mu\nu}$ are all omitted here. 
The scaling dimension of conserved currents $l_A = l_B = 3$ have been
used. The momentum space version of the OPE is \cite{Mueller-PRD98}
\begin{eqnarray}
 i \int d^4 x \; e^{- iq_2 \cdot x} \;  T\left\{ J^\nu(x) J^\mu(0)
					 \right\}
 & \simeq & \eta^{\mu\nu} \sum_n 
  \frac{(2\pi)^2 \Gamma\left(\frac{l_n+j_n-2}{2} \right)}
       {4^{2-\frac{\tau_n}{2}} \Gamma\left(3-\frac{\tau_n}{2} \right)}
  c_n \; 
  \frac{(-2i)^{j_n} q_2^{\rho_1} \cdots q_2^{\rho_{j_n}}} 
       {(q^2_2)^{\frac{\tau_n}{2}-1} (q^2_2)^{j_n}}
    \nonumber  \\
 && \!\!\!\!\!\!\!\!\!\!\!\!\!\!\!\!\!\!\!\!\!\!\!\!\!\!
 {}_2 F_1 \left( \frac{l_n + j_n}{2}, \frac{l_n+j_n}{2} -1, l_n + j_n; 
    \frac{- 2i q_2 \cdot \partial}{q_2^2} \right) \overline{\cal O}_n(0) + \cdots,
 \label{eq:conf-OPE-2side}
\end{eqnarray}
or equivalently \cite{KK-Mueller-07}, 
\begin{eqnarray}
 i \int d^4(x-y) e^{-i q \cdot (x-y)} \; T\left\{ J^\nu(x) J^{\mu}(y) \right\} 
  & \simeq & \eta^{\mu \nu} \sum_n 
  \frac{(2\pi)^2 \Gamma\left(\frac{l_n+j_n-2}{2} \right)}
       {4^{2-\frac{\tau_n}{2}} \Gamma\left(3-\frac{\tau_n}{2} \right)}
  c_n \; 
  \frac{(-2i)^{j_n} q^{\rho_1} \cdots q^{\rho_{j_n}}} 
       { (q^2)^{\frac{\tau_n}{2}-1}  (q^2)^{j_n}}
  \nonumber \\
 && \!\!\!\!\!\!\!\!\!\!\!\!\!\!\!\!\!\!\!\!\!\!\!\!\!\!\!\!\!\!\!\!\!\!\!\!\!\!\!\!\!\!\!\!\!\!\!\!\!\!\!\!\!\!\!\!\!\!\!\!\!\!\!
 {}_2 F_1 \left( \frac{l_n + j_n-2}{4}, \frac{l_n+j_n}{4}, \frac{l_n + j_n}{2}; 
    \left(\frac{i q \cdot \partial}{q^2}\right)^2 \right)
   \overline{\cal O}_n \left(\frac{x+y}{2} \right)
   + \cdots.
\label{eq:conf-OPE-center}
\end{eqnarray}
Either in the form of (\ref{eq:conf-OPE-2side}) or
(\ref{eq:conf-OPE-center}), the primary operators $\overline{\cal O}_n$
and corresponding coefficients $c_n$ are renormalized multiplicatively. 

\subsection{AdS/CFT Approach}
\label{ssec:idea-simply}

In AdS/CFT correspondence, Type IIB string theory on AdS$_5 \times W$ with 
a 5-dimensional Einstein manifold $W$ corresponds to a gauge theory on 
$\R^{3,1}$ with an exact conformal symmetry; theories with an exact conformal 
symmetry, however, are qualitatively different from the QCD in the real world. 
But the Type IIB string on a geometry that is close to AdS$_5 \times W$, but 
with confining end in the infrared, may be used to extract qualitative 
lesson on strongly coupled gauge theories with confinement, which are not 
qualitatively different from the QCD.

In a dual pair of a CFT and a string theory on a background AdS$_5 \times W$,
primary operators of the CFT are in one to one correspondence with string 
states on AdS$_5$, and their correlation functions can be 
calculated by using the wavefunctions of the string states on AdS$_5$.
When the background geometry is changed from AdS$_5 \times W$ to 
some warped geometry that is nearly AdS$_5$ with an end in the infrared, 
then the wavefunctions might be used to calculate matrix elements of 
the corresponding ``primary'' operators in an almost conformal theory. 
The correspondence between the operators and string states can be made 
precise, because they are both classified in terms of representation 
of the conformal algebra, which is shared by both of the dual theories. 

In order to determine GPD $\overline{H}$ in gravitational dual descriptions, 
it is therefore sufficient to determine wavefunctions of string states 
corresponding to the ``primary'' operators of interest. 
Although there are plenty of literature discussing the correspondence between 
the (superconformal) primary operators and string states at the supergravity 
level, it is known that the flavor-singlet twist-2 operators (labeled by 
the spin $j$) correspond to the stringy excitations with arbitrary high spin 
$j$ that are in the same trajectory as graviton \cite{GKP-02, BPST-06}. 
Our task is therefore to determine the wavefunctions of such string states. 
Needless to say, one has to fix all the gauge degrees of freedom associated 
with string component fields (not just the general coordinate invariance 
associated with the graviton) before working out the mode decomposition. 
Furthermore, wavefunctions need to be grouped together properly so that 
they form an irreducible representation of the conformal group, in order 
to establish correspondence with a primary operator of the gauge theory 
side, which also forms an irreducible representation of the conformal 
group along with its descendants.  

It will be clear by the end of this article that all of such 
technical works is necessary and essential for the purpose of 
extracting skewness dependence of GPD. 

There are two different (but equivalent) ways to study the DVCS 
$\gamma^* + h \rightarrow \gamma^{(*)} + h$ amplitude and GPD in gravitational 
dual descriptions.
One is to determine the hadron matrix elements of spin-$j$ primary 
operators by using appropriate wavefunctions; GPD $\bar{H}$ is obtained 
by the inverse Mellin transform of the matrix elements. Using the Wilson 
coefficients that are governed by the conformal symmetry 
(see (\ref{eq:conf-OPE-center})), the DVCS amplitude will also be obtained. 
Conversely, the other way is to calculate disc/sphere amplitude directly, 
with the vertex operators given (approximately) by using the wavefunctions 
associated with the target hadron (see sections \ref{sec:set-up} 
and \ref{sec:cubic-sft}). 
We will identify the structure of conformal OPE in the expression for the 
$\gamma^*+h \rightarrow \gamma^{(*)}+h$ scattering amplitude in gravity dual 
(see (\ref{eq:V1-m=0-j}, \ref{eq:V1-m=0}, \ref{eq:V1-m})), 
with the Wilson coefficient for the ``twist-2'' operators precisely as 
predicted by conformal symmetry (\ref{eq:conf-OPE-center}). That makes it 
also possible to read out hadron matrix elements, and to extract the GPD.  
In these approaches, one can hope to work also for higher twist 
contributions, in principle, but we are not ambitious enough to 
do that in this article.  
In this article, we will proceed in the latter approach.

\section{Gravity Dual Settings}
\label{sec:set-up}

A number of warped solutions to the Type IIB string theory has been 
constructed, and they are believed to be dual to some strongly coupled 
gauge theories. When the geometry is close to AdS$_5 \times W$ with 
some 5-dimensional Einstein manifold $W$, with weak running of the 
AdS radius along the holographic radius, the corresponding gauge 
theory will also have approximate conformal symmetry, and 
the gauge coupling constant runs slowly. If the ``AdS$_5 \times W$'' 
geometry has a smooth end at the infra red as in
\cite{Klebanov-Strassler}, then the dual gauge theory will end up 
with confinement. Gravitational backgrounds in the Type IIB string
theory with the properties we stated above all provide a decent framework 
of studying qualitative aspects of non-perturbative information
associated with gluons/Yang--Mills theory on 3+1 dimensions. 

In studying the $h+\gamma^{*} \rightarrow h + \gamma$ scattering
process in gravitational dual, we need a global symmetry to be gauged 
weakly, just like QED for QCD. In Type IIB D-brane constructions of gauge
theories that have gravity dual, U(1) subgroups of an R-symmetry 
or a flavor symmetry on D7-branes can be used as the models of the 
electromagnetic U(1) symmetry. Therefore, we have in mind 
gravity dual models on a background that is approximately 
``AdS$_5 \times W$'' with a non-trivial isometry group on $W$, or 
with a D7-brane configuration on it, as in \cite{PS-02-DIS}.

Our interest, however, is not so much in writing down an exact 
mathematical expression based on a particular gravity dual model 
that is equivalent to a particular strongly coupled gauge theory, 
but more in extracting qualitative information of partons in hadrons 
of confining gauge theories in general. It is therefore more suitable 
for this purpose to use a simplified set-up that carries common (and essential)
features of the Type IIB models that we described above. 
Throughout this article, we assume pure AdS$_5 \times W$ metric background, 
\begin{align}
 \label{eq:AdS5timesS5}
  ds^2  =G_{MN}dx^M dx^N&=g_{mn}dx^m dx^n +R^2 (g_{W})_{ab}d\theta^a d\theta^b,\\
 \label{eq:hwb}
  g_{mn}dx^m dx^n&=e^{2A(z)}(\eta_{\mu\nu}dx^\mu dx^\nu+dz^2), \;\;\;
 e^{2A(z)}=\frac{R^2}{z^2};  
\end{align}
that is, we ignore the running effect, and we do not specify the 
5-dimensional manifold $W$.
The dilaton vev is simply assumed to be constant, $e^\phi = g_s$. 
Confining effect---the infra-red end of this geometry---can be 
introduced, for example, by sharply cutting off the AdS$_5$ space at 
$z = \Lambda^{-1}$ (hard wall models), or by similar alternatives 
(soft wall models). We are not committed to a particular implementation 
of the infra-red cut-off in this article (see discussion in 
section \ref{ssec:confine}), except in a couple of places 
where we write down some concrete expressions for illustrative purposes 
(sections \ref{ssec:D=0-cancellation} and \ref{ssec:numerical}).
The energy scale $\Lambda$ associated with (any form of implementation of) 
the infra-red cut-off corresponds to the confining energy scale in the 
dual gauge theories. 
When we consider (simplified version of the) models with D7-branes for 
flavor, we assume that the D7-brane worldvolume wraps on a 3-cycle on $W$, 
and extends all the way down to the infra-red end of the holographic radius 
$z$; i.e., all of $0 \leq z \leq \Lambda^{-1}$. This corresponds to 
assuming massless quarks. In this article, we will not pay attention to 
physics where spontaneous chiral symmetry breaking is essential. 

As we stated earlier, we would like to work out the 
$h + \gamma^{*} \rightarrow h+\gamma^{(*)}$ scattering amplitude by 
using the gravity dual models. This is done by summing up 
sphere / disc amplitudes, along with those with higher genus worldsheets. 
We will restrict our attention to kinematical regions where saturation 
is not important (i.e., large $q^2$ and/or not too small $x$, and 
large $N_c$).  That allows us to focus only on sphere / disc amplitudes, 
with insertion of four vertex operators corresponding to the incoming
and outgoing hadron $h$ and (possibly virtual) photon $\gamma$.

As a string-based model of the target hadron $h$ (that is $\SO(3,1)$ 
scalar), we have in mind either a scalar ``glueball''\footnote{By
``glueball'', we only mean a bound state of fields in super Yang--Mills
theory. } that has non-trivial R-charge, or a scalar meson made of 
matter fields. The former corresponds to a vertex operator 
(in the $(-1,-1)$ picture)
\begin{equation}
 V(p) = : e^{i p_{\mu} \cdot X^{\hat{\mu}} }
    \psi^m \tilde{\psi}^n g_{mn} \Phi(Z; m_n) Y(\Theta):,
\end{equation} 
where $Y(\Theta)$ is a ``spherical harmonics'' on $W$, and 
the latter to 
\begin{equation}
 V(p) = : e^{i p_\mu \cdot X^{\hat{\mu}} }  \psi \Phi(Z; m_n):,
\label{eq:vx-op-D7-scalar}
\end{equation}
where $\psi$ corresponds to the D7-brane fluctuations in its transverse
directions. $\Phi(Z)$ is the wavefunction on AdS$_5$, with the argument 
promoted to the field on the world sheet \cite{BPST-06}. Vertex operators above 
are approximate expressions in the $(\alpha'/R^2) \sim 1/\sqrt{\lambda}$
expansion (e.g., \cite{RW}) in a theory formulated with a non-linear sigma 
model given by (\ref{eq:AdS5timesS5}). If we are to employ the hard-wall 
implementation of the infra-red boundary, with the AdS$_5$ metric in the bulk 
without modification, then the wavefunction $\Phi(Z;m_n)$ is of the form
\begin{equation}
 \sqrt{t_h}\Phi(z;m_n) = 2\Lambda z^2
   \frac{J_{\Delta-2}(j_{\Delta-2,n}\Lambda z)}
        {\left| J'_{\Delta-2}(j_{\Delta-2,n}) \right|}.
\label{eq:hadron-wvfc}
\end{equation}
This wavefunction is that of the $n$-th lightest hadron corresponding 
to some scalar operator with conformal dimension $\Delta = \Delta_\phi$;
the hadron mass $m_n = j_{\Delta-2,n}\Lambda$ is given by the $n$-th
zero of the Bessel function $J_{\Delta -2}$. We will comment on 
the normalization factor $\sqrt{t_h}$ in later sections, though 
it disappears from the expression for physical observables. 

The ``photon'' current in the correlation function/matrix element 
$T^{\nu\mu}$ in the gauge theory description corresponds to insertion of 
vertex operators associated with non-normalizable wavefunctions,
rather than with the normalizable wavefunctions (\ref{eq:hadron-wvfc}) 
for the target hadron state. If we are to employ an R-symmetry current 
as the string-based model of the QED current, then the corresponding 
closed string vertex operator is 
\begin{equation}
 V(q) = : e^{i q_\mu \cdot X^{\hat{\mu}} } v_a(\Theta) A_m(Z; q)
     (\psi^a \tilde{\psi}^m + \psi^m \tilde{\psi}^a):,
 \label{eq:vx-op-Killing-vector}
\end{equation}
with some Killing vector $v_a \partial/\partial \theta^a$ on $W$. The
vertex operator in the case of D7-brane U(1) current is 
\begin{equation}
 V(q) = : e^{i q_\mu \cdot X^{\hat{\mu}} } A_m(Z; q) \psi^m :.
\label{eq:vx-op-D7-vector}
\end{equation}
The wavefunction $A_m(Z; q)$ on AdS$_5$ is of the form 
\begin{align}
A_{\mu}(z; q)  = &
  \left[ \delta_\mu^{\; \hat{\kappa}}
       - \frac{q_\mu q^{\hat{\kappa}} }{q^2} \right]
 \epsilon_\kappa(q) (qz) K_1(qz) + &
 q_\mu \frac{q^{\hat{\kappa}} \epsilon_\kappa(q)}{2 q^2}
 (qz)^2 K_2(qz), \label{eq:photon-wvfc-pre} \\
A_z (z; q)  = & &
 \!\!\! -i \partial_z  \frac{q^{\hat{\kappa}} \epsilon_\kappa(q)}{2 q^2}
 (qz)^2 K_2(qz).
\label{eq:photon-wvfc}
\end{align}
Rationale for our choice of the terms proportional to 
$(q \cdot \epsilon)$ will be explained later on in 
the appendix \ref{ssec:vector}, but those terms should 
not be relevant in the final result, because of the gauge invariance 
of $T^{\nu\mu}$. When the infra-red boundary is implemented by the hard
wall, $K_1(qz)$ should be replaced by
$K_1(qz)+[K_0(q/\Lambda)/I_0(q/\Lambda)] I_1(qz)$, and 
$K_2(qz)$ by arbitrary linear combination of $K_2(qz)$ and $I_2(qz)$.

It is not as easy to calculate the sphere/disc amplitudes in practice,
however. It has been considered that the parton contributions to 
$\gamma^{*} + h \rightarrow \gamma^{(*)} + h$ scattering is given by 
amplitude with states in the leading trajectory with arbitrary high spin 
being exchanged \cite{BPST-06}. Those fields are not scalar on AdS$_5$
but come with multiple degree of freedom associated with polarizations.
Such polarization of higher spin fields propagating on AdS$_5$ needs to 
be treated properly---including such issues as covariant derivatives and 
kinetic mixing among different polarizations (diagonalization of the 
Virasoro generator $L_0$)---in gravity dual descriptions, in order to 
be able to discuss skewness dependence of GPD / DVCS amplitude.  
Direct impact of the curved background geometry can be implemented 
through the non-linear sigma model on the world sheet, but one has to 
define the vertex operators as a composite operator properly in such an 
interacting theory. Ramond--Ramond background is an essential ingredient 
in making the warped background metric stable, yet non-zero 
Ramond--Ramond background cannot be implemented in the NSR formalism. 

Instead of world-sheet calculation in the NSR formalism in implementing 
the effect of curved background (\ref{eq:AdS5timesS5}), we use 
string field theory action on flat space in this article, and make it 
covariant. 
Because the gravity dual set-up of our interest is in the Type IIB 
string theory, we are thus supposed to use superstring field theory 
for closed string and open string. In order to avoid technical 
complications associated with the interacting superstring field
theories, however, we employ a sort of toy-model approach by using 
the cubic string field theory for bosonic string theory.

In our toy-mode approach, we deal with the cubic string field theory 
on AdS$_5$ ($\times$ some internal compact manifold),
and ignore instability of the background geometry. 
The probe photon in this toy-model gravity dual set-up will be the 
massless vector state of the bosonic string theory with the
wavefunction (\ref{eq:photon-wvfc-pre}, \ref{eq:photon-wvfc}). The target 
hadron can be any scalar states, say, the tachyon, with the 
wavefunction (\ref{eq:hadron-wvfc}). We are to construct a 
toy-model amplitude of the $h+\gamma^{*} \rightarrow h+\gamma^{(*)}$ 
scattering, by using the 2-to-2 scattering of the massless photon and 
some scalar in the bosonic string on the AdS$_5$ background. 
In short, this is to maintain the spirit of the set-up in \cite{PS-02-DIS, 
BPST-06}, use the bosonic cubic string field theory to compute and obtain 
something concrete, from which qualitative lessons are to be extracted 
for the set-up of our interest. 

Clearly one of the cost of this approach (without technical complexity
of interacting superstring field theory) is that we have to restrict our 
attention to the Reggeon exchange (flavor non-singlet) amplitude.  
The amplitude constructed in this way is certainly not faithful to 
the equations of the Type IIB string theory, either.
Since our motivation is not in constructing yet another 
exact solution to superstring theory, however, we still expect 
that this (flavor non-singlet) toy-mode amplitude in bosonic string 
still maintains some fragrance of hadron scattering amplitude to be 
calculated in superstring theory. 
This discussion continues to section \ref{ssec:super-Pomeron}.

\section{Cubic String Field Theory}
\label{sec:cubic-sft}

Section \ref{ssec:cubic-action} summarizes technical details of 
cubic string field theory that we need in later sections. 
We then proceed in section \ref{ssec:Veneziano} to explain an idea 
of how to reproduce disc amplitude only 
from string-field-theory $t$-channel amplitude, using photon--tachyon 
scattering on a flat spacetime background as an example.  
This idea of constructing amplitude is generalized in 
section \ref{sec:organize} for scattering on a warped spacetime, and we will 
see that this construction of the amplitude allows us to 
cast the amplitude almost immediately into the form of conformal 
OPE (\ref{eq:conf-OPE-2side}, \ref{eq:conf-OPE-center}).  

\subsection{Action of the Cubic SFT on a Flat Spacetime}
\label{ssec:cubic-action}

The action of the cubic string field theory (cubic SFT) is given by
\cite{SFT-Witten-cubic} 
\begin{eqnarray}
 S  & = &  - \frac{1}{2\alpha'} \int \left(
    \Phi * Q_B \Phi +  \frac{2}{3} g_o \Phi * \Phi * \Phi \right), 
 \label{eq:SFT-action} \\
& = & -\frac{1}{2\alpha'}\left(
  \Phi\cdot Q_B \Phi +\frac{2g_o}{3}\Phi \cdot \Phi * \Phi \right), 
 \label{eq:SFT-action-prev}
\end{eqnarray}
where $g_o$ is a coupling constant of mass dimension $(1-D/2)$, where 
$D=26$ is the spacetime dimensions of the bosonic string theory.\footnote{ 
The sign of the interaction term is just a matter of convention,
because field redefinition for all the component fields $\Phi \rightarrow
- \Phi$ is always possible. Under this redefinition, however, covariant
derivative can be either 
$\partial_m - i \rho(A_\mu)$ or $\partial_m+i\rho(A_m)$. The sign
convention above is for $\partial_m-i\rho(A_m)$, following the
convention of section 6.5 of Polchinski's textbook.} 
The string field $\Phi$ is, as a ket state, expanded in terms of the 
Fock states as in 
\begin{align}
 \Phi = |\Phi \rangle=&
\phi(x)|\downarrow \rangle 
+(A_M(x)\alpha^M_{-1}+C(x)b_{-1}+\bar C(x)c_{-1})|\downarrow \rangle 
 \notag
 \\
& + 
 \left(
 f_{MN}(x)\frac{1}{\sqrt{2}}\alpha_{-1}^M\alpha_{-1}^N
+ig_{M}(x)\frac{1}{\sqrt{2}}\alpha_{-2}^M
+h(x)b_{-1}c_{-1}+\cdots \right)|\downarrow\rangle, 
\label{eq:string-field-comp-exp}
\end{align}
with component fields 
$\phi, A_M, C, \bar{C}, f_{MN}, g_M, h, \cdots$; we have already chosen the
Feynman--Siegel gauge here. We will eventually be interested only in 
the states with vanishing ghost number, $N_{\text{gh}} = 0$, because 
states with non-zero ghost number do not appear in the $t$-channel / 
$s$-channel exchange for the disc amplitude. 

The Hilbert space of one string state is spanned by the Fock states 
given (in this gauge) by 
\begin{align}
 \prod_{a=1}^{h_a} \alpha_{-n_a}^{M_a} \prod_{b=1}^{h_b} b_{-l_b}
 \prod_{c=1}^{h_c} c_{-m_c} |\downarrow\rangle, 
\label{eq:Fock-bos-string}
\end{align}
with $1\le  n_1\le n_2 \le \dots \le n_{h_a}$,
$1\le  l_1< l_2 < \dots < n_{h_b}$ and $1\le  m_1< m_2 < \dots < m_{h_c}$.
Let us use 
$Y := \left\{ \left\{ n_a \right\}\mbox{'s}, \; \left\{ l_b
\right\}\mbox{'s}, \; \left\{ m_c \right\}'s \right\}$ as the 
label distinguishing different Fock states of string on a flat spacetime.
Mass of these Fock states are determined by 
\begin{equation}
\alpha' k^2 + (N^{(Y)} - 1)  = 0, \qquad  
N^{(Y)} = \sum_{a=1}^{h_a} n_a + \sum_{b=1}^{h_b} l_b + \sum_{c=1}^{h_c} m_c.
\label{eq:level-bos-string}
\end{equation}
A component field corresponding to a Fock state may be further decomposed 
into multiple irreducible representation of the Lorentz group, but at least, 
the rank-$h_a$ totally symmetric traceless tensor representation 
is always contained. 
Fock states of particular interest to us are the ones in the leading
trajectory: $Y = \left\{ 1^N, 0, 0 \right\}$, so that 
all $n_a$'s are $1$, $h_b = h_c = 0$, and $N^{(Y)} = h_a$. 
The totally symmetric traceless tensor component field of these
states are denoted by $(N !)^{-1/2} A^{(Y)}_{M_1 \cdots M_{h_a}}$. 

The kinetic term---the first term of (\ref{eq:SFT-action}, 
\ref{eq:SFT-action-prev})---is written down in terms of the component
fields as follows:
\begin{align}
 -\frac{1}{2\alpha'} & \Phi\cdot Q_B \Phi=
\frac{1}{2}\int d^{26}x \; 
 {\rm tr} \left[
\phi(x) \left(\partial^2+\frac{1}{\alpha'} \right) \phi(x)+
A_M(x)\partial^2A^M(x)+
\right. \notag
\\
&\left.
f_{MN}(x) \left(\partial^2-\frac{1}{\alpha'}\right)f^{MN}(x)+
g_M(x) \left(\partial^2-\frac{1}{\alpha'}\right) g^M(x)-
h(x) \left( \partial^2-\frac{1}{\alpha'} \right) h(x) + \cdots
\right].
\label{eq:cubic-sft-26D-kin}
\end{align}
The totally symmetric tensor component field of the Fock states in the 
leading trajectory $Y = \left\{ 1^N,0,0\right\}$ has a kinetic term
\begin{align}
 \frac{1}{2} \int d^{26}x \; \tr \left[ A^{M_1\dots M_j}\left(\partial^2
 -\frac{N-1}{\alpha '} \right) A_{M_1\dots M_j} \right].
\label{eq:cubic-sft-26D-kin-ldg-trj}
\end{align}
The cubic string field theory action in the Feynman--Siegel gauge has 
two nice properties; first, the kinetic terms of those Fock states do 
not mix in the flat spacetime background, and second, 
the second derivative operators are simply given by d'Alembertian, 
without complicated restrictions or mixing among various polarizations 
in the component fields. 

The second term of the action (\ref{eq:SFT-action}, 
\ref{eq:SFT-action-prev}) gives rise to interactions involving three 
component fields. Interactions involving Fock states with small excitation 
level $N$ are \cite{SFT-interaction}
\begin{align}
 - \frac{1}{2\alpha'}\frac{2g_o}{3}\Phi \cdot \Phi * \Phi= - 
 \int d^{26}x \frac{g_o \lambda_{\rm sft}}{3\alpha'}\hat E& \left(
  \tr \left[ \phi^3(x) \right]
  + \sqrt{\frac{8\alpha'}{3}} 
      \tr \left[ (-i A_M) \left(\phi \overleftrightarrow{\partial}^M
                                  \phi  \right)\right]   \right.
  \notag  \\
& \quad \left.  -\frac{8\alpha'}{9\sqrt{2}}
    \tr \left[ f_{MN}\left( \phi \overleftrightarrow{\partial}^M
                   \overleftrightarrow{\partial}^N \phi \right) \right]
-\frac{5}{9\sqrt{2}}\tr \left[f^M_M\phi^2 \right]
\right.
\notag
\\&\quad +
\left.
  \frac{2\sqrt{\alpha'}}{3}\tr \left[ (\partial_Mg^M)\phi^2 \right] 
 -\frac{11}{9} \tr \left[h \phi^2 \right]
\right) + \cdots,
\label{eq:cubicSFT-interaction-A}
\end{align}
where $\lambda_{\rm sft}=3^{9/2}/2^6$ \cite{SFT-Hata-lec.note}, 
$\overleftrightarrow{\partial^M}=\left({\overrightarrow{\partial^M}-\overleftarrow{\partial^M}}\right)$,
and
\begin{align}
 \hat E = \exp\left[-2\alpha'\ln\sfrac{2}{3^{3/4}}(\partial_{(1)}^2+\partial_{(2)}^2+\partial_{(3)}^2)\right].
\end{align}
The $\partial^2_{(1,2,3)}$ means taking derivatives of the 1st, 2nd, 
and 3rd field.\footnote{Concretely,
\begin{align}
 \hat E A(x)B(x)C(x) =
 \left[ \left(\frac{27}{16} \right)^{\frac{\alpha'}{2} \partial^2 } A(x) \right]
 \left[ \left(\frac{27}{16} \right)^{\frac{\alpha'}{2} \partial^2 } B(x) \right]
 \left[ \left(\frac{27}{16} \right)^{\frac{\alpha'}{2} \partial^2 } C(x) \right]
\notag.
\end{align}
}

Interactions involving totally symmetric leading trajectory states are 
also of interest to us. 
The tachyon--tachyon--$Y=\{1^N,0,0\}$ cubic coupling with
$N$-derivatives is given by 
\begin{equation}
 - \frac{g_o \lambda_{\rm sft}}{\alpha'} \int d^{26}x \; \hat{E} \; 
  {\rm tr} \left[ A^{(Y)}_{M_1 \cdots M_N}
                   \left( \phi (-i \overleftrightarrow{\partial}^{M_1})
                         \cdots (-i \overleftrightarrow{\partial}^{M_N})      
                          \phi \right) 
      	   \right] 
  \left( \frac{8\alpha'}{27} \right)^{\frac{N}{2}}   
  \frac{1}{\sqrt{N!}}
\label{eq:cubicSFT-interaction-B}
\end{equation}
in the interaction part of the action. 
The photon (the level-1 state)--photon--$Y=\{1^N,0,0 \}$ coupling 
in the cubic string field theory includes  
\begin{eqnarray}
&& - \frac{g_o \lambda_{\rm sft}}{\alpha'} \int d^{26}x \; \hat{E} \; 
  {\rm tr} \left[ A^{(N)}_{M_1 \cdots M_N}
                  \left( A_L (-i \overleftrightarrow{\partial}^{M_1})
                         \cdots (-i \overleftrightarrow{\partial}^{M_N})
                         A_K \right)
  \left( \frac{8\alpha'}{27} \right)^{\frac{N}{2}}   
      \frac{\eta^{KL} \frac{16}{27}}{\sqrt{N!}}+ \cdots 
  \right],
\label{eq:cubicSFT-interaction-C}
\end{eqnarray}
where we kept only the terms that have $N$-derivatives and are
proportional to $\eta^{KL}$, as they are necessary in deriving 
(\ref{eq:tch-leading-tp}).

\subsection{Cubic SFT Scattering Amplitude and $t$-Channel Expansion}
\label{ssec:Veneziano}

Before proceeding to study the 
$h + \gamma^{*} \rightarrow h + \gamma^{(*)}$ scattering amplitude 
by using the cubic string field theory on the warped spacetime
background, let us remind ourselves how to obtain $t$-channel 
operator product expansion from the amplitude calculation 
based on string field theory, by using tachyon--photon scattering 
on the flat spacetime as an example.   

\begin{figure}[tbp]
  \begin{center}
\begin{tabular}{cccc}
 \includegraphics[scale=0.35]{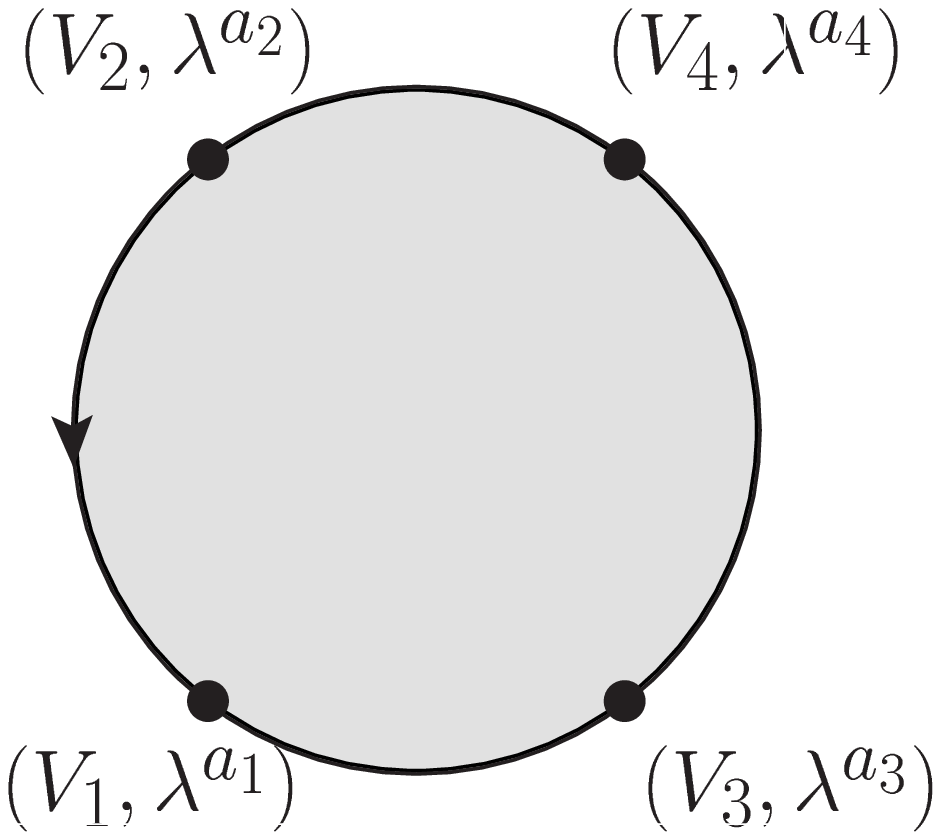} &
 \includegraphics[scale=0.35]{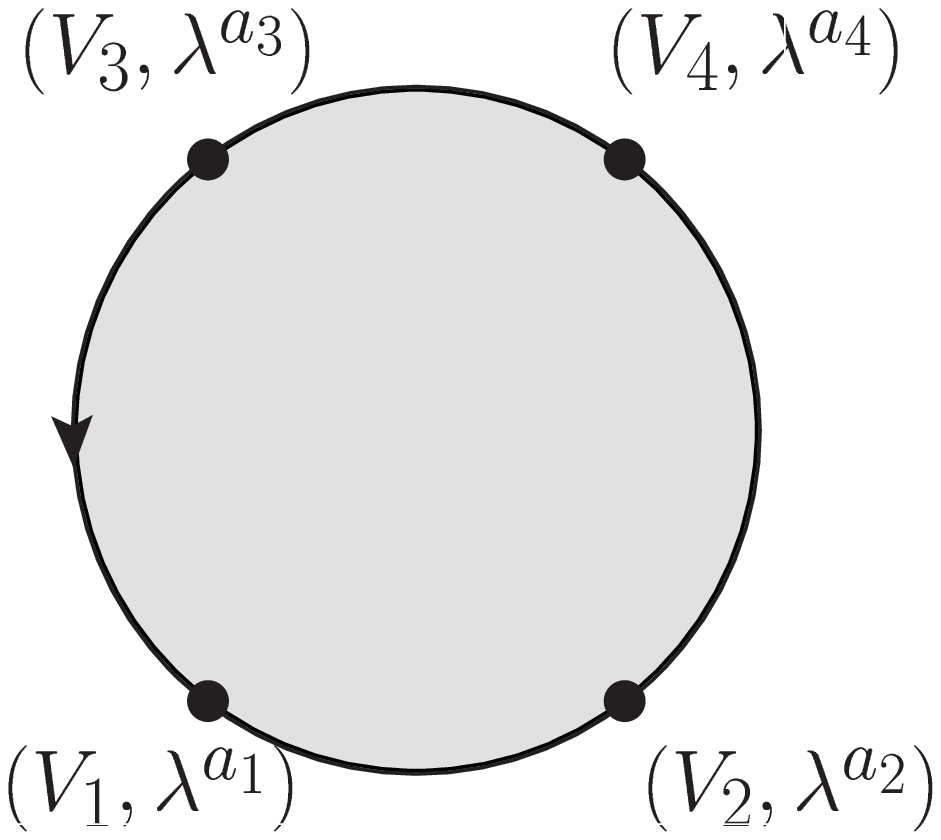} &
 \includegraphics[scale=0.35]{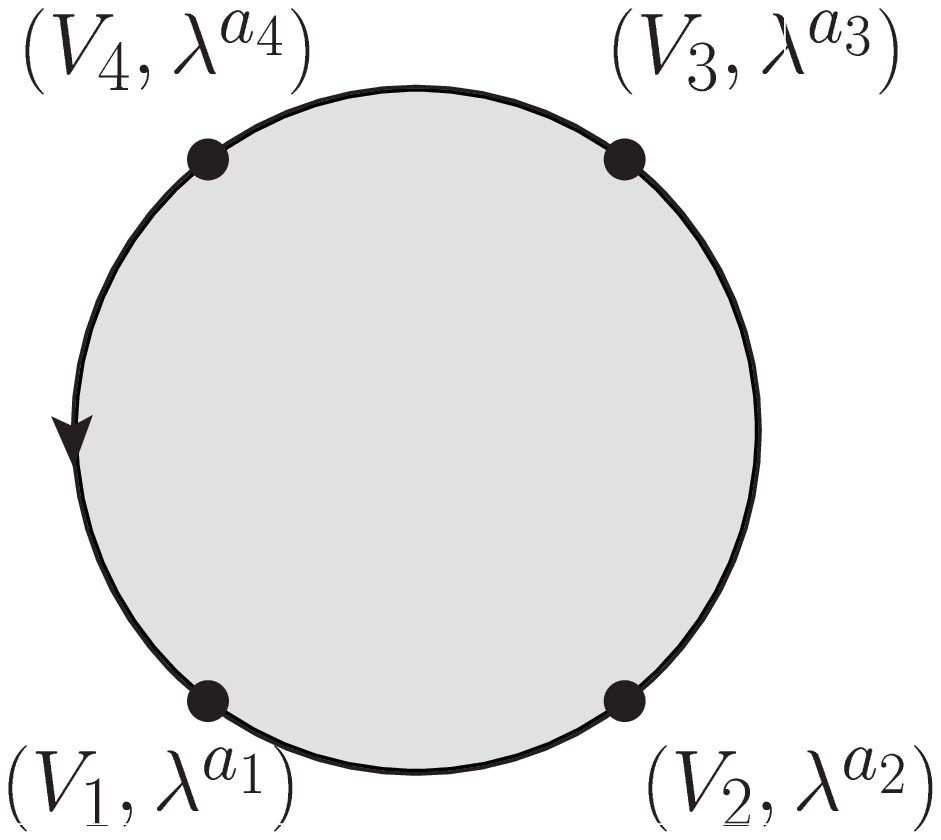} &
 \includegraphics[scale=0.35]{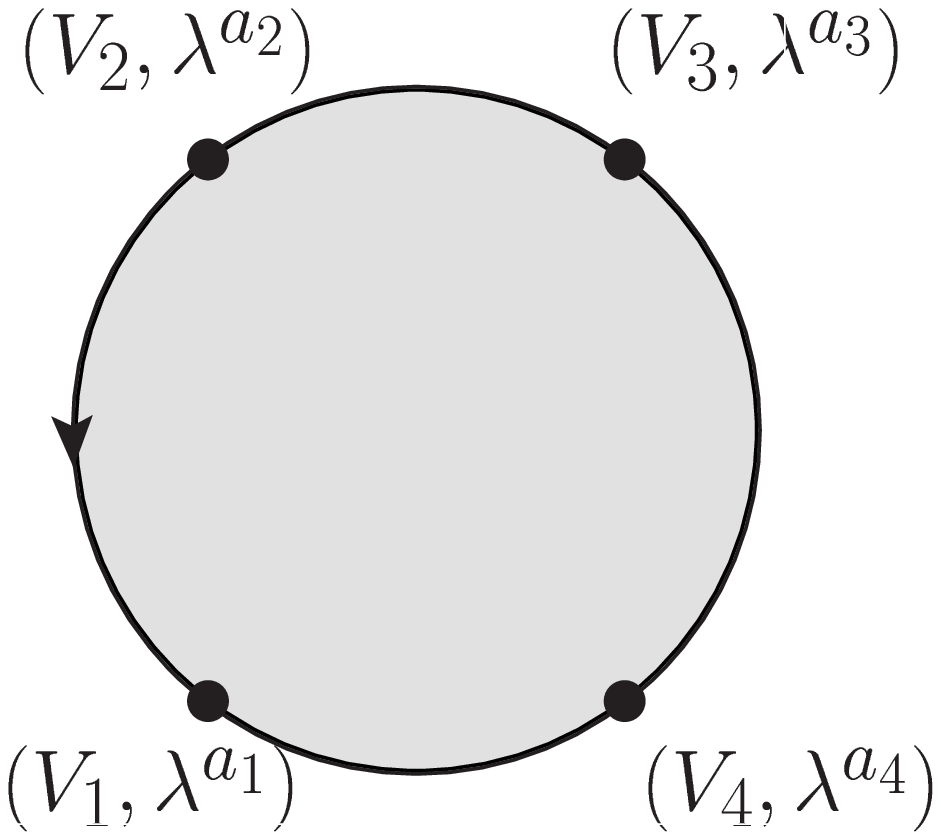}   \\
(a) 1342 & (b) 1243 & (c) 1234 & (d) 1432
\end{tabular}
\caption{Disc amplitudes with two photon vertex operators ($V_1$ and
   $V_2$) and two tachyon vertex operators ($V_3$ and $V_4$) inserted. 
Kinematical amplitudes given by the disc amplitudes above are multiplied 
by the Chan--Paton factors 
$\tr [\lambda^{a_1} \lambda^{a_3} \lambda^{a_4} \lambda^{a_2}]$ in (a), 
$\tr [\lambda^{a_1} \lambda^{a_2} \lambda^{a_4} \lambda^{a_3}]$ in (b), 
$\tr [\lambda^{a_1} \lambda^{a_2} \lambda^{a_3} \lambda^{a_4}]$ in (c) 
and $\tr [\lambda^{a_1} \lambda^{a_4} \lambda^{a_3} \lambda^{a_2}]$ in
(d), respectively. The two disc amplitudes (a, b) become 
${\cal M}_{\rm Ven}(s,t)$, while (c, d) ${\cal M}_{\rm Ven}(u,t)$.}
\label{fig:disc-tp-tp}
  \end{center}
\end{figure}

Let us consider the disc amplitude of tachyon--photon scattering. 
The vertex operators labeled by $i=1,2$,
$V_i = :\epsilon^{i}_M \partial X^M e^{i k_i \cdot X }:$, are for photon
incoming ($i=1$) and outgoing ($i=2$) states, which come with 
Chan--Paton matrices $\lambda^{a_i}$.
Tachyon incoming $(i=3)$ and outgoing $(i=4)$ states correspond to 
vertex operators $V_i = : e^{i k_i \cdot X} :$ with Chan--Paton matrices 
$\lambda^{a_i}$. The photon--tachyon scattering amplitude  
$A + \phi \rightarrow A + \phi$ in bosonic open 
string theory (Veneziano amplitude) is given by\footnote{
Here, $p := (k_3 - k_4)/2$, averaged momentum of tachyon before and after
the scattering, just like in (\ref{kinematics1}). } 
\begin{eqnarray}
{\cal M}_{\rm Ven}(s,t) & = & - \left(\frac{g_o^2}{\alpha'}\right) 
    \frac{\Gamma(-\alpha' t-1) \Gamma(-\alpha' s-1)}
      {\Gamma(-\alpha'(s+t) -1)}
   \epsilon_{M}(k_2) \epsilon_{N}(k_1)  \nonumber \\
 & & \times \left\{
  \left[ \eta^{MN} - \frac{k_1^M k_2^N}{k_1 \cdot k_2} \right]
  (\alpha' s + 1)  \right.  \label{eq:Veneziano-amplitude-tp}\\
 & & \left. \quad + 2\alpha' 
  \left( \left[p^M - k_1^M \frac{k_2 \cdot p}{k_1 \cdot k_2}\right]
       - \frac{k_2^M}{2} \right)
  \left( \left[p^N - k_2^N \frac{k_1 \cdot p}{k_2 \cdot k_1}\right]
       - \frac{k_1^N}{2} \right)(\alpha' t+1) 
  \right\},    \nonumber 
\end{eqnarray}
which is to be multiplied by the Chan--Paton factor 
${\rm Tr}\left[\lambda^{a_2}\lambda^{a_4}\lambda^{a_3}\lambda^{a_1} 
 +\lambda^{a_4}\lambda^{a_2}\lambda^{a_1}\lambda^{a_3}\right]$.
(see Figure \ref{fig:disc-tp-tp}~(a,~b).)
If the Chan--Paton matrices of a pair of incoming and outgoing vertex
operators, $\lambda^{a_1}$ and $\lambda^{a_2}$, commute with each
other,\footnote{Just like in the case both $\lambda^{a_1}$ and
$\lambda^{a_2 }$ are an $N_F \times N_F$ matrix 
$\diag(2/3, -1/3,-1/3)$.} 
then the Chan--Paton factors from the diagrams 
Figure~\ref{fig:disc-tp-tp}~(c,~d) are the same, and the total 
kinematical part of the amplitude for this Chan--Paton factor becomes 
${\cal M}_{\rm Ven}(s,t) + {\cal M}_{\rm Ven}(u,t)$.

Let us stay focused on ${\cal M}_{\rm Ven}(s,t)$ alone for now. 
The amplitude proportional to $\eta^{MN}$ can be expanded, as 
is well-known, as a sum only of $t$-channel poles:\footnote{It is also
possible to expand this as a sum only of $s$-channel poles; that's the
celebrated $s$-$t$ duality of the Veneziano amplitude. } 
\begin{eqnarray}
 \frac{g_o^2}{\alpha'}
 \frac{\Gamma(-\alpha' t-1) \Gamma(-\alpha' s)}
      {\Gamma(-\alpha'(s+t)-1)} & = &
 \frac{g_o^2}{\alpha'} \int_0^1 dx \;
  x^{-\alpha' t-2}(1-x)^{-\alpha's-1},
 \label{eq:Veneziano-str-fcn-tp}
 \\
 & = &  \frac{g_o^2}{\alpha'}
  \sum_{N=0}^{\infty} \frac{-1}{\alpha' t - (N-1)} 
   \frac{(\alpha's+1)\cdots (\alpha' s+N)}{N!}.
\label{eq:pole-expansion-of-Veneziano-Amplitude-tp}
\end{eqnarray}

\begin{figure}[tbp]
  \begin{center}
\begin{tabular}{ccc}
 \includegraphics[scale=0.5]{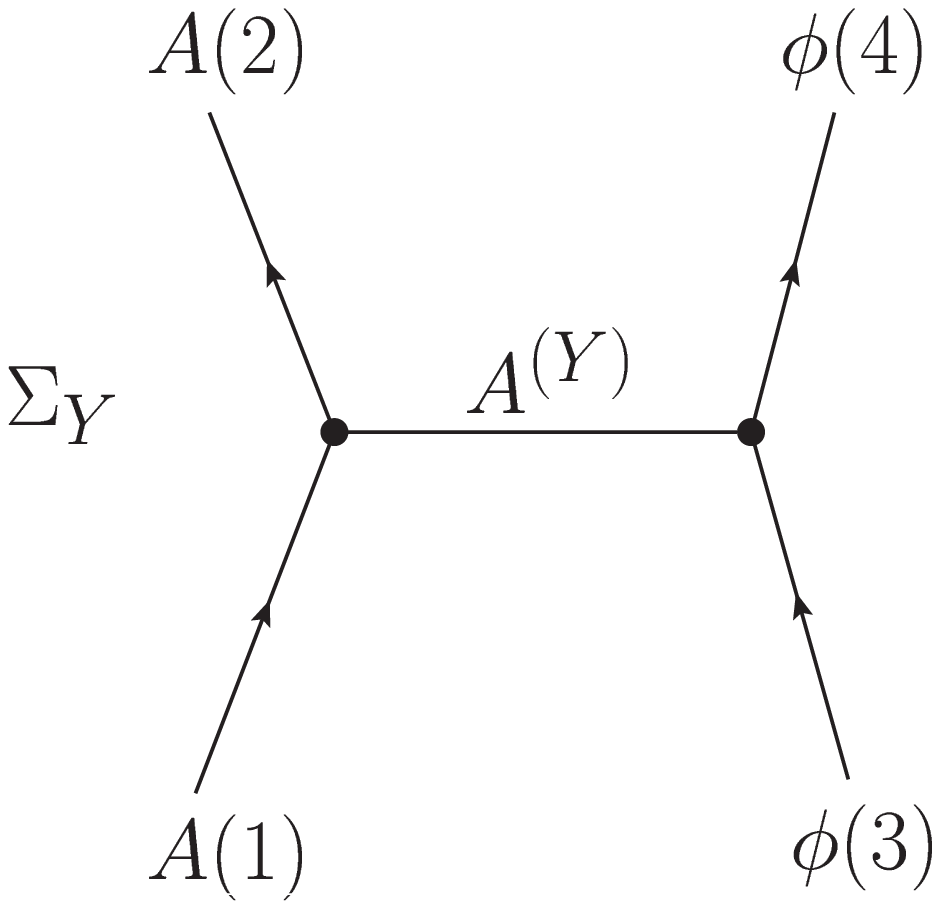} & $\qquad$ &
 \includegraphics[scale=0.5]{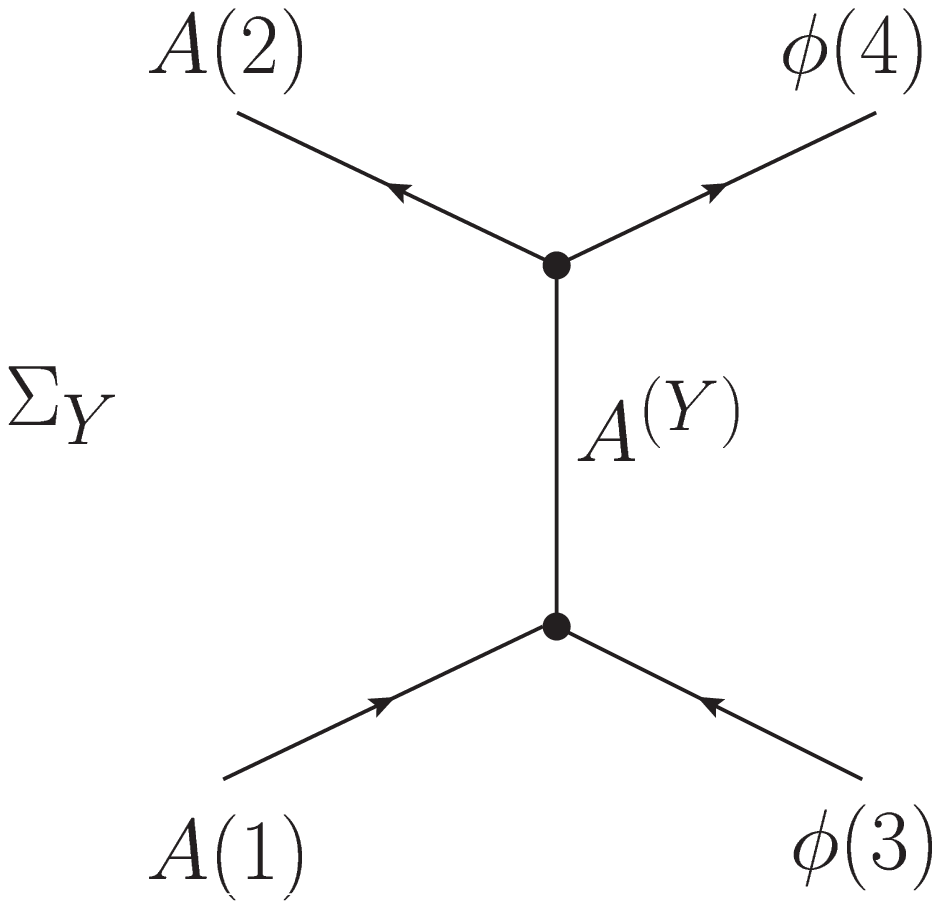} \\
\end{tabular}
\caption{\label{fig:SFT-TandS}Two types of diagrams contribute to the 
photon--tachyon scattering amplitude ${\cal M}_{\rm Ven}(s,t)$ 
in cubic string field theory: 
$t$-channel exchange of one string states labeled by $Y$ (left), 
and $s$-channel exchange (right). }
  \end{center}
\end{figure}

The Veneziano amplitude (\ref{eq:Veneziano-amplitude-tp}) can also be 
obtained in cubic string field theory \cite{SFT-Giddings-Veneziano}.
In the cubic SFT, the scattering amplitude consists of two pieces, 
a collection of $t$-channel exchange diagrams and that of $s$-channel 
diagrams (Figure~\ref{fig:SFT-TandS}).
\begin{align}
 &{\cal M}_\text{Ven}(s,t)= \sum_Y {\cal M}^{(t)}_Y(s,t)
                          + \sum_Y {\cal M}^{(s)}_Y(s,t).
\label{eq:Veneziano-in-SFT}
\end{align}
%
%
Infinitely many one string states (\ref{eq:Fock-bos-string}) with zero 
ghost number ($h_b = h_c$)---labeled by $Y$---can be exchanged 
in the $t$-channel or in the $s$-channel, and the corresponding 
contributions are in the form of 
\begin{align}
&{\cal M}^{(t)}_Y= \frac{f^{(t)}_Y(s,t)}{-\alpha't-1+N^{(Y)}}, \quad
{\cal M}^{(s)}_Y= \frac{f^{(s)}_Y(t,s)}{-\alpha's-1+N^{(Y)}},
\label{eq:pole-SandT}
\end{align}
where $f^{(t)}_Y$ and $f^{(s)}_Y$ are regular function at finite $s$ and $t$; 
$N^{(Y)}$ is the excitation level (\ref{eq:level-bos-string}) of 
a component field $A^{(Y)}$.

Because both the world-sheet calculation (\ref{eq:Veneziano-amplitude-tp}, 
\ref{eq:pole-expansion-of-Veneziano-Amplitude-tp}) and the cubic SFT
calculation (\ref{eq:Veneziano-in-SFT}, \ref{eq:pole-SandT}) are the
same thing, ${\cal M}_{\rm Ven}(s,t)$ in both approaches should be completely 
the same functions of $(s,t)$. 
Therefore, for an arbitrary given value 
of $s$, the residue of all the poles in the complex $t$-plane should be
the same. We also know that the Veneziano amplitude can be expanded
purely in the infinite sum of $t$-channel poles with $t$-independent 
residues. This means that the full Veneziano 
amplitude (\ref{eq:Veneziano-amplitude-tp}) can be reproduced just from 
the $t$-channel cubic SFT amplitude\footnote{The $t$-channel and
$s$-channel amplitudes of the cubic SFT, $\sum_Y {\cal M}_Y^{(t)}$ and 
$\sum_Y {\cal M}_Y^{(s)}$ correspond to the integration over $[0, 1/2]$
and $[1/2, 1]$ in (\ref{eq:Veneziano-str-fcn-tp}), 
respectively \cite{SFT-Giddings-Veneziano}. 
Thus, $\sum_Y {\cal M}_Y^{(s)}$ does not contain a pole in $t$.} 
$\sum_Y {\cal M}_Y^{(t)}(s,t)$, by the following procedure:
\begin{equation}
 \sum_Y \frac{f^{(t)}_Y(s,t)}{-\alpha' t - 1 + N^{(Y)}} \longrightarrow 
 \sum_Y \frac{f^{(t)}_Y(s, (N^{(Y)}-1)/\alpha')}{-\alpha' t-1+N^{(Y)}} = 
  {\cal M}_{\rm Ven}(s,t).
  \label{eq:SFT-Veneziano-prescr}
\end{equation}
%

\begin{figure}[tbp]
  \begin{center}
\begin{tabular}{cccc}
 \includegraphics[scale=0.4]{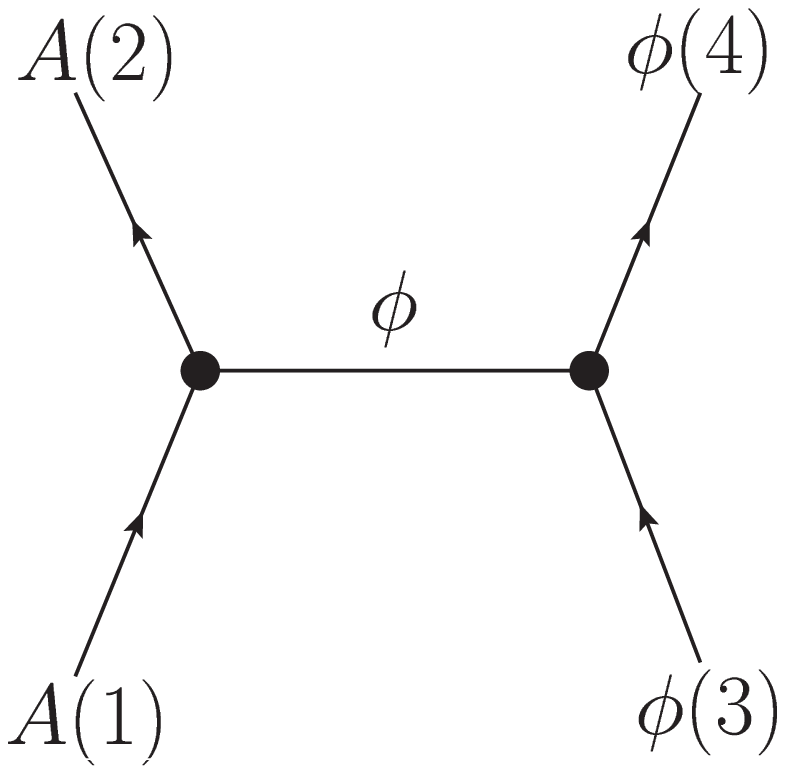} &
 \includegraphics[scale=0.4]{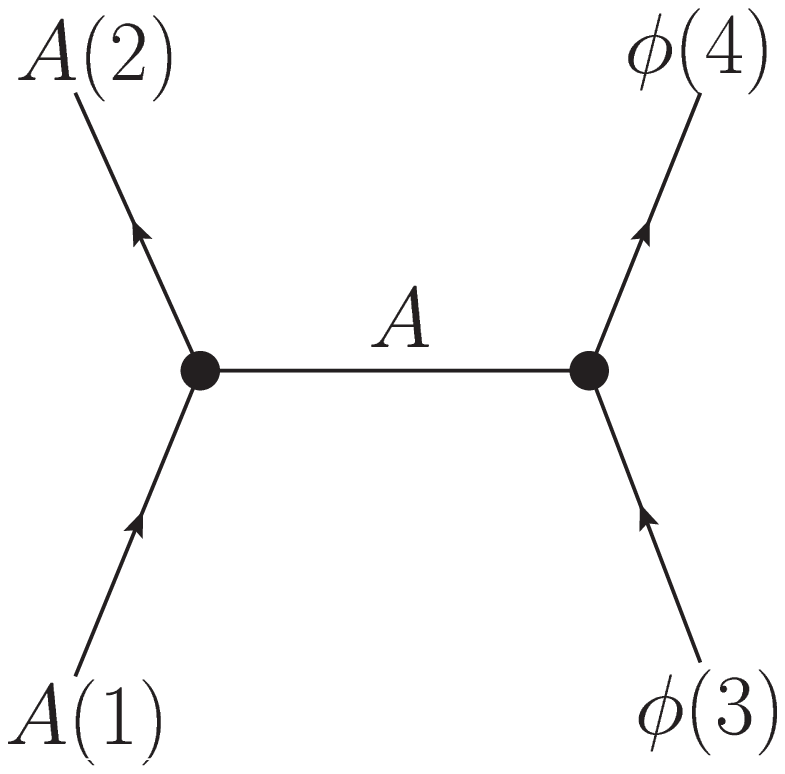} &
 \includegraphics[scale=0.4]{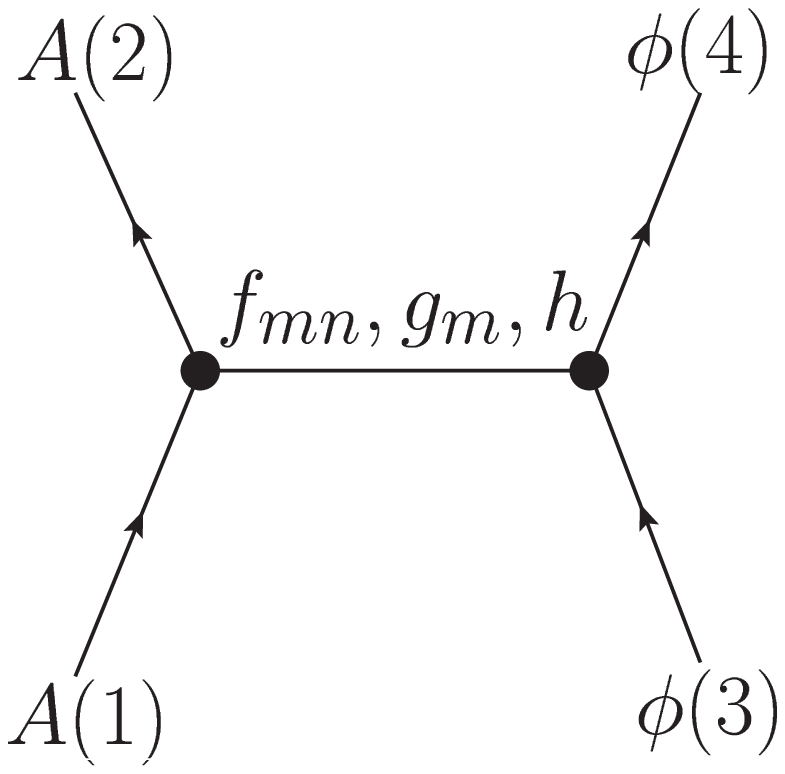} \\
 (a) & (b) & (c) 
\end{tabular}
\caption{\label{fig:SFT-tch-diag} $t$-channel exchange diagrams for 
$A + \phi \rightarrow A + \phi$ scattering in the cubic string field
   theory. The tachyon ($N=0$), photon ($N=1$) and level-2 states are 
exchanged in the diagrams (a), (b) and (c), respectively.}
  \end{center}
\end{figure}

To see that this prescription really works, let us take a look at 
the amplitudes of $t$-channel exchange of one string states with 
small excitation level $N^{(Y)}=0,1,2$. Focusing on the 
amplitude of $A + \phi \rightarrow A + \phi$ proportional to 
$\eta^{MN}$, we find that the tachyon exchange in the $t$-channel 
(Figure~\ref{fig:SFT-tch-diag} (a)) gives rise to the amplitude 
\cite{SFT-Thorn}
\begin{align}
 {\cal M}_{\phi}^{(t)}(s,t) =
\left(\frac{g_o \lambda_{\rm sft}}{\alpha'}\right)^2
\left(\frac{2}{3^{3/4}}\right)^{-2\alpha' t -2\alpha' t + 4}
\frac{-1}{t+1/\alpha'} = 
  \frac{g_o^2}{\alpha'}
  \left( \frac{27}{16} \right)^{\alpha' t+1}
  \frac{-1}{\alpha' t+1},
\end{align}
which is obtained simply by using the $\phi$-$\phi$-$\phi$ vertex rule 
(\ref{eq:cubicSFT-interaction-A}) and $A$-$A$-$\phi$ vertex rule 
(\ref{eq:cubicSFT-interaction-B}). 
The prescription (\ref{eq:SFT-Veneziano-prescr}) turns this amplitude 
into 
\begin{align}
 \longrightarrow
 {\cal M}_\phi(s,t)  = \frac{g_o^2}{\alpha'}\frac{-1}{\alpha' t+1},
\end{align}
which reproduces the $N=0$ term of
(\ref{eq:pole-expansion-of-Veneziano-Amplitude-tp}). 

The $t$-channel exchange of level $N^{(Y)} = 1$ excited states can also 
be calculated in the cubic string field theory 
(Figure~\ref{fig:SFT-tch-diag}~(b)). 
The amplitude proportional to $\eta^{MN}$ is 
\begin{equation}
 {\cal M}^{(t)}_A(s,t)  = 
  \frac{g_o^2}{\alpha'}\left(\frac{27}{16}\right)^{\alpha't}
  \frac{-1}{\alpha't} \; 
  \left[ \frac{\alpha'(s-u)}{2} \right], 
\end{equation}
where $(s-u) = (k_{(1)} - k_{(2)}) \cdot (k_{(4)} - k_{(3)})$. 
Using the relation $\alpha' (s+t+u) = -2$ in the tachyon--photon 
scattering to eliminate $u$ in favor of $s$ and $t$, and following 
the prescription (\ref{eq:SFT-Veneziano-prescr})---which is to 
exploit $\alpha' t = 0$ in the numerator, 
this amplitude is replaced by \cite{SFT-Thorn}
\begin{equation}
\longrightarrow 
{\cal M}_A(s,t) = \frac{g_o^2}{\alpha'} 
   \frac{-(\alpha' s + 1)}{\alpha' t}.
\end{equation}
Once again, this reproduces the level $N=1$ contribution to the 
Veneziano amplitude (\ref{eq:pole-expansion-of-Veneziano-Amplitude-tp}).

Similar calculation for level-2 state exchange can be carried out 
(Figure~\ref{fig:SFT-tch-diag}~(c)). 
Using the vertex rule in (\ref{eq:cubicSFT-interaction-A}) 
for the [level-2]-$\phi$-$\phi$ couplings, and also the 
interactions among [level-2]-$A$-$A$ coupling in the literature, 
the cubic SFT $t$-channel amplitude is given by \cite{SFT-Thorn}
\begin{eqnarray}
 {\cal M}_f^{(t)}(s,t) & = &
  \frac{g_o^2}{\alpha'}\left(\frac{27}{16}\right)^{\alpha't-1}
  \frac{-1}{\alpha't-1} \;
  \left[ \frac{ (\alpha'(s-u))^2 }{8}
       - \frac{5(\alpha' t + 2)}{16 \cdot 2}
       + \frac{ 490 }{ 16^2 \cdot 2} \right],  \\
 {\cal M}_g^{(t)}(s,t) & = &
  \frac{g_o^2}{\alpha'}\left(\frac{27}{16}\right)^{\alpha't-1}
  \frac{-1}{\alpha't-1} \;
  \left[ - \frac{36 \; \alpha' t} {16^2} \right] ,  \\
 {\cal M}_h^{(t)}(s,t) & = &
  \frac{g_o^2}{\alpha'}\left(\frac{27}{16}\right)^{\alpha't-1}
  \frac{-1}{\alpha't-1} \; 
  \left[ - \frac{11^2}{16^2} \right]. 
\end{eqnarray}
After using $\alpha'u = - \alpha'(s+t) - 2$ to eliminate $u$ in favor of
$s$ and $t$, and further following the prescription
(\ref{eq:SFT-Veneziano-prescr}) [$\alpha' t \rightarrow 1$ in the 
numerator], one will see that the level $N^{(Y)} = 2$ amplitude turns into 
\begin{align}
\longrightarrow 
 \left( {\cal M}_f + {\cal M}_g + {\cal M}_h \right)(s,t) =
\frac{g_o^2}{\alpha'}\frac{-1}{\alpha' t-1}\left[
\frac{(\alpha's)^2+3(\alpha' s)+2}{2} \right].
\end{align}
Once again, this is precisely the same as the $N=2$ contribution 
to the Veneziano amplitude (\ref{eq:pole-expansion-of-Veneziano-Amplitude-tp}).

Contributions from the $t$-channel exchange of states in the leading
trajectory can also be examined systematically. 
Using the vertex rule (\ref{eq:cubicSFT-interaction-B}, 
\ref{eq:cubicSFT-interaction-C}) involving the states in the leading
trajectory ($Y=\left\{1^N,0,0 \right\}$), one finds that the amplitude 
proportional to $\eta^{MN}$ is   
\begin{equation}
 {\cal M}_{\left\{1^N,0,0 \right\} }^{(t)} \simeq \frac{g_o^2}{\alpha'}
   \left(\frac{27}{16}\right)^{\alpha' t -(N-1)}
   \frac{-1}{\alpha' t-(N-1)} 
   \frac{\left(\alpha' (s-u)/2 \right)^N}{N!},
\label{eq:tch-leading-tp}
\end{equation}
where we maintained only the terms with the highest power of either $s$
or $u$.
After using the kinematical relation $\alpha'(s+t+u)+2 = 0$ to eliminate 
$u$ in favor of $s$ and $t$, and following the prescription
(\ref{eq:SFT-Veneziano-prescr}) [$\alpha' t \rightarrow (N-1)$ 
in the numerator], we obtain the large-$(\alpha's)$ leading power contribution to
the $N$-th term of (\ref{eq:pole-expansion-of-Veneziano-Amplitude-tp}) 
with the correct coefficient. 

We have therefore seen that the prescription (\ref{eq:SFT-Veneziano-prescr})
allows us to use the $t$-channel exchange amplitude in the cubic 
string field theory to construct the full disc scattering amplitude.
In section \ref{sec:organize}, this prescription is extended for the 
disc scattering amplitudes on a spacetime with curved background metric, 
which is the situation of real interest in the context of hadron scattering. 

\section{Mode Decomposition on AdS$_5$}
\label{sec:mode-fcn}

Let us now proceed to work out mode decomposition of 
the totally symmetric (traceless) component field on the warped
spacetime. The correspondence between the primary operators of the 
conformal field theory on the (UV) boundary and wavefunctions on 
AdS$_5$ is made clear in this section. The Pomeron/Reggeon wavefunctions 
are obtained as a {\it holomorphic} function of the spin variable $j$, since  
we need to do so for the further inverse Mellin transformation.
The wavefunctions will then be used also to construct 
the cattering amplitude of $h+\gamma^{*} \rightarrow h+\gamma^{(*)}$ 
and GPD in sections \ref{sec:organize} and \ref{sec:model}.

Let the bilinear (free) part of the (bulk) action of a rank-$j$ tensor 
field on AdS$_5$ to be\footnote{\label{fn:closed-kin-normalization}
The dimensionless constant $t_{Ay}$ 
is something like $N_c^2$ for a mode obtained by reduction of closed 
string component fields in higher dimensions. 
More comment on $t_{Ay}$ for open string is found in 
footnote \ref{fn:open-kin-normalization}.}
\begin{eqnarray}
 S_{\rm eff.~kin.} & = & - \frac{1}{2} \frac{t_{Ay}}{R^3}\int d^4 x \int dz \sqrt{-g(z)} 
  g^{m_1 n_1} \cdots g^{m_j n_j} \nonumber \\
  & & \quad 
  \left[ 
    g^{m_0 n_0} (\nabla_{m_0} A^{(y)}_{m_1 \cdots m_j})
                (\nabla_{n_0} A^{(y)}_{n_1 \cdots n_j}) + 
    \left(\frac{c_y}{R^2} + \frac{N^{(y)}_{\rm eff.}}{\alpha'}\right) 
       A^{(y)}_{m_1 \cdots m_j} A^{(y)}_{n_1 \cdots n_j}
  \right],  
\label{eq:cubic-sft-kin-5D}
\end{eqnarray}
where we assume that kinetic mixing between different fields is either 
absent or sufficiently small. 
Here, the dimensionless parameter $N^{(y)}_{\rm eff.}$ is 
$N^{(Y)}-1$ for an $N^{(Y)} \in {\bf Z}_{\geq 0}$ for bosonic open string 
($j \leq N^{(Y)}$), which would be $4(N^{(Y)}-1)$ for an 
$N^{(Y)} \in {\bf Z}_{\geq 1}$ for closed string ($j \leq 2N^{(Y)}$).
This field is regarded as a reduction of some field with some 
``spherical harmonics'' on the internal manifold,\footnote{The internal 
manifold would be a five-dimensional one, $W$, for
closed string modes in Type IIB, and a three-cycle for open string
states on the flavor D7-branes. For sufficiently small $x$, however, 
amplitudes of exchanging modes with non-trivial ``spherical harmonics'' 
on these internal manifolds are relatively suppressed, and we are not 
so much interested.} and hence $j \leq h_a$ in general. 
Another dimensionless coefficient $c_y$ may contain a contribution from the 
``mass'' associated with the ``spherical harmonics'' over the internal 
manifold, and also include ambiguity (which is presumably of order
unity) associated with making d'Alembertian of the flat metric
background covariant.\footnote{The ambiguity in $c_y/R^2$ includes 
insertion of the curvature tensor, 
\begin{equation}
  \left( [\nabla_M, \nabla_N ]\right)_Q^{\; Q'} = 
  - \Gamma^{Q'}_{QN,M}+\Gamma^{Q'}_{QM, N}
  + \Gamma^{L}_{QM} \Gamma^{Q'}_{LN} 
  - \Gamma^{L}_{QN} \Gamma^{Q'}_{LM} 
  = \frac{\delta^{Q'}_{\; M} g_{QN} - \delta^{Q'}_{\; N} g_{QM}}{R^2},
\end{equation}
which vanishes in flat space. Depending on details of how it is 
inserted, the value of $c_y$ may not be the same for all the individual 
irreducible components of $SO(4,1)$ in a rank-$j$ tensor field 
$A_{m_1 \cdots m_j}$.} 

The equation of motion (in the bulk part)\footnote{There is also IR boundary 
part of the equation motion. We will come back to this issue 
in section \ref{ssec:confine}.} then becomes
\begin{equation}
 g^{m_1 m_2} (\nabla_{m_1} \nabla_{m_2} A^{(y)}_{n_1 \cdots n_j} )
  - \left(\frac{c_y}{R^2}+ \frac{N^{(y)}_{\rm eff.}}{\alpha'}\right)
    A^{(y)}_{n_1 \cdots n_j} = 0. 
\label{eq:bulk-EL-eq}
\end{equation}
Solutions to this equation of motion can be obtained from 
solutions of the following eigenmode equation\footnote{The 
differential operator $\nabla^2 := g^{mn} \nabla_m \nabla_n$ 
is Hermitian under the measure 
$d^4x  dz \sqrt{-g(z)} g^{m_1n_1} \cdots g^{m_j n_j}$. } 
\begin{equation}
 \nabla^2 A_{m_1 \cdots m_j} =
   - \frac{{\cal E}}{R^2} A_{m_1 \cdots m_j},
  \label{eq:Eigenequation}
\end{equation}
by imposing the on-shell condition 
\begin{equation}
 \frac{({\cal E} +c_y)}{\sqrt{\lambda}}+ N^{(y)}_{\rm eff.} = 0.
\label{eq:mass-shell}
\end{equation}
We will work out the eigenmode decomposition for rank-$j$ tensor 
fields in the following, where we only have to work for separate $j$, 
without referring to the mass parameter.\footnote{There are many states 
with the same value of $j$, but with different $c_y$ and $N^{(y)}_{\rm eff.}$.} 

The eigenmode wavefunctions are used not just for construction of  
solutions to the equation of motions, but also in constructing 
the Reggeon exchange contributions to the 
$h+\gamma^{*} \rightarrow h+\gamma^{(*)}$ scattering amplitude. 
The propagator is proportional to 
\begin{equation}
 \frac{-i}{\frac{ {\cal E} +c_y }{\sqrt{\lambda}}
                  + N^{(y)}_{\rm eff.} - i\epsilon}
  \frac{\alpha' R^3}{t_{Ay}}.
\label{eq:propagator-idea}
\end{equation}

The mode equation for a rank-$j$ tensor field $A_{m_1 \cdots m_j}$ on 
AdS$_5$ is further decomposed into those of irreducible representations of
$SO(4,1)$. For simplicity of the argument, we only deal with the mode 
equations for the totally symmetric (and traceless) rank$-j$ tensor fields. 
Namely, 
\begin{equation}
 A_{m_1 \cdots m_j} = A_{m_{\sigma(1)} \cdots m_{\sigma(j)}} \qquad
  \qquad  {\rm for~}{}^{\forall} \sigma \in \mathfrak{S}_j.
\end{equation}
We call them spin-$j$ fields. 

The eigenmode equation (\ref{eq:Eigenequation}) for a totally symmetric 
spin $j$ field can be decomposed into $j+1$ pieces, labeled by
$k=0, \cdots, j$:
\begin{eqnarray}
&& \left( (R^2 \Delta_j) 
- \left[(2k+1)j-2k^2+3k\right] \right) A_{z^k \mu_1 \cdots \mu_{j-k} } 
   \nonumber  \\ 
&& + 2zk \partial^{\hat{\rho}} A_{z^{k-1}\rho\mu_1\cdots \mu_{j-k}}
+ k(k-1) A^{\hat{\rho}}_{z^{k-2}\rho \mu_1 \cdots \mu_{j-k} } 
   \nonumber \\
&& - 2z (D[A_{z^{k+1} \cdots}])_{\mu_1 \cdots \mu_{j-k}}
+ (E[A_{z^{k+2} \cdots}])_{\mu_1 \cdots \mu_{j-k}} = -{\cal E} A_{z^k \mu_1
\cdots \mu_{j-k}}.  
\label{eq:Eigenequation-4+1}
\end{eqnarray}
Here, 
\begin{equation}
A_{z^k \mu_1 \cdots \mu_{j-k}} := A_{\underbrace{z \cdots z}_k \mu_1
 \cdots \mu_{j-k}},   
\end{equation}
and can be regarded as a rank-$(j-k)$ totally symmetric tensor of 
$\SO(3,1)$ Lorentz group.\footnote{
The $\SO(3,1)$ indices with $\hat{}$ in the superscript, 
such as ${}^{\hat{\rho}}$ in $\partial^{\hat{\rho}}$, 
are raised by the 4D Minkowski metric $\eta^{\rho\sigma}$ from 
a subscript ${}_\sigma$, not by the 5D warped metric $g^{mn}$.
} $D[ a ]$ and $E[ a ]$ are operations 
creating totally symmetric rank-$(r+1)$ and rank-$(r+2)$ tensors of 
$\SO(3,1)$, respectively, from a totally symmetric rank-$r$ tensor 
of $\SO(3,1)$, $a$; 
\begin{eqnarray}
 \left(D[ a ]\right)_{\mu_1 \cdots \mu_{r+1}} & := & \sum_{i=1}^{r+1}
   \partial_{\mu_i} a_{\mu_1 \cdots \check{\mu_i} \cdots
   \mu_{r+1}}, 
\label{eq:def-D} \\ 
 \left(E[ a ]\right)_{\mu_1 \cdots \mu_{r+2}} & := & 
  2 \sum_{p < q} \eta_{\mu_p \mu_q}
  a_{\mu_1 \cdots \check{\mu_p} \cdots \check{\mu_q} \cdots
  \mu_{r+2}}.
\label{eq:def-E}
\end{eqnarray} 
The differential operator $\Delta_j$ in the first term is defined, 
as in \cite{BPST-06}, by 
\begin{eqnarray}
 R^2 \Delta_j & := & R^2 z^{-j} \left[ 
    \left(\frac{z}{R}\right)^5 \partial_z 
        \left[ \left(\frac{R}{z}\right)^3 \partial_z \right] 
                           \right] z^j
  + R^2 \left(\frac{z}{R}\right)^2 \partial^2, \nonumber \\  
  & = & z^2 \partial_z^2 + (2j-3)z\partial_z +j(j-4) + z^2 \partial^2.
\end{eqnarray}
The eigenmode equation (\ref{eq:Eigenequation},
\ref{eq:Eigenequation-4+1}) is a generalization of the 
``Schr\"{o}dinger equation'' of \cite{BPST-06} determining the Pomeron 
wavefunction. As we will see, the single-component Pomeron wavefunction 
discussed in \cite{BPST-06} etc. corresponds 
to (\ref{eq:Pomeron-wvfc-1compnt})---that of $(n,l,m) = (0,0,0)$ 
eigenmode in our language, and the Schr\"{o}dinger equation to
(\ref{eq:EGeq-q=not0-n=0}, \ref{eq:EGeq-q=0-n=0-app}); 
there are other eigenmodes, whose
wavefunctions are to be determined in the following.  

In the following sections \ref{ssec:eigen-q=0}--\ref{ssec:eigen-q=not0}, 
we simply state the results of the eigenmode decomposition of
(\ref{eq:Eigenequation}, \ref{eq:Eigenequation-4+1}) for spin-$j$
fields. More detailed account is given in the 
appendix \ref{sec:appendix-mode-fcn}.

\subsection{Eigenvalues and Eigenmodes for $\Delta^\mu = 0$}
\label{ssec:eigen-q=0}

Because of the 3+1-dimensional translational symmetry in $\nabla^2$, 
solutions to the eigenmode equations can be classified by the
eigenvalues of the generators of translation, $(-i\partial_\mu)$.
Until the end of section \ref{ssec:eigen-q=not0}, we will focus on eigenmodes 
in the form of 
\begin{equation}
 A_{m_1 \cdots m_j}(x,z) = e^{i \Delta \cdot x} A_{m_1 \cdots m_j}(z;\Delta),
\end{equation}
and study the eigenmode equation (\ref{eq:Eigenequation}) separately 
for different eigenvalues $\Delta^\mu$. 

The eigenmode equation for $\Delta^\mu=0$ and that for 
$\Delta^\mu \neq 0$ are qualitatively different, and need separate 
study. The eigenmodes for $\Delta^\mu \neq 0$ will be presented in 
section \ref{ssec:eigen-q=not0} 
(and appendix \ref{ssec:appendix-eigen-q=not0}); we begin 
in section \ref{ssec:eigen-q=0} 
(and appendix (\ref{ssec:appendix-eigen-q=0})) with the eigenmode equation  
for $\Delta^\mu = 0$, which is also regarded as an approximation of 
the eigenmode equation for $\Delta^\mu \neq 0$ in the asymptotic UV 
boundary region (at least $\Delta z \ll 1$, and maybe $z$ is as small as $R$).

For now, we relax the traceless condition on the spin-$j$ field $A_{m_1
\cdots m_j}$ ($m_i = 0,1,\cdots,3,z$), and we just assume that the
rank-$j$ tensor field $A_{m_1 \cdots m_j}$ is totally
symmetric.\footnote{This only makes the following presentation more 
far reaching; in the end, it is quite easy to identify which eigenmodes 
fall into the traceless part within $A_{m_1 \cdots m_j}$. 
See (\ref{eq:egval-l=0}--\ref{eq:coeff-l=0-odd}) at the end of 
section \ref{ssec:eigen-q=0}.}
Consider the following decomposition of the space of $z$-dependent 
field configuration $A_{m_1 \cdots m_j}(z; \Delta = 0)$:
\begin{equation}
 A_{z^k \mu_1 \cdots \mu_{j-k}}(z; \Delta^\mu = 0) =
 \sum_{N=0}^{[(j-k)/2]} \left(E^N[ a^{(k,N)}] \right)_{\mu_1 \cdots \mu_{j-k}};
\label{eq:block-dcmp-q=0}
\end{equation}
here, $\left(a^{(k,N)}(z; \Delta^\mu = 0)\right)_{\mu_1 \cdots
\mu_{j-k-2N}}$ is a rank-$(j-k-2N)$ totally symmetric tensor of
$\SO(3,1)$, and satisfies the 4D-traceless condition,  
\begin{equation}
 \eta^{\hat{\mu}_1 \hat{\mu}_2} a^{(k,N)}_{\mu_1 \cdots \mu_{j-k-2N}} = 0.
\label{eq:cond-4D-traceless}
\end{equation}
Thus, the field configuration can be described by 
$a^{(k,N)}$'s with $0\leq k \leq j$, $0\leq N \leq [(j-k)/2]$.  
These components form groups labeled by $n = 0, \cdots, j$, 
where the $n$-th group consists of $a^{(k,N)}$'s with 
$k+2N = n$; they are all rank-$(j-n)$ totally symmetric tensors of 
$\SO(3,1)$; let us call the subspace spanned by the components 
in this $n$-th group as the $n$-th subspace. 
The eigenmode equation for $\Delta^\mu = 0$ becomes block diagonal 
under the decomposition into these subspaces labeled by $n = 0, \cdots, j$. 
(See (\ref{eq:Eigen-eq-q=0-app}) in the appendix.) 
Therefore, the eigenmode equation for $\Delta^\mu = 0$ can be 
studied separately for the individual diagonal blocks.

The $n$-th diagonal block contains $[n/2] + 1$ components, and hence 
there are $[n/2]+1$ eigenmodes. Let ${\cal E}_{n,l}$ ($l=0, \cdots,
[n/2]$) be the eigenvalues in the $n$-th diagonal block. 
The corresponding eigenmode wavefunction is of the form 
\begin{equation}
 \left(a^{(k,N)}(z; \Delta^\mu = 0)\right)_{\mu_1 \cdots\mu_{j-n}} 
= c_{k,l,n} \; \left(\epsilon^{(n,l)} \right)_{\mu_1 \cdots \mu_{j-n}} 
            \; z^{2-j-i\nu},
\end{equation}
where $\epsilon^{(n,l)}$ is a $z$-independent $k$-independent 
rank-$(j-n)$ tensor of $\SO(3,1)$. $c_{k,l,n} \in \R$. 
In the eigenmode equation for $\Delta^\mu = 0$, the eigenmode
wavefunctions are all in a simple power of $z$, and the power is
parameterized by $i\nu$ ($\nu \in \R$). 
The eigenvalues ${\cal E}_{n,l}$ are functions of $\nu$; once the
mass-shell condition (\ref{eq:mass-shell}) is imposed, the eigenmodes 
turn into solutions of the equation of motion, and $i\nu$ is determined 
by the mass parameter. 

The eigenmodes with smaller $(n,l)$ are as follows:
\begin{eqnarray}
 {\cal E}_{0,0} = (j + 4 + \nu^2), & \qquad & 
a^{(0,0)}(z)_{\mu_1 \cdots \mu_{j}} = 
 \epsilon^{(0,0)}_{\mu_1 \cdots \mu_{j}} \; z^{2-j-i\nu}, \\
{\cal E}_{1,0} = (3j + 5 + \nu^2), & \qquad & 
a^{(1,0)}(z)_{\mu_1 \cdots \mu_{j-1}} = 
 \epsilon^{(1,0)}_{\mu_1 \cdots \mu_{j-1}} z^{2-j-i\nu}, \\
 {\cal E}_{2,0} =(5j+4+\nu^2), & \qquad &
\left( \begin{array}{c}
 a^{(0,1)}(z)_{\mu_1 \cdots \mu_{j-2}} \\
 a^{(2,0)}(z)_{\mu_1 \cdots \mu_{j-2}}
       \end{array}\right) =
 \left( \begin{array}{c}
  1 \\ -4j
	\end{array}\right) \; 
 \epsilon^{(2,0)}_{\mu_1 \cdots \mu_{j-2}} \; z^{2-j-i\nu}, \\
 {\cal E}_{2,1} = (j+2+ \nu^2), & \qquad & 
\left( \begin{array}{c}
 a^{(0,1)}(z)_{\mu_1 \cdots \mu_{j-2}} \\
 a^{(2,0)}(z)_{\mu_1 \cdots \mu_{j-2}}
       \end{array}\right) =
 \left( \begin{array}{c}
  1 \\ 2
	\end{array}\right) \; 
 \epsilon^{(2,1)}_{\mu_1 \cdots \mu_{j-2}} \; z^{2-j-i\nu}.
\end{eqnarray}
Empirically, the $j$-dependence of the eigenvalues in the $n$-th 
diagonal block appear to be 
${\cal E}_{n,l} = ((2n+1-4l) j + \nu^2 + O(1))$ ($l = 0,\cdots,[n/2]$), 
[see (\ref{eq:bgn-of-empirical-egval}--\ref{eq:end-of-empirical-egval}) 
in the appendix for more samples of the eigenvalues]
and we promote this $j$-dependence to a rule of the labeling of 
the eigenmodes with $l$.

The eigenmode with $l=0$ is found in any one of the diagonal
blocks ($n = 0, \cdots, j$). Its eigenvalue is 
\begin{equation}
  {\cal E}_{n,0} = (2n+1)j + 2n - n^2 + 4+\nu^2, 
\label{eq:egval-l=0}
\end{equation}
and 
\begin{eqnarray}
 c_{2\bar{k},0,2\bar{n}} = (-)^{\bar{k}} 4^{\bar{k}}
   \frac{ \bar{n}! }{ (\bar{n}-\bar{k})! }
   \frac{ (j-\bar{n}+1)! }{ (j-\bar{n}-\bar{k}+1)! }, & \qquad & 
   (n=2\bar{n}, \bar{k}=0,\cdots,\bar{n}),
  \label{eq:coeff-l=0-even} \\
 c_{2\bar{k}+1,0,2\bar{n}+1} = (-)^{\bar{k}} 4^{\bar{k}}
   \frac{ \bar{n}! }{ (\bar{n}-\bar{k})! }
   \frac{ (j-\bar{n})! }{ (j-\bar{n}-\bar{k})! }, & \qquad &
   (n = 2\bar{n}+1, \bar{k}=0, \cdots, \bar{n}). 
 \label{eq:coeff-l=0-odd}
\end{eqnarray}
These $(n,l) = (n,0)$ eigenmodes are characterized 
by the 5D-traceless condition
\begin{equation}
 g^{m_1 m_2} A_{m_1 \cdots m_j} = 0. \nonumber 
\end{equation}
Thus, the eigenmodes within the 5D-traceless (and totally
symmetric) component---spin-$j$ field---for $\Delta^\mu=0$ 
are labeled simply by $n = 0, \cdots, j$.

\subsection{Mode Decomposition for non-zero $\Delta_\mu$}
\label{ssec:eigen-q=not0}

\subsubsection{Diagonal Block Decomposition for the $\Delta^\mu \neq 0$ Case}

The eigenmode equation 
(\ref{eq:Eigenequation}, \ref{eq:Eigenequation-4+1}) 
is much more complicated in the case of $\Delta^\mu \neq 0$, 
because of the 2nd and 4th terms in (\ref{eq:Eigenequation-4+1}). 
The eigenmode equation is still made block diagonal for an appropriate 
decomposition of the space of field $A_{m_1\cdots m_j}(z; \Delta^\mu)$.

Consider a decomposition 
\begin{equation}
 A_{z^k \mu_1 \cdots \mu_{j-k}}(z; \Delta^\mu)  = 
 \sum_{s=0}^{j-k} \sum_{N=0}^{[s/2]} 
   \left( \tilde{E}^N D^{s-2N} [ a^{(k,s,N)} ] \right)_{\mu_1 \cdots \mu_{j-k}},
\label{eq:block-dcmp-q=not0-tilde}  
\end{equation}
where a new operation $a \mapsto \tilde{E}[a]$ on a totally symmetric 
$\SO(3,1)$ tensor $a$, 
\begin{equation}
  \left(\tilde{E}[ a ]\right)_{\mu_1 \cdots \mu_{r+2}}  :=  
  2 \sum_{p < q}
   \left( \eta_{\mu_p \mu_q} -
          \frac{ \partial_{\mu_p} \partial_{\mu_q} }{ \partial^2 }
  \right)
  a_{\mu_1 \cdots \check{\mu_p} \cdots \check{\mu_q} \cdots
  \mu_{r+2}},
\label{eq:def-E-tilde}
\end{equation}
is used. $a^{(k,s,N)}$'s are totally symmetric 4D-traceless 
(i.e. (\ref{eq:cond-4D-traceless})) rank-$(j-k-s)$ tensor fields 
of $\SO(3,1)$ that satisfies an additional condition, 
the 4D-transverse condition:
\begin{equation}
 \partial^{\hat{\rho}}
  \left( a^{(k,s,N)} \right)_{\rho \mu_2 \cdots \mu_{j-k-s} }
 = i \Delta^{\hat{\rho}} 
  \left( a^{(k,s,N)} \right)_{\rho \mu_2 \cdots \mu_{j-k-s} } = 0.
\label{eq:cond-4D-transverse}
\end{equation} 
The space of field configuration $A_{m_1 \cdots m_j}(z; \Delta^\mu)$
is now decomposed into $a^{(k,s,N)}$'s with $0 \leq k \leq j$, 
$0 \leq s \leq j-k$, $0 \leq N \leq [s/2]$; these components form 
groups labeled by $m = 0, \cdots, j$, 
where the $m$-th group consists of $a^{(k,s,N)}$'s with 
$k+s = m$; they are all rank-$(j-m)$ totally symmetric 4D-traceless and 
4D-transverse tensors of $\SO(3,1)$; let us call the subspace spanned by 
the components in this $m$-th group as the $m$-th subspace. 
The eigenmode equation for $\Delta^\mu \neq 0$ becomes block diagonal 
under the decomposition into these subspaces labeled by $m = 0, \cdots, j$. 
The eigenmode equation for the $m$-th sector is given by 
(\ref{eq:EGeq-q=not0-matrix-m}) in the 
appendix \ref{ssec:appendix-eigen-q=not0}.
The $m$-th subspace should have 
\begin{equation}
  \sum_{s=0}^m \left([s/2] + 1 \right) 
\end{equation}
eigenmodes.

Eigenvalues ${\cal E}$ are determined in terms of the characteristic 
exponent in the expansion of the solution in power series of $z$; 
let the first term in the expansion be $z^{2-j-i\nu}$; the eigenvalues 
are functions of $\nu$ then.  Because the indicial equation at 
the regular singular point $z \simeq 0$ allows us to determine 
the eigenvalues in terms of $\nu$, the eigenvalues in the case of 
$\Delta^\mu \neq 0$ cannot be different from the ones 
we have already known in the $\Delta^\mu = 0$ case. 
In the $m$-th diagonal block, the eigenvalues consist of 
${\cal E}_{n,l}$ with $0 \leq n \leq m$, $0 \leq l \leq [n/2]$.

To summarize, the eigenmodes in the totally symmetric rank-$j$ tensor 
field of $\SO(4,1)$ are labeled by $(n,l,m)$ and $\Delta^\mu$ and
$\nu$. Their eigenvalues ${\cal E}_{n,l}$ depend only on $n$ and $l$
(with $0 \leq n \leq j$ and $0 \leq l \leq [n/2]$) and $\nu$. 
Corresponding eigenmodes are denoted by 
\begin{eqnarray}
 A(x,z)^{n,l,m; \Delta, \nu}_{z^k \mu_1 \cdots \mu_{j-k}} & = &
  \nonumber \\
 e^{i \Delta \cdot x} 
   A^{n,l,m}_{z^k \mu_1 \cdots \mu_{j-k}}(z; \Delta^\mu, \nu) & = &
 e^{i \Delta \cdot x} \sum_{N=0}^{[s/2]} 
    \tilde{E}^{N} D^{s-2N}[\epsilon^{(n,l,m)}] 
    \frac{b^{(j-m)}_{s,N}}{\Delta^{s-2N}}
  \Psi^{(j);s,N}_{i\nu;n,l,m}(-\Delta^2,z).
\label{eq:Pomeron-wvfc-general-expans}
\end{eqnarray}
$\epsilon^{(n,l,m)}$ is a ($z$-independent) totally symmetric 
4D-traceless 4D-transverse rank-$(j-m)$ tensor of $\SO(3,1)$, and 
all the $s$'s appearing in the expression above are understood as
$s=m-k$. $b^{(r)}_{s,N}$ is a
constant whose definition is given in (\ref{eq:def-of-b}) in the appendix.

\subsubsection{Single Component Pomeron Wavefunction}

The Pomeron wavefunction that has been discussed in the literature
(e.g. \cite{BPST-06}) does not look as awful 
as (\ref{eq:Pomeron-wvfc-general-expans}). To our knowledge, 
the Pomeron wavefunction in the literature in the context of hadron
high-energy scattering has been a single component one, $\Psi_{i\nu}(t,z)$. 
How is $A^{n,l,m}_{m_1 \cdots m_j}(z; \Delta^\mu, \nu)$ related to
$\Psi_{i\nu}(-\Delta^2; z)$?

In the block diagonal decomposition of the eigenmode equation, 
there is only one subspace where the diagonal block is $1 \times 1$.
That is the $m=0$ subspace, which consists only of $a^{(0,0,0)}$.
The eigenmode equation is 
\begin{equation}
 \left[\Delta_j - \frac{j}{R^2} \right] a^{(0,0,0)}(z; \Delta^\mu) 
 = - \frac{{\cal E}}{R^2} \; a^{(0,0,0)}(z; \Delta^\mu).
\label{eq:EGeq-q=not0-n=0}
\end{equation}
This equation, as well as (\ref{eq:EGeq-q=0-n=0-app}) in the 
$\Delta^\mu = 0$ case, corresponds to the ``Schr\"{o}dinger equation'' 
in \cite{BPST-06} determining the Pomeron wavefunction. 
It should be noted, however, that we consider that $\nabla^2$ is 
the operator relevant to the eigenmode decomposition\footnote{Thus, 
the propagator (\ref{eq:propagator-idea}) uses the eigenvalue of 
$\nabla^2$, rather than that of $\Delta_j$. The eigenvalue ${\cal E}$ 
of $\nabla^2$ in the $m=0$-th subspace is $(j+4+\nu^2)$ as in 
(\ref{eq:egval-nlm=000}), instead of $(4+\nu^2)$. 
Reference \cite{BPST-06} uses a mode 
$h_{mn} \propto z^{-2} (\eta_{\mu\nu}, \delta_{zz})$ of the spin-2 field to 
fix the details of (\ref{eq:Eigenequation}, \ref{eq:mass-shell}) 
and (\ref{eq:EGeq-q=not0-n=0}).
This $h_{mn}\propto z^{-2}(\eta_{\mu\nu}, \delta_{zz})$ mode, however, corresponds 
to the $(n,l)=(2,1)$ mode of the spin-$j=2$ field 
in (\ref{eq:graviton-wavefnc}), rather than the 5D-traceless 5D-transverse 
mode $(n,l)=(0,0)$. The eigenvalue ${\cal E}_{2,1}=(2+j+\nu^2)$ with $j=2$ 
becomes $(4+\nu^2)$, though.} 
rather than $\Delta_j$; furthermore the operator 
$\nabla^2$ and $\Delta_j$ has a simple relation 
$\nabla^2 = \Delta_j - j/R^2$ only on this $m=0$-th subspace 
of a totally symmetric rank-$j$ tensor field of $\SO(4,1)$.

The eigenvalue is 
\begin{equation}
 {\cal E}_{0,0} = (j+4+\nu^2),  
\label{eq:egval-nlm=000}
\end{equation}
when we define the first term in the power series expansion of $z$ to be 
$z^{2-j-i\nu}$. The eigenmode wavefunction is 
\begin{eqnarray}
 a^{(0,0,0)}(z; \Delta^\mu)_{\mu_1 \cdots \mu_j} & = &  
 \epsilon^{(0,0,0)}_{\mu_1 \cdots \mu_j} \;
 \Psi^{(j)}_{i\nu}(- \Delta^2, z), \\
\Psi^{(j)}_{i\nu}(-\Delta^2,z) & := & 
  \frac{2}{\pi}\sqrt{\frac{\nu \sinh (\pi \nu)}{2R}} \; 
  e^{(j-2)A} K_{i\nu}(\Delta z).
\label{eq:Pomeron-wvfc-1compnt}
\end{eqnarray}
The normalization factor is determined \cite{BPST-06}\footnote{
The Pomeron wavefunction in \cite{NW-first} was of the form 
(\ref{eq:normalization-w-bdry-m=0-Drchlt}), which 
becomes (\ref{eq:Pomeron-wvfc-1compnt}) in the limit of 
$\Lambda \rightarrow 0$, while keeping $z$ and $\Delta^\mu$ fixed. 
}
so that it satisfies the normalization condition\footnote{
The normalization condition is generalized 
to (\ref{eq:wvfc-normalization-cond}) later on.}
\begin{equation}
 \int d^4 x \int dz \sqrt{-g(z)} e^{-2jA} \; [e^{i \Delta \cdot x}
  \Psi^{(j)}_{i\nu}(-\Delta^2,z)] \; [\Psi^{(j)}_{i\nu'}(-\Delta^{'2}, z)
  e^{-i \Delta' \cdot x}]
  = (2\pi)^4 \delta^4(\Delta-\Delta') \; \delta(\nu-\nu').
\end{equation}
The single component Pomeron/Reggeon wavefunction 
$\Psi^{(j)}_{i\nu}(-\Delta^2, z)$ is now understood as 
$\Psi^{(j);0,0}_{i\nu; 0,0,0}(-\Delta^2,z)$.

\subsubsection{5D-Traceless 5D-Transverse Modes}

The eigenmode equation (\ref{eq:Eigenequation}) for a totally symmetric 
rank-$j$ tensor field of $\SO(4,1)$ should be closed within its 
5D-traceless component. The subspace of 5D-traceless component is 
characterized by the 5D-traceless condition
\label{page:5D-trl-trv}
\begin{equation}
 g^{m_1 m_2} A_{m_1 \cdots m_j} (z; \Delta^\mu) = 0.
 \label{eq:cond-5D-traceless}
\end{equation}
The fact that the Hermitian operator $\nabla^2$ maps this subspace to
itself implies that the eigenmode equation of $\nabla^2$ is block
diagonal, when the space of (not-necessarily 5D-traceless) 
$A_{m_1 \cdots m_j}$ is decomposed into the sum of the 5D-traceless
subspace and its orthogonal complement. Collection of the eigenmodes 
with $l=0$ correspond to the subspace of 5D-traceless field configuration.

Similarly, one can think of a subspace of field configuration satisfying 
both the 5D-traceless condition (\ref{eq:cond-5D-traceless}) and 
the 5D-transverse condition
\begin{equation}
 g^{n m_1} \nabla_n A_{m_1 m_2 \cdots m_j} = 0.
\label{eq:cond-5D-transverse}
\end{equation}
Obviously this is a subspace of the subspace of 5D-traceless modes we 
discussed above. Since the Hermitian operator $\nabla^2$ on AdS$_5$ 
maps this new subspace also to itself, the eigenmode equation of 
$\nabla^2$ should also become block diagonal, when the subspace of 
5D-traceless modes is decomposed into this new subspace and its orthogonal 
complement. 

As we will see in the appendix \ref{ssec:appendix-5D-trltrv}, 
there is only one such mode satisfying this set of conditions 
(\ref{eq:cond-5D-traceless}, \ref{eq:cond-5D-transverse}) 
in each one of the $m$-th diagonal block. Thus, the combination 
of the 5D-traceless and 5D-transverse conditions allows us to 
determine an eigenmode completely. This mode turns out to be 
$(n,l,m) = (0,0,m)$ (for $0 \leq m \leq j$). Put differently, 
the eigenmodes with the eigenvalue 
${\cal E}_{n,l} = {\cal E}_{0,0} = (j+4+\nu^2)$ are characterized 
by the traceless and transverse conditions on AdS$_5$.

The eigenmode wavefunctions of the 5D-traceless transverse modes 
$(n,l,m) = (0,0,m)$ are (see the appendix \ref{ssec:appendix-5D-trltrv}) 
\begin{equation}
  \Psi^{(j);s,N}_{i\nu;0,0,m}(-\Delta^2,z) =
    \sum_{a=0}^N (-)^a {}_N C_a 
      \left(\frac{z^3 \partial_z z^{-3}}{\Delta}\right)^{s-2a} 
        \left[(z \Delta)^m \Psi^{(j);0,0}_{i\nu;0,0,0}(-\Delta^2, z)
	\right] \times N_{j,m}.
\label{eq:00m-mode-fcn}
\end{equation}
$N_{j,m}$ is a dimensionless normalization constant. We choose it to
be\footnote{Note that $N_{j,m} = 1$, if $m=0$.} 
\begin{equation}
 N_{j,m}^{-2} = {}_jC_m
  \frac{\Gamma(j+1-i\nu)}{\Gamma(j+1-m-i\nu)}
  \frac{\Gamma(j+1+i\nu)}{\Gamma(j+1-m+i\nu)} 
  \frac{\Gamma(3/2+j-m)}{2^m\Gamma(3/2+j)}
  \frac{\Gamma(2+2j)}{\Gamma(2+2j-m)}, 
\end{equation}
so that the eigenmode wavefunctions are normalized as in 
\begin{eqnarray}
&& \int d^4 x \int_0 dz \sqrt{-g(z)} g^{m_1 n_1} \cdots g^{m_j n_j} \; 
   A^{n,l,m; \Delta, \nu}_{m_1 \cdots m_j}(x,z) \; 
   A^{n',l',m'; \Delta',\nu'}_{n_1 \cdots n_j}(x,z) \nonumber \\ 
&=& (2\pi)^4 \delta^4(\Delta+\Delta^{'}) \; \delta(\nu - \nu') \; 
    \delta_{n,n'} \delta_{l,l'} \delta_{m,m'} \;
    \left[ \epsilon^{(n,l,m)}(\Delta) \right] \cdot
    \left[ \epsilon^{(n',l',m')}(\Delta') \right].
\label{eq:wvfc-normalization-cond}
\end{eqnarray}
Here, $[\epsilon^{(n,l,m)}] \cdot [\epsilon^{'(n,l,m)}] :=
 \epsilon^{(n,l,m)}_{\mu_1 \cdots \mu_{j-m}}
 \epsilon^{'(n,l,m)}_{\nu_1 \cdots \nu_{j-m}}
 \eta^{\hat{\mu}_1 \hat{\nu}_1 } \cdots 
 \eta^{\hat{\mu}_{j-m} \hat{\nu}_{j-m} }$.

\subsubsection{Propagator}

The propagator of the totally symmetric rank-$j$ tensor field 
[resp. spin-$j$ field] on AdS$_5$ is given by summing up propagators 
of the $(n,l,m)$ modes [resp. $(n,l,m)$ modes with $l=0$]. 
For the purpose of writing down the propagator of a given $(n,l,m)$ eigenmode, 
it is convenient to introduce the following notation:
\begin{equation}
 A^{n,l,m; \Delta,\nu}_{m_1 \cdots m_j}(x,z) = 
 \left[A^{n,l,m; \Delta,\nu}_{m_1 \cdots m_j}(x,z) 
 \right]^{\hat{\kappa}_1 \cdots \hat{\kappa}_{j-m}} 
    \epsilon^{(n,l,m)}_{\kappa_1 \cdots \kappa_{j-m}} = 
 e^{i \Delta \cdot x}
 \left[A^{n,l,m}_{m_1 \cdots m_j}(z; \Delta^\mu, \nu) 
 \right]^{\hat{\kappa}_1 \cdots \hat{\kappa}_{j-m}} 
    \epsilon^{(n,l,m)}_{\kappa_1 \cdots \kappa_{j-m}}.
\end{equation}
With this notation, the propagator of the $(n,l,m)$ mode is given by 
\begin{eqnarray}
 G(x,z; x',z')_{m_1 \cdots m_j;n_1 \cdots n_j}^{(n,l,m)} & = & 
 \int \frac{d^4 \Delta}{(2\pi)^4} \int_0^{\infty} d\nu \; 
\frac{- i 
      P^{(j-m)}_{ \rho_1 \cdots \rho_{j-m}; \; \sigma_1 \cdots \sigma_{j-m} } }
     {\frac{{\cal E}_{n,l} + c}{\sqrt{\lambda}} + N_{\rm eff.} - i \epsilon}  
    \frac{\alpha' R^3}{t_y} \nonumber \\ 
&&
   \left[A^{n,l,m; \Delta,\nu}_{m_1 \cdots
    m_j}(x,z)\right]^{\hat{\rho}_1\cdots \hat{\rho}_{j-m}}
   \left[A^{n,l,m; -\Delta,\nu}_{n_1 \cdots n_j}(x',z')
   \right]^{\hat{\sigma}_1 \cdots \hat{\sigma}_{j-m}}.
\label{eq:Pomeron-propagator-nlm}
\end{eqnarray}
Here, $P^{(j-m)}_{\rho_1 \cdots \rho_{j-m}; \; \sigma_1 \cdots \sigma_{j-m}}$ 
is a polarization tensor generalizing 
$\eta_{\rho\sigma} - \partial_\rho \partial_\sigma /\partial^2$; 
when an orthogonal basis 
$\epsilon_a(q) \cdot \epsilon_b(-q) = \delta_{a,b} D_a$ of rank-$r$ 
4D-traceless 4D-transverse tensors is given, 
\begin{equation}
 P^{(r)}_{\mu_1 \cdots \mu_r; \; \nu_1 \cdots \nu_r} := \sum_a \frac{1}{D_a} 
  \epsilon(q)_{a; \; \mu_1 \cdots \mu_r}
  \epsilon(-q)_{a; \; \nu_1 \cdots \nu_r}.
\label{eq:projection-000-r}
\end{equation}
An alternative characterization of this 
$P^{(r)}_{\mu_1 \cdots \mu_r; \nu_1 \cdots \nu_r}$ is given by a 
combination of the two following conditions: one is  
\begin{equation}
 P^{(r)}_{\mu_1 \cdots \mu_r; \; \nu_1 \cdots \nu_r} 
  \epsilon_a^{\hat{\nu}_1 \cdots \hat{\nu}_r}  =  
   \epsilon_{a;\mu_1 \cdots \mu_r},  
\end{equation}
and the other is that  $P^{(r)}_{\mu_1 \cdots \mu_r; \; \nu_1 \cdots \nu_r}$ 
be also a totally symmetric 4D-transverse 4D-traceless tensor 
with respect to $(\mu_1 \cdots \mu_r)$ for any choice of 
$(\nu_1 \cdots \nu_r)$. Its explicit form (\ref{eq:4D-tensor-projection-op}) 
given in the appendix is useful for practical computations.

\subsection{Representation in the Dilatation Eigenbasis}
\label{ssec:repr-dilatation}

It is an essential process in the application of AdS/CFT correspondence 
to classify solutions to the equation of motions on the gravity dual 
background (AdS$_5$) into irreducible representations of the conformal 
group $\SO(4,2)$ (or possibly its supersymmetric extension). 
In the CFT description, primary operators are in one to one
correspondence with (highest weight) irreducible representations of 
the conformal group, and it is believed that one can establish an 
one-to-one correspondence between i) a primary operator in the CFT
description and ii) a group of solutions to the equation of motion forming 
an irreducible representation in the gravity dual description. 
Once this correspondence is given, then hadron matrix elements of 
the primary operators in a (nearly conformal) field theory can be
calculated by using the corresponding solutions to the equation 
of motions (wavefunctions) on AdS$_5$.
Note that the hadron matrix elements of primary operators are all 
that remains unknown in the formulation of conformal operator product
expansion (\ref{eq:conf-OPE-center}). 

Let $P_\mu$, $K_\mu$, $L_{\mu\nu}$ and $D$ denote the generators 
of the unitary operators of the conformal group transformation on the
Hilbert space. They satisfy the following commutation relations:
\begin{eqnarray}
& [ D, P_\mu ] = i P_\mu,
& [ P_\rho, L_{\mu\nu} ] = i ( \eta_{\rho\mu} P_\nu- \eta_{\rho\nu}P_\mu
), \\
& [ D, K_{\mu} ] = -i K_{\mu}, 
& [K_\rho, L_{\mu\nu}] = i (\eta_{\rho\mu} K_\nu- \eta_{\rho\nu} K_\mu), 
\end{eqnarray}
\begin{align}
 & [ P_\mu, K_\nu ] = -2i ( \eta_{\mu\nu} D + L_{\mu\nu} ), \\ 
 & [L_{\mu\nu}, L_{\rho\sigma}] = i (
    \eta_{\nu\rho} L_{\mu \sigma} - \eta_{\nu \sigma} L_{\mu \rho}
    - \eta_{\mu\rho} L_{\nu \sigma} + \eta_{\mu \sigma} L_{\nu \rho} ).
\end{align}

When such a conformal symmetry exists in a conformal field theory 
on 3+1 dimensions, those generators have a representation as
differential operators on fields on $\R^{3,1}$; those differential 
operators are denoted by ${\cal P}_\mu$, ${\cal K}_\mu$, 
${\cal L}_{\mu\nu}$ and ${\cal D}$. The generators and the differential
operators on a CFT are in the following relation:
\begin{equation}
 [ {\cal O}(x), P_\mu] = {\cal P}_\mu {\cal O}(x), \quad 
 [ {\cal O}(x), K_{\mu} ] = {\cal K}_\mu {\cal O}(x), \quad 
 [ {\cal O}(x), D ] = {\cal D} {\cal O}(x), \cdots, 
\end{equation}
and those differential operators acts on primary operators as follows:
\begin{eqnarray}
 {\cal D} \overline{\cal O}_n(x)  & = & -i (x \cdot \partial + l_n)
      \overline{\cal O}_n(x), \\
 {\cal L}_{\mu\nu} \overline{\cal O}_n(x) & = &
    \left(i(x_\mu \partial_\nu - x_\nu \partial_\mu) + [S_{\mu\nu}]
    \right) \overline{\cal O}_n(x), \\
 {\cal P}_\mu \overline{\cal O}_n(x)& = &
       -i \partial_\mu \overline{\cal O}_n(x), \\
 {\cal K}_\mu \overline{\cal O}_n(x)& = &
    \left( -i(2x_\mu x\cdot \partial - x^2 \partial_\mu )
           -i 2l_n x_\mu
           - x^{\nu} [S_{\mu\nu}]
   \right) \overline{\cal O}_n(x),
\end{eqnarray}
where $[S_{\mu\nu}]$ is a finite dimensional representation of
$\SO(3,1)$ generators satisfying the same commutation relation as 
$L_{\mu\nu}$'s. Thus, for a primary operator $\overline{\cal O}_n(x)$, 
$\overline{\cal O}_n(x = 0)$ plays the role of the highest weight state 
\begin{equation}
 [ \overline{\cal O}_n(0), K_\mu ] = 0, \qquad 
 [ \overline{\cal O}_n(0), D] = -i l_n \overline{\cal O}_n(0); 
\end{equation}
all other states in the highest weight state
representation---descendants---are generated by applying 
$[ \bullet, P_\mu]$ multiple times; the whole representation, therefore, 
is spanned by a collection of 
\begin{equation}
 \left\{ \overline{\cal O}_n(0), \; \partial_\mu \overline{\cal O}_n(0), \; 
         \partial_\mu \partial_\nu \overline{\cal O}_n(0), \; \cdots \right\};
\end{equation}
it is also equivalent to a collection of $\overline{\cal O}(x = x_0)$ with 
arbitrary $x_0 \in \R^{3,1}$.

In the preceding sections, we have worked on solutions to the eigenmode 
equation on AdS$_5$; once the mass-shell condition (\ref{eq:mass-shell}) 
is imposed, they become solutions to the equation of motion. 
They are obtained as an eigenmode of the spacetime translation in 3+1 
dimensions, $(-i \partial^\mu) = \Delta^\mu$. Under the conformal group 
$\SO(4,2)$, which contains Lorentz $\SO(3,1)$ symmetry, however, 
an irreducible representation has to include solutions 
with all kinds of eigenvalues $\Delta^\mu$. 

In the case of a scalar field on AdS$_5$, one can think of the following 
linear combination $G(x,z; x_0; R_0)$ (for some $R_0 \ll \Delta^{-1}$):
\begin{equation}
 G(x,z; x_0) =
 \frac{i}{\pi^2}\frac{\Gamma(l_n)}{\Gamma(l_n-2)}
     R_0^{l_n-4}
     \left(\frac{z}{z^2 + (x-x_0)^2}\right)^{l_n} =
 \int \frac{d^4 \Delta}{(2\pi)^4} e^{i \Delta \cdot (x-x_0)}
     \frac{(\Delta z)^2 K_{l_n-2}(\Delta z) }{ (\Delta R_0)^2 K_{l_n-2}(\Delta R_0) }.
\end{equation}
The factor $[e^{i \Delta \cdot x} (\Delta z)^2 K_{l_n-2}(\Delta z)]$ 
in the integrand on the right hand side is a solution to the equation 
of motion of a scalar field on AdS$_5$ whose mass-square 
$M^2_{\rm eff.}$ is given by 
$l_n -2 = i \nu = \sqrt{4 + M^2_{\rm eff.} R^2}$. The coefficient 
of the linear combination, 
$e^{-i \Delta \cdot x_0} [(\Delta R_0)^2 K_{l_n-2}(\Delta R_0)]^{-1}$, is 
chosen so that the integrand behaves as 
\begin{equation}
 e^{i \Delta \cdot (x-x_0)} \left(\frac{z}{R_0}\right)^{4-l_n} 
\end{equation}
at $0 \leq z \ll \Delta^{-1}$. The space of solutions to the equation of
motion $G(x,z;x_0)$ parameterized by $x_0 \in \R^{3,1}$ is alternatively 
spanned by derivatives of $G(x,z;x_0)$ with respect to $x^\mu_0$ 
at $x^\mu_0 = 0$. It is easy to see that this basis 
\begin{equation}
 \left\{ G(x,z;0), \; \partial^{(x_0)}_\mu G(x,z; 0), \;
         \partial^{(x_0)}_\mu \partial^{(x_0)}_\nu G(x,z;0), 
    \; \cdots \right\}
\end{equation}
is an eigenbasis under the action of dilatation, 
${\cal D} := i (z \partial_z + x \cdot \partial)$, and their 
weights are $-i l_n$, $-i (l_n + 1)$, $-i (l_n + 2), \cdots$,
respectively. Correspondence between scalar field wavefunctions on 
AdS$_5$ and scalar primary operators of the dual CFT is established 
in this way \cite{AdSCFT-review}. 

Let us now generalize the discussion above slightly, to construct 
an analogue of $G(x,z; x_0)$ for a spin-$j$ field $A_{m_1 \cdots m_j}$ 
on AdS$_5$, from which the dilatation eigenbasis is constructed. 
To this end, note that all the $(0,0,m)$-modes ($m = 0, \cdots, j$) 
have the leading $z^{2-j-i\nu}$ term in the power series expansion only 
in the $A_{z^0 \mu_1 \cdots \mu_j}$ component, not in any other 
$A_{z^k \mu_1 \cdots \mu_{j-k}}$ components\footnote{
Use (\ref{eq:00m-mode-fcn}).} with $k > 0$. It is possible 
to choose $\epsilon^{(0,0,m)}(\Delta^\mu)$ properly so that  
\begin{equation}
 \sum_{m=0}^j \left[A^{0,0,m;\Delta,\nu}_{\mu_1 \cdots \mu_j} (x,z) 
 \right]^{\hat{\kappa}_1 \cdots \hat{\kappa}_{j-m}}
 \epsilon^{(0,0,m)}_{\kappa_1 \cdots \kappa_{j-m}}
 e^{-i \Delta \cdot x_0}  \simeq 
 e^{i \Delta \cdot (x-x_0)} \left(\frac{z}{R_0}\right)^{2-j-i\nu}
   \epsilon_{\mu_1 \cdots \mu_j}
\end{equation}
in the region near the UV boundary $z \ll \Delta^{-1}$, where 
$\epsilon_{\mu_1 \cdots \mu_j}$ is a $\Delta^\mu$-independent 
4D-traceless totally symmetric rank-$j$ tensor of $\SO(3,1)$;  
the condition on $\epsilon^{(0,0,m)}(\Delta^\mu)$ is 
\begin{eqnarray}
  \epsilon_{\mu_1 \cdots \mu_j}  & = &  
   \left(\frac{R_0}{R}\right)^{2-j} \!\!\!\!\! K_{i\nu}(\Delta R_0)
   \frac{2}{\pi}\sqrt{ \frac{\nu \sinh (\pi \nu)}{2R} } \nonumber \\
&& \sum_{m=0}^j \frac{ N_{j,m} \Gamma(m-j-i\nu)}{\Gamma(-j-i\nu)}
\sum_{N=0}^{[m/2]} \frac{b^{(j-m)}_{m,N}}{\Delta^{m-2N}} 
  \left(\tilde{E}^N D^{m-2N}[\epsilon^{(0,0,m)}]\right)_{\mu_1 \cdots \mu_j} .
\end{eqnarray}
It is possible to invert this relation by 
using (\ref{eq:extract-SO(3,1)-tensor}) and write down 
$\epsilon^{(0,0,m)}(\Delta^\mu)$ in terms of 
$\epsilon_{\mu_1 \cdots \mu_j}$, though we will not present the result here. 
What really matters to us is that 
$\epsilon^{(0,0,m)}(\lambda \Delta) = 
 \epsilon^{(0,0,m)}(\Delta) \lambda^{i\nu}$.
With $\epsilon^{(0,0,m)}$'s satisfying the condition above, one can see
that the following linear combination of solutions to the equation of motion, 
\begin{equation}
 G_{m_1 \cdots m_j}(x,z; x_0) := \int \frac{d^4 \Delta}{(2\pi)^4} 
  \sum_{m=0}^j
    \left[ A^{0,0,m;\Delta,\nu}_{m_1 \cdots m_j}(x,z)
    \right]^{\hat{\kappa}_1 \cdots \hat{\kappa}_{j-m}}
    \epsilon^{(0,0,m)}(\Delta)_{\kappa_1 \cdots \kappa_{j-m}}
   e^{-i \Delta \cdot x_0}, 
\end{equation}
has a property 
\begin{equation}
 G_{m_1 \cdots m_j}( \lambda x, \lambda z; \lambda x_0) = 
\lambda^{-(2+j+i\nu)} G_{m_1 \cdots m_j}(x,z; x_0).
\end{equation}
$i\nu$ is determined by the mass parameter on AdS$_5$, once the 
mass shell condition (\ref{eq:mass-shell}) is imposed.
Therefore, $G_{m_1 \cdots m_j}(x,z;0)$ is an eigenstate of dilatation, 
and so are the derivatives of $G_{m_1 \cdots m_j}(x,z;x_0)$ with respect
to $x_0^\mu$ at $x_0^\mu = 0$. All the derivatives combined forms a 
dilatation eigenbasis in the space of solutions to the equation of
motion of a spin-$j$ field. 

It is now clear that the eigenmodes with $(n,l,m) = (0,0,m)$ 
($0 \leq m \leq j$) and arbitrary $\Delta^\mu$ as a whole---modes that 
satisfy the 5D-traceless and 5D-transverse conditions 
(\ref{eq:cond-5D-traceless}, \ref{eq:cond-5D-transverse})---forms 
an irreducible representation of the conformal group. 
If one is interested purely in the matrix element of a spin-$j$ 
primary operator $\overline{\cal O}_n(x_0 = 0)$ of an approximately 
conformal gauge theory, then the matrix element can be calculated 
by using the wavefunction $G_{m_1 \cdots m_j}(x,z; 0)$.
Note that the $m=0$ mode alone, where the Pomeron/Reggeon wavefunction 
has a single component as in \cite{BPST-06}, cannot reproduce all the 
matrix element associated with matrix elements of spin-$j$ primary 
operators. 


\subsection{Confinement Effect}
\label{ssec:confine}

{\bf Top down approach}:
QCD in the real world is not a conformal gauge theory, 
but it has a mass gap in the hadron spectrum due to confinement. 
Confinement of a nearly conformal strongly coupled gauge theory 
is realized in its gravitational dual description in the form of 
a nearly AdS geometry with a minimum value in the warp factor.

Klebanov--Strassler geometry of Type IIB string 
theory \cite{Klebanov-Strassler} will be one of the most popular 
background geometries of that kind. The Klebanov--Strassler geometry 
is not dual to a confining gauge theory that is asymptotically free, 
however; it is dual to a gauge theory that is confining in the infrared, 
but its 't Hooft couplings become stronger and stronger toward ultraviolet. 
Such geometries as Klebanov--Strassler are not truly dual to the QCD 
of the real world, but still one will be able to learn a lot from studying 
the mode-decomposition on such geometries. 

Mode decomposition can be carried out, once we know the background 
configuration and the action of the bilinear fluctuations around the 
background; we do not need interactions of stringy fields. 
Thus, it will be a doable task, at least at the supergravity level. 
Reduction over the $W_5 = T^{1,1}$ geometry has been worked out in the 
literature, and one is left to translate the smoothness condition of mode 
functions at the tip of the deformed warped conifold into the language of 
boundary conditions on a warped 4+1-dimensional spacetime.\footnote{
\label{fn:bdry}Such geometries typically are often in the form of 
$\R^{3,1} \times W'_n$ which nearly remains constant around the tip of 
the throat $r=0$, and a shrinking $(5-n)$-cycle with the metric  
$ds^2 = dr^2 + r^2 (d\Omega_{5-n})^2$.
For simplicity, let $n=4$ and $d\Omega_1 = d \theta$. 
A scalar field $\phi(r,\theta)$ with smooth configuration in the coordinate 
$(r \cos \theta, r\sin\theta)$ is decomposed into 
$\sum_k e^{i k \theta} \phi_k(r)$, when the mode $\phi_k(r)$ needs to be
in the form of $r^k \times {\rm fcn}(r^2)$. Thus, 
$\partial_r[ r^{-k}\phi_k(r) ] = 0$ at $r=0$.} 
The authors do not find a reason not 
to work on it, except that it will take extra time to do so.  

In this article, however, we set higher priority in getting a broader 
perspective on the subject ranging from string theory to hadron physics, 
and avoid taking too much time to solve technical problems in string theory. 
Instead, we discuss in the following, two temporary approaches of 
implementing the confinement effects; one is an effective-theory model 
building approach, and the other is phenomenological approach. 
We will proceed with the phenomenological approach in the following 
sections, although we understand that the topdown approach above 
will eventually replace/backup/verify the phenomenological approach 
to be adopted in this article. The following ``effective theory model building 
approach'' is not used in this article, but we present it here, because 
it helps us understand physical meaning (hidden assumptions) of 
the phenomenological approach.

{\bf Effective Theory Model Building Approach}: 
The hard-wall model and its variations are introduced in order to mimic 
the presence of minimum value of the warped factor, mass gap and 
nearly AdS background geometry. It remains simple enough so that 
analytic results are obtained in a relatively short amount of time, 
though we cannot discuss stability of the geometry or theoretical 
consistency of string theory.

With this philosophy in mind, one could think of implementing the 
confining effect in the form of 
\begin{equation}
 S = \int d^4 x \int_0^{1/\Lambda} dz \sqrt{-g(z)} \; {\cal L}_{\rm bulk}
  + \int d^4 x \sqrt{-g}|_{z=1/\Lambda} \; {\cal L}_{\rm bdry},
\end{equation}
where the background geometry remains to be AdS$_5$, and the holographic 
radius $z$ is cut-off at $z = \Lambda^{-1}$.
Note that different choices of ${\cal L}_{\rm bdry}$ lead to different 
physics; to be more precise, different choices of 
$({\cal L}_{\rm bulk}, {\cal L}_{\rm bdry})$ modulo partial integration should 
be regarded as different models. 
It is reasonable to have such freedom in the choice of effective-theory 
models, because we know that there are more than one holographic backgrounds 
of Type IIB string theory that are dual to confining gauge theories. 
Such constraints as $\SO(3,1)$ symmetry unbroken global symmetry
of a strongly coupled gauge theory, however, are very weak 
in constraining ${\cal L}_{\rm bdry}$.

Once a model is fixed, then the Euler--Lagrange equation of 
this theory not only includes the equation of motion in the 
bulk (\ref{eq:bulk-EL-eq})=(\ref{eq:Eigenequation}+\ref{eq:mass-shell})  
but also boundary conditions at $z=1/\Lambda$.
Different models (i.e., different ${\cal L}_{\rm bdry}$) predict different 
Pomeron/Reggeon wavefunctions.

We require that $\SO(3,1)$ symmetry is preserved even in 
${\cal L}_{\rm bdry}$. Boundary conditions might introduce mixing between 
the eigenmode decomposition determined in the bulk, in principle, but 
the unbroken $\SO(3,1)$ symmetry excludes mixing between 
$\SO(3,1)$-irreducible tensors of different ranks. This observation still 
does not exclude mixing among $(n,l,m)$-modes of a spin-$j$ totally symmetric 
field on AdS${}_5$ with a common $m$, but different $(n,l)$'s.  

{\bf Phenomenological Approach}: As another alternative approach, 
one can think of a phenomenological approach, which is to start from a small 
number of parameters, and let the physical consequences constrain those 
parameters. When one finds that reasonable physical consequences cannot be 
available under a given set of parameters, then a little more parameters 
will be introduced so that more freedom is available. 

As one of the simplest trial parametrizations of the confining effect, 
we make a following changes in the mode functions 
$\Psi^{(j);0,0}_{i\nu;0,0,m}(\Delta, z)$: 
\begin{equation}
 K_{i\nu}(\Delta z) \longrightarrow
 \left[ K_{i\nu}(\Delta z) 
  + \frac{\pi}{2}\frac{c^{(j)}_{i\nu;0,0,m}}{\sin(\pi i \nu)} I_{i\nu}(\Delta z)
 \right]   =: ``K_{i\nu}(\Delta z)''.
\label{eq:modified-K-inu}
\end{equation}
$c^{(j)}_{i\nu;0,0,m}$'s, which may depend on $\Delta^2$ and $\Lambda$, are 
the parameters we introduce. An implicit assumption here is that 
the confining effect does not introduce mixing among modes with 
different $(n,l,m)$'s. Under this assumption, however, the parametrization 
above does not lose any generality; once the ratio between 
the $K_{i\nu}(\Delta z)$ wave and $I_{i\nu}(\Delta z)$ is given for 
$\Psi^{(j);0,0}_{i\nu;0,0,m}(-\Delta^2,z)$, there is no freedom left for 
the other $\Psi^{(j);s,N}_{i\nu; 0,0,m}(-\Delta^2,z)$ functions 
($(s,N) \neq (0,0)$) of 
the same $(n,l,m)=(0,0,m)$ mode, because the relation among them 
is completely fixed by the equation of motion in the bulk.
In section \ref{ssec:D=0-cancellation}, we will carry out a test of 
whether this simple parametrization works well or not. 

When the infrared boundary is introduced in the holographic background 
geometry, the normalization of the Pomeron/Reggeon wavefunction also 
needs to be changed. 
In the case of $(n,l,m)=(0,0,0)$ mode, with the Dirichlet boundary 
condition at the IR boundary $z = 1/\Lambda$, for example, 
the wavefunction $\Psi^{(j);0,0}_{i\nu;0,0,0}=\Psi^{(j)}_{i\nu}(-\Delta^2,z)$ 
was given the following normalization \cite{BPST-06, NW-first}:
\begin{eqnarray}
 \Psi^{(j)}_{i\nu}(- \Delta^2, z) & = & e^{(j-2)A}
  \frac{2}{\pi} \sqrt{ \frac{\nu \sinh(\pi\nu)}{2R} } 
  \sqrt{ \frac{ I_{i\nu}(x_0)}{I_{-i\nu}(x_0)} }
  \left[ K_{i\nu}(\Delta z)
     - \frac{K_{i\nu}(x_0)}{I_{i\nu}(x_0)} I_{i\nu}(\Delta z),
  \right], \label{eq:normalization-w-bdry-m=0-Drchlt}
\end{eqnarray}
with an extra factor $\sqrt{I_{i\nu}(x_0)/I_{-i\nu}(x_0)}$,  
where $x_0 := \Delta/\Lambda$. This result is generalized as follows. 
By repeating the same argument as in the appendix \ref{sssec:normalization}, 
one finds that the normalization factor $N_{j,m}$ should be replaced by 
\begin{equation}
  N_{j,m} \longrightarrow N_{j,m} \times \frac{1}{\sqrt{1-c^{(j)}_{i\nu;0,0,m}}}.
\label{eq:normalization-w-bdry}
\end{equation}
The Dirichlet boundary condition for the $m=0$ mode above corresponds 
to $(1-c^{(j)}_{i\nu;0,0,0})=[I_{-i\nu}(x_0)/I_{i\nu}(x_0)]$; the modified 
normalization (\ref{eq:normalization-w-bdry-m=0-Drchlt}) is a special case 
of (\ref{eq:normalization-w-bdry}).
The mode functions are defined, so far, for $\nu \geq \R$, since 
the eigenvalue ${\cal E}_{0,0}=4+j+\nu^2$ depends only on $\nu^2$. 
When the modefunction is analytically continued to the $\nu<0$ region, 
the mode function for $-\nu$ should be the same as $+\nu$. 
From this observation, it follows that 
\begin{equation}
(1-c^{(j)}_{-i\nu;0,0,m}) = (1-c^{(j)}_{i\nu;0,0,m})^{-1}.
  \label{eq:IR-effect-nu-asymmetry}
\end{equation}
%

\section{Organizing the Scattering Amplitude on AdS$_5$}
\label{sec:organize}

\subsection{``Effective'' String Field Action on AdS$_5$}

If we are to start from Type IIB string theory on 10-dimensions 
with a background that is approximately AdS$_5 \times W_5$ (except near
the infrared boundary), one can think of an effective theory on AdS$_5$ after 
carrying out ``spherical harmonics'' mode decomposition on $W_5$.
As we have already discussed in section \ref{sec:mode-fcn} 
how to construct propagators in such an effective theory, we would now 
like to construct the scattering amplitude.

For this purpose, we need interaction among string fields, and we 
turn to the cubic string field theory, which we reviewed already in 
section \ref{sec:cubic-sft}. This allows us to write down a concrete 
expression for the scattering amplitude.  
Clearly the biggest drawback of this approach is in the fact that 
no stable background geometry AdS$_5 \times W_{21}$ is known 
in bosonic string theory for some 21-dimensional internal manifold $W_{21}$. 
In the following, we will construct an ``effective'' action on AdS$_5$ 
by carrying out dimensional reduction of the cubic string field theory 
action, as if there exists an AdS$_5 \times W_{21}$ solution to 
the bosonic string theory. This is not meant to claim that we obtain 
such an action as an effective theory of the bosonic string theory, 
but to use it as a starting point in constructing a toy-model 
scattering amplitude of a hadron and a (virtual) photon that may 
still carry some fragrance of interaction structure in superstring
theory. 

Let us start off by clarifying the relation between the normalization of 
string component fields in (\ref{eq:string-field-comp-exp},
\ref{eq:cubic-sft-26D-kin}, \ref{eq:cubic-sft-26D-kin-ldg-trj}) and 
that of the component fields in (\ref{eq:cubic-sft-kin-5D}). 
All the component fields in (\ref{eq:string-field-comp-exp}) are 
normalized so that they have canonically normalized kinetic terms 
in the action in the 26-dimensional spacetime. Now, we make them 
dimensionless by redefinition $\phi \rightarrow g_o^{-1} \phi$, 
$A_M \rightarrow g_o^{-1} A_M$, etc. All the terms in the cubic string 
field theory---both the kinetic terms and interactions---will then have 
$(1/g_o^2)$ as an overall factor. When a mode decomposition of the
following form is assumed for the component in this new normalization, 
\begin{equation}
 \phi(x,z,\theta) = \sum_y \phi^{(y)}(x,z) Y_y (\theta), \quad 
 A_{M}(x,z,\theta) = \left\{  \begin{array}{ll}
	  \sum_y A^{(y)}_m (x,z) Y_y(\theta) & M=m=0,\cdots,3,z \\
          0 & M = 5, \cdots, 25.
		  \end{array}\right.
\end{equation}
Similarly decomposition holds for spin-$h_a$ fields 
$A_{M_1 \cdots M_{h_a}}(x,z,\theta)$;  
we take spherical harmonics $Y_y(\theta)$ (labeled by $y$) to be 
dimensionless, so that the component fields on AdS$_5$ such as 
$\phi^{(y)}(x,z)$, $A^{(y)}_m(x,z)$, $A^{(y)}_{m_1 \cdots m_{h_a}}(x,z)$
are also dimensionless. 

The overall coefficient of the ``effective'' action on AdS$_5$ then
becomes a dimension-$(+3)$ parameter 
\begin{equation}
 \frac{ {\rm vol} (W_{21})}{2(g_o)^2} \times {\cal O}(1),
\label{eq:prefactor-reduction-W21}
\end{equation}
which is to be identified with the overall coefficient $t_y/(2 R^3)$ 
in (\ref{eq:cubic-sft-kin-5D}). Reduction of interaction terms 
(\ref{eq:cubicSFT-interaction-A}, \ref{eq:cubicSFT-interaction-B}, 
\ref{eq:cubicSFT-interaction-C}) also yield the same overall factor 
(\ref{eq:prefactor-reduction-W21}) apart from possibly order one 
factor coming from overlap integration of spherical harmonics over 
the internal manifold. Because the amplitudes from exchanging states 
with higher spherical harmonics are suppressed in small-$x$ DIS and 
DVCS (e.g. \cite{NW-first}), we will be interested only in the interactions 
involving $\phi^{(y)}$--$\phi^{(y)}$--[intermediate-states] and 
$A_m^{(y)}$--$A^{(y)}_m$--[intermediate-states] cubic couplings, with 
the intermediate states having spherical harmonics $Y(\theta) = 1$.
The overall factor of the cubic interactions then becomes precisely the 
same as that of the kinetic terms of $\phi^{(y)}$ and $A_m^{(y)}$.

For this reason, we write down the following interaction terms for 
the ``effective'' action on AdS$_5$:
\begin{align}
 S_{\rm eff.~int.}  = 
 - \frac{t_{\phi y} \lambda_{\rm sft}}{3\alpha' R^3}
 \int d^{4}x dz \sqrt{-g(z)} \hat E& \left(
  3 \tr \left[ \phi^2_y \phi \right]
  + \sqrt{\frac{8\alpha'}{3}} 
      \tr \left[ (-i A_m) \left(\phi_y \overleftrightarrow{\nabla}^m
                                  \phi_y  \right)\right]   \right.
  \notag  \\
& \quad \left.  -\frac{8\alpha'}{9\sqrt{2}}
    \tr \left[ f_{mn}\left( \phi_y \overleftrightarrow{\nabla}^m
                   \overleftrightarrow{\nabla}^n \phi_y \right) \right]
-\frac{5}{9\sqrt{2}}\tr \left[f^m_m\phi^2_y \right]
\right. \notag
\\&\quad +
\left.
  \frac{2\sqrt{\alpha'}}{3}\tr \left[ (\nabla_m g^m)\phi^2_y \right] 
 -\frac{11}{9} \tr \left[h \phi^2_y \right]
\right) + \cdots.
\end{align}
Fields without a label $y$ are to be used for the intermediate states 
exchanged in the $t$-channel (in the sense that we explained 
in section \ref{ssec:Veneziano}); $\phi_y$ are for the incoming and 
outgoing states. 
Partial derivatives have been replaced by covariant derivatives on 
AdS$_5$. Similarly, all other
interactions such as (\ref{eq:cubicSFT-interaction-B},
\ref{eq:cubicSFT-interaction-C}) in 26-dimensions also give rise to 
their corresponding cubic interactions on AdS$_5$. 
Certainly such a choice of ``effective'' action on AdS$_5$ will be 
one of the most likely (and simple enough) set-ups that may still 
maintain some aspects of scattering amplitude in string theory, 
although top-down justification is not given. 

We will only sum up $t$-channel amplitudes where $Y_y(\theta) = 1$ modes 
of the stringy states in the leading Reggeon/Pomeron trajectory are 
exchanged, because that constitutes the dominant contribution in 
the small $x$ scattering. Thus, three point interactions of such modes 
with incoming and outgoing tachyon states are necessary, which 
we write down as follows, 
\begin{equation}
\Delta S_{\rm eff.int.} = - \frac{t_{h} \lambda_{\rm sft}}{R^3 \alpha'}
 \int d^{4}x dz \sqrt{-g(z)} \; \hat{E} \; 
  \tr \left[ A^{(Y)}_{m_1 \cdots m_N}
                   \left( \phi \overleftrightarrow{\nabla}^{m_1}
                         \cdots \overleftrightarrow{\nabla}^{m_N}      
                          \phi \right) 
      	   \right] 
  \left( \frac{8\alpha'}{27} \right)^{\frac{N}{2}}   
  \frac{(-i)^N}{\sqrt{N!}}, 
\end{equation}
by keeping only the $Y_y(\theta)=1$ modes and replacing derivatives 
in (\ref{eq:cubicSFT-interaction-B}) by covariant derivatives.
The normalization constant $t_{\phi y}$ for the target hadron kinetic term 
is now simply written as $t_h$, as we will only have to pay attention to 
individual choices of target hadrons (individual choices of $Y_y(\theta)$) 
in the external states. 
Similarly, we also need interaction of the same group of modes 
with the incoming and outgoing photon states, which we write down as follows: 
\begin{eqnarray}
\Delta S_{\rm eff.~int.} &=&
   - \frac{t_{\gamma} \lambda_{\rm sft}}{R^3 \alpha'}
        \int d^{4}x dz \sqrt{-g(z)}\; \hat{E} \; \\
&& \quad
   {\rm Tr} \left[ A^{(N)}_{m_1 \cdots m_N}
                  \left( A_l (-i \overleftrightarrow{\nabla}^{m_1})
                         \cdots (-i \overleftrightarrow{\nabla}^{m_N})
                         A_k \right)
  \left( \frac{8\alpha'}{27} \right)^{\frac{N}{2}}   
      \frac{g^{kl} \frac{16}{27}}{\sqrt{N!}}+ \cdots 
  \right],  \nonumber 
\end{eqnarray}
following the same procedure by starting from 
(\ref{eq:cubicSFT-interaction-C}).
We have retained only the terms that have $N$-derivatives and are proportional 
to $\eta^{kl}$, as they are necessary in determining the 
``twist-2'' contributions to the structure function $V_1$. 
Since we only need the normalization constant $t_{A y}$ of  
the kinetic term of the external state only for the spherical harmonics 
$Y(\theta)=1$, we no longer need to refer to the choice of spherical 
harmonics; $t_{A y}$ is therefore rewritten as $t_\gamma$.

\subsection{External States Wavefunction}

The vertex operator insertions in the world-sheet calculation 
are replaced by appropriate external state wavefunctions in amplitude 
calculations based on string field theories. 

First, the insertions of vertex operator of the form (\ref{eq:vx-op-D7-vector})
for the U(1) currents on flavor D7-branes are replaced by wavefunctions 
for the massless vector field in the bosonic string theory. 
We use the wavefunctions for the incoming state $\gamma^{*}(q_1)$ and 
outgoing state $\gamma^{(*)}(q_2)$
\begin{eqnarray}
 A^{\rm in}_m(x_\gamma, z_\gamma) = R \int \frac{d^4 q_1}{(2\pi)^4}
     e^{i q_1 \cdot (x_\gamma - (\bar{x} - (\Delta x)/2)}
       A_m(z_\gamma; q_1), 
\label{eq:photon-wvfc-xz-1} \\
 A^{\rm out}_m(x_\gamma, z_\gamma) = R \int \frac{d^4 q_2}{(2\pi)^4}
     e^{-i q_2 \cdot (x_\gamma -(\bar{x}+ (\Delta x)/2))}
       A_m(z_\gamma; q_2),
\label{eq:photon-wvfc-xz-2}
\end{eqnarray}
where $A_m(z; q)$ on the right-hand sides are the wavefunctions given in
(\ref{eq:photon-wvfc}).
A factor $R$ is inserted here, because we adopted a normalization
convention so that $A^{\rm (in/out)}_m(x,z)$ on AdS$_5$ is
dimensionless.\footnote{\label{fn:open-kin-normalization}
$A_m(x,z)$ is often normalized so that 
it has mass dimension $(+1)$, and hence this factor $R$ is not necessary then. 
In the case the gauging of a global symmetry of a strongly coupled gauge 
theory is realized in the form of flavor D7-brane, the natural reduction 
of the 7-brane action on a three-cycle leads to the form of 
\begin{equation}
 S_{\rm eff.} \sim - \frac{N_c}{R} \int d^4x dz \sqrt{-g(z)} F_{mn} F^{mn};
\end{equation}
the external state wavefunction (\ref{eq:photon-wvfc-xz-1}, \ref{eq:photon-wvfc-xz-2}) without the factor $R$ can be used in such cases. 
In the presentation adopted in this section, where bosonic string is used  
and the gauge field is assigned zero mass dimension (like other higher 
spin fields), the factor $R$ is included in (\ref{eq:photon-wvfc-xz-1}, 
\ref{eq:photon-wvfc-xz-2}), and the kinetic term of $F_{mn} F^{mn}$ has the 
coefficient $t_{\gamma}/R^3$ instead. Thus, we can think of $t_{\gamma}$ as 
something like $N_c$.} 
The arguments of the electromagnetic current insertions  
$T\{J^\nu(x) J^\mu(y)\}$---coordinates in the boundary theory $x$ and 
$y \in \R^{3,1}$------are now denoted by $\bar{x}+(\Delta x)/2$ and 
$\bar{x}- (\Delta x)/2$, respectively. 

The vertex operators (\ref{eq:vx-op-D7-scalar}) for the target hadron are 
replaced by wavefunctions of the form 
\begin{equation}
 \phi^{\rm in}(x_h, z_h) = e^{i p_1 \cdot x_h} \Phi(z_h; m_n), \qquad 
 \phi^{\rm out}(x_h, z_h) = e^{-i p_2 \cdot x_h} \Phi(z_h; m_n), 
\end{equation}
where $\Phi(z; m)$'s on the right-hand sides are the wavefunction given
by (\ref{eq:hadron-wvfc}). The first one is for the incoming state, and 
the second for the outgoing hadron. 

\subsection{Leading Trajectory Contribution to the Compton Tensor}

When the target hadron is to be identified with some Kaluza--Klein 
state of the tachyon of the bosonic string theory, then 
$\Delta_\phi -2 = \sqrt{4 + c -\sqrt{\lambda}}$ is not real valued for 
$\lambda \gg 1$. We treat this $\Delta_\phi -2$ as if it were real valued, 
until last moment. Since our true interest is in scattering amplitude 
in Type IIB string theory, or in hadron scattering in the real world, 
this problem is absent in such situations, and we do not bother about this 
issue.  

Let us combine all the pieces together to organize an amplitude 
of photon-tachyon scattering given by $t$-channel exchange of 
leading trajectory spin-$j$ state reduced to AdS$_5$ with 
$Y_y(\theta) = 1$. Such an amplitude---denoted by 
$i {\cal M}^{(t)}_{(N_{\rm eff.}=j,j)}$---consists of $t$-channel 
exchange of all the eigenmodes labeled by $(n,l,m)$.
We will further focus on contributions from $(n,l,m) = (0,0,m)$.
It is given by 
\begin{eqnarray}
i {\cal M}^{(t)}_{(j,j);(0,0,m)}
& \simeq &
    \frac{-i t_{\gamma}}{R^3 \alpha'}
         \int d^4 x_\gamma dz_\gamma \sqrt{-g(z_\gamma)}
              J^{\gamma\gamma}_{k_1 \cdots k_j; p q} g^{pq} 
              \left( g^{k_1 r_1} \cdots g^{k_j r_j}\right)(z_\gamma)
              \left(\frac{\alpha'}{2}\right)^{j/2} e^{-2A(z_\gamma)} \nonumber \\
& & \frac{-i t_h}{R^3 \alpha'}
         \int d^4 x_h dz_h \sqrt{-g(z_h)}
              J^{hh}_{l_1 \cdots l_j}
              \left( g^{l_1 s_1} \cdots g^{l_j s_j}\right)(z_h)
              \left(\frac{\alpha'}{2}\right)^{j/2} e^{-2A(z_h)} \nonumber \\
&& \frac{1}{j!} \left(\frac{27}{16}\right)^{\alpha' t - (j-1)}
          \left[ e^{2A(z_\gamma)} e^{2A(z_h)}
                 G^{(0,0,m)}(x_\gamma,z_\gamma; x_h, z_h)_{r_1 \cdots
		 r_j; s_1 \cdots s_j} \right];    
\end{eqnarray}
just like in the amplitude calculation in section \ref{ssec:Veneziano}, 
this amplitude is meant to be the coefficient of 
${\rm Tr}[\lambda^{\gamma 2} \lambda^{\gamma 1} \lambda^{h 1} \lambda^{h 2}]$. 
$J^{\gamma \gamma}$ and $J^{hh}$ above are given by the external state 
wavefunctions as follows:
\begin{eqnarray}
 J^{\gamma\gamma}_{k_1 \cdots k_j; pq}(x_\gamma, z_\gamma) & = &
    (-i)^j \left[A^{\rm out}_p \overleftrightarrow{\nabla}_{k_1} \cdots 
         \overleftrightarrow{\nabla}_{k_j}
                 A^{\rm in}_q \right](x_\gamma, z_\gamma), \\ 
 J^{hh}_{l_1 \cdots l_j}(x_h, z_h) & = &
    (-i)^j \left[\phi^{\rm in} \overleftrightarrow{\nabla}_{l_1} \cdots 
         \overleftrightarrow{\nabla}_{l_j} \phi^{\rm out} \right](x_h, z_h). 
\end{eqnarray}
Here, $\phi^{\rm in/out}(x_h,z_h)$ are both of mass dimension $(-1)$, 
and $A^{\rm in/out}_m(x_\gamma, z_\gamma)$ of mass dimension 
$(+3)+ {\rm dim}[\epsilon_\mu]$. From this expression, 
one can see that the first line has mass dimension 
$(+6)+2 \times{\rm dim}[\epsilon_\mu]$, the second line $(-2)$, and 
the last line $0$. Thus, $i{\cal M}^{(t)}_{(j,j);(0,0,m)}$ is 
a function of $p_1^{\kappa}$, $p_2^{\kappa}$, $\bar{x}^\kappa$ and 
$\Delta x^{\kappa}$ of mass dimension $4 + 2 \times {\rm dim}[\epsilon_\mu]$. 
This is precisely the property expected for 
\begin{equation}
(i)^2 \bra{h(p_2)} T \left\{ J^\nu(\bar{x}+(\Delta x)/2)
  J^\mu(\bar{x}-(\Delta x)/2) \right\} \ket{h(p_1)}
 \epsilon^1_\mu \epsilon^{2*}_\nu . 
\label{eq:current-current-ME}
\end{equation}
Its Fourier transform with respect to $(\Delta x)^\mu$ becomes 
$(iT^{\mu\nu}) \times e^{-i \bar{x} \cdot (p_2-p_1)}$. 

If we carry out integration over $d^4 x_\gamma$, $d^4 x_h$ and 
$d^4 (\Delta x)$ first, then the three integration variables 
$\Delta^\mu$ in (\ref{eq:Pomeron-propagator-nlm}) and $q_{1,2}$ 
in (\ref{eq:photon-wvfc-xz-1}, \ref{eq:photon-wvfc-xz-2}) are
determined in terms of the input $p^\mu_{1,2}$ and $q^\mu$; we have 
$\Delta^\mu := (p_2 - p_1)^\mu$, $q_2^\mu = (q - \Delta/2)^\mu$
and $q_1 := (q + \Delta/2)^\mu$. As a result, it follows that 
\begin{eqnarray}
 \!\!\!\!\!\!\!\!\!\!\!  \!\!\!\!\!\!\!\!\!\!\!  
[T^{\mu \nu} \epsilon^1_\mu \epsilon^{2*}_\nu]^{(t)} & = &
\int d^4 (\Delta x) \; e^{-i q \cdot (\Delta x)} 
  \; {\cal M}^{(t)}_{(j,j);(0,0,m)}|_{\bar{x}=0} \nonumber \\
  & \simeq & 
    \frac{t_{\gamma}}{R^3 \alpha'}
         \int dz_\gamma \sqrt{-g(z_\gamma)}
              \bar{J}^{\gamma\gamma}_{k_1 \cdots k_j; p q} R^2 g^{pq} 
              \left( g^{k_1 r_1} \cdots g^{k_j r_j}\right) 
              \left(\frac{\alpha'}{2}\right)^{j/2} 
   \nonumber \\
& & \frac{t_h}{R^3 \alpha'}
         \int dz_h \sqrt{-g(z_h)}
              \bar{J}^{hh}_{l_1 \cdots l_j}
              \left( g^{l_1 s_1} \cdots g^{l_j s_j}\right) 
              \left(\frac{\alpha'}{2}\right)^{j/2} 
    \nonumber \\
&& \frac{1}{j!} \left(\frac{27}{16}\right)^{\alpha' t - (j-1)}
          \frac{R^3 \alpha'}{t_{(j,j,1)}}
               \int_0^{\infty} d\nu \; 
               \frac{P^{(j-m)}_{ \rho_1 \cdots \rho_{j-m};
                         \sigma_1 \cdots \sigma_{j-m}}} 
                    { \frac{ {\cal E}_{0,0} + c_y}{\sqrt{\lambda}} +
		    N_{\rm eff.} - i \epsilon } \nonumber \\
&&
        \left[ A^{0,0,m}_{r_1 \cdots r_j}(z_\gamma; - \Delta,\nu)
        \right]^{\hat{\rho}_1 \cdots \hat{\rho}_{j-m}}    
        \left[ A^{0,0,m}_{s_1 \cdots s_j}(z_h; \Delta,\nu)
        \right]^{\hat{\sigma}_1 \cdots \hat{\sigma}_{j-m}},     
\label{eq:Compton-j-00m-t}
\end{eqnarray}
where 
\begin{eqnarray}
 \bar{J}^{\gamma\gamma}_{k_1 \cdots k_j; pq}(z_\gamma) & = &
    (-i)^j \left[ A_p(z_\gamma; -q_2) 
                     \overleftrightarrow{\nabla}_{k_1} \cdots 
                     \overleftrightarrow{\nabla}_{k_j}
                  A_q(z_\gamma; q_1) \right] , \\ 
 \bar{J}^{hh}_{l_1 \cdots l_j}(z_h) & = &
    (-i)^j \left[\Phi(z_h; p_1) \overleftrightarrow{\nabla}_{l_1} \cdots 
         \overleftrightarrow{\nabla}_{l_j} \Phi(z_h; -p_2) \right].  
\end{eqnarray}
Although momentum vectors are used in the second arguments of the 
external state wavefunctions $A$ and $\Phi$ here, instead of their 
Lorentz-invariant momentum square, this is only to remind ourselves 
of the sign when $\nabla$'s act on the wavefunctions. 

The expression (\ref{eq:Compton-j-00m-t}) is meant to be 
a part of the $t$-channel contribution to the Compton tensor, 
$[T^{\mu \nu} \epsilon^1_\mu \epsilon^{2*}_\nu]^{(t)}$, and we should 
obtain the full contribution to the Compton tensor 
$[T^{\mu \nu} \epsilon^1_\mu \epsilon^{2*}_\nu]$ after employing 
the prescription (\ref{eq:SFT-Veneziano-prescr}). 
At least this prescription tells us to set the factor 
$(27/16)^{[\alpha' t - (j-1)]}$ in the 4th line to 
$(27/16)^{{\cal O}(1/\sqrt{\lambda})} \simeq 1$. 
Now, we claim that this is the only necessary change under this prescription, 
so far as the amplitude of $(0,0,m)$-mode exchange is concerned. 

To see this, remember that, prior to applying the prescription 
(\ref{eq:SFT-Veneziano-prescr}), we need to rewrite the residues 
of the $t$-channel poles in terms only of Mandelstam variables $s$ and $t$, 
not of $u$. Let us take an expression 
$[\Phi_h \overleftrightarrow{\nabla}_m \Phi_h] g^{mn} 
[A_\gamma \overleftrightarrow{\nabla}_n A_\gamma]$ as an example, which 
captures the feature of contraction of SO(4,1) indices 
in (\ref{eq:Compton-j-00m-t}). In the scattering 
$\phi(P_1)+A(Q_1) \longrightarrow \phi(P_2)+A(Q_2)$ with $P_{1,2}$ 
and $Q_{1,2}$ ``momenta'' $\sim$ derivatives in 5-dimensions, 
$(s-u) \sim (P_1+P_2) \cdot (Q_1 + Q_2)$ is converted to 
$(2s+t)$ in the following steps:
\begin{eqnarray}
 & &  (P_1+P_2) \cdot (Q_1 + Q_2) = (2P_1 + (P_2-P_1)) \cdot (Q_1+Q_2), 
   \nonumber \\
& = & (2P_1) \cdot (Q_1+Q_2) + (Q_1-Q_2) \cdot (Q_1+Q_2)
 = (2P_1) \cdot (2Q_1 + (Q_2-Q_1)) + (Q_1)^2 - (Q_2)^2,  \nonumber \\
& = & (2P_1) \cdot (2Q_1 + (P_1-P_2)) + (Q_1)^2-(Q_2)^2
 = (4 P_1 \cdot Q_1) + (-2 P_1 \cdot P_2) + 2(P_1)^2 + (Q_1)^2 - (Q_2)^2;  
   \nonumber 
\end{eqnarray}
each one of the steps above is regarded as either one of partial 
integration in $dx_\gamma dz_\gamma$, one in $dx_h dz_h$, or rewriting 
$(P_2-P_1)$ by $(Q_1-Q_2)$ or vice versa.
The last procedure is to pass a derivative on one side of the propagator 
to the other. Because of the 5D-transverse condition characterizing the 
$(0,0,m)$ modes, such terms proportional to $\nabla$ drop out from the 
amplitudes exchanging the $(0,0,m)$-modes. Noting that the 
prescription (\ref{eq:SFT-Veneziano-prescr}) modifies the 
$-2(P_1 \cdot P_2) \sim t$ term above into the propagator mass, and that 
this term appeared only after passing a derivative $\nabla$ through the 
propagator, we see that the term which would have been affected by the 
prescription (\ref{eq:SFT-Veneziano-prescr}) has already dropped out indeed.

\subsubsection{Casting the Amplitude into the form of OPE}

So far, the (virtual) photon and target hadron have been treated 
equally in the scattering amplitude. We are interested, however, in the 
$h+\gamma^* \longrightarrow h+\gamma^{(*)}$ scattering 
in the regime of generalized Bjorken scaling, where 
\begin{equation}
 |(q^2)|, (q \cdot p), |(q_1 \cdot \Delta)|, |(q_2 \cdot p)| \gg 
|\Delta^2|, m_h^2, \Lambda^2,
\label{eq:Bjorken-general} 
\end{equation}
while the ratio among $(q \cdot p)$, $(q^2)$ and $(q \cdot \Delta)$, 
namely, $x$ and $\eta$, are kept finite. It is, thus, desirable to 
rewrite the scattering amplitude (structure functions) in a form that 
fits to the conformal OPE. To do this, we follow a prescription that has 
been used in the study of DIS in holographic models. 

Let us focus on the following factors that appear in the 3rd and 4th
lines of (\ref{eq:Compton-j-00m-t}):
\begin{equation}
 \int_0^{\infty} d\nu \;
        \left[ A^{0,0,m}_{r_1 \cdots r_j}(z_\gamma; - \Delta,\nu)
        \right]^{\hat{\rho}_1 \cdots \hat{\rho}_{j-m}}    
        \left[ A^{0,0,m}_{s_1 \cdots s_j}(z_h; \Delta,\nu)
        \right]^{\hat{\sigma}_1 \cdots \hat{\sigma}_{j-m}} \times [
        \cdots ].     
\label{eq:prop-rewrite-tmp}
\end{equation}
The last factor $[ \cdots ]$ denotes the remaining $\nu$-dependence 
(denominator) in the integrand; we only need to remember that 
${\cal E}_{0,0} = (4+j+\nu^2)$, and hence it is even under the change 
$\nu \rightarrow - \nu$.

We begin with the case $m=0$. The expression (\ref{eq:prop-rewrite-tmp})
for the $m=0$ case becomes 
\begin{align}
& \int_0^\infty d\nu \;
   [ \Psi_{i\nu; 0,0,0}^{(j);0,0}(-\Delta^2; z_\gamma) ]
   [ \Psi_{i\nu; 0,0,0}^{(j);0,0}(-\Delta^2; z_h) ] \times [ \cdots ], 
\label{eq:propagator-rewrite-a} \\
 = & \frac{2}{\pi^2 R} \int_0^\infty d\nu \; 
    \frac{\nu \sinh(\pi \nu)}{(1-c^{(j)}_{i\nu;0,0,0})} \; 
   [e^{(j-2)A(z_\gamma)} ``K_{i\nu}(\Delta z_\gamma)'']
   [e^{(j-2)A(z_h)} ``K_{i\nu}(\Delta z_h)''] 
   \times [ \cdots ]   \nonumber 
\end{align}
multiplied by a factor 
$[   \delta_{r_1}^{\; \hat{\rho}_1} \cdots
       \delta_{r_j}^{\; \hat{\rho}_j}
       \delta_{s_1}^{\; \hat{\sigma}_1} \cdots
       \delta_{s_j}^{\; \hat{\sigma}_j} ]$.
Using the fact that 
$K_{i\nu}(x) = i\pi/2 \times (I_{i\nu}(x)-I_{-i\nu}(x))/[\sinh(\pi
\nu)]$, the $\nu$-integral above can be rewritten 
as  
\begin{eqnarray}
 \frac{1}{\pi R} \int_{-\infty}^{+\infty} d\nu \; i \nu 
   [e^{(j-2)A(z_\gamma)} I_{i\nu}(\Delta z_\gamma)] 
   [``K_{i\nu}(\Delta z_h)'' e^{(j-2) A(z_h)}] \times [ \cdots ], 
\label{eq:propagator-rewrite-c}
\end{eqnarray}
where we used the relation (\ref{eq:IR-effect-nu-asymmetry}).
This expression is more convenient than (\ref{eq:propagator-rewrite-a});  
this is because i) the $z_\gamma$ integration is dominated in the region 
$q z_\gamma \lesssim 1$, due to the photon external state wavefunctions 
containing $K_1(q_{1,2}z)$, ii) $I_{i\nu}(\Delta z_\gamma)$ decreases 
rapidly toward positive $i\nu$, for $qz_\gamma \lesssim 1$ and 
$q \gg \Delta$ (generalized Bjorken scaling (\ref{eq:Bjorken-general})), 
and iii) the rapidly decreasing $I_{i\nu}(\Delta z_\gamma)$ 
in the lower half of the complex $\nu$-plane allows us to close the 
$\nu$-integration contour through the large-radius lower half complex 
$\nu$-plane (see \cite{NW-first} and literatures therein).

It is straightforward to generalize this treatment for all other 
$m \neq 0$ modes. 
Note that the Pomeron/Reggeon wavefunction 
$[A^{0,0,m}_{m_1 \cdots m_j}(z; \Delta,  \nu)]
 ^{\hat{\rho}_1 \cdots \hat{\rho}_{j-m}}$ for $m \neq 0$ is obtained
from that of $m=0$ by multiplying $(\Delta z)^m$ and $N_{j,m}$ (which is
even in $\nu$), applying
differential operators in $z$ and manipulating Lorentz indices. 
Obviously the order of such manipulations on the wavefunction and 
the procedure from (\ref{eq:propagator-rewrite-a}) to 
(\ref{eq:propagator-rewrite-c}) can be exchanged.

Therefore, the contribution to the Compton tensor from the leading 
trajectory spin-$j$ state $(0,0,m)$ mode is 
\begin{eqnarray}
&& (T^{\mu\nu} \epsilon^1_\mu \epsilon^{2*}_\nu)_{(j,j);(0,0,m)}
 \nonumber \\
 & \simeq &
\frac{1}{j!} \frac{t_\gamma \sqrt{\lambda}}{t_y \pi}
    \left(\frac{\alpha'}{2}\right)^j 
   \int_{-\infty}^{+\infty} \!\!\!\! d\nu \; 
               \frac{P^{(j-m)}_{ \rho_1 \cdots \rho_{j-m};
                         \sigma_1 \cdots \sigma_{j-m}}} 
                    { \frac{ {\cal E}_{0,0} + c_y}{\sqrt{\lambda}} +
		    N_{\rm eff.} - i \epsilon } 
      i\nu \\ 
& &  \frac{R^2}{R^3}
         \int dz_\gamma \sqrt{-g(z_\gamma)}
              \bar{J}^{\gamma\gamma}_{k_1 \cdots k_j; p q} g^{pq} 
              \left( g^{k_1 r_1} \cdots g^{k_j r_j}\right) 
        \left[ 
             \bar{\bar{A}}^{0,0,m}_{r_1 \cdots r_j}(z_\gamma; - \Delta,\nu)
        \right]^{\hat{\rho}_1 \cdots \hat{\rho}_{j-m}}    
\nonumber \\
& & \frac{t_h}{R^3}
         \int dz_h \sqrt{-g(z_h)}
              \bar{J}^{hh}_{l_1 \cdots l_j}
              \left( g^{l_1 s_1} \cdots g^{l_j s_j}\right) 
        \left[ 
               \bar{A}^{0,0,m}_{s_1 \cdots s_j}(z_h; \Delta,\nu)
        \right]^{\hat{\sigma}_1 \cdots \hat{\sigma}_{j-m}},     
 \nonumber 
\label{eq:Compton-j-00m-full}
\end{eqnarray}
where $\bar{A}$ and $\bar{\bar{A}}$ are obtained from $A$ by removing 
the factor $(2/\pi)\sqrt{[\nu \sinh(\pi \nu)/2R]}$ 
in (\ref{eq:Pomeron-wvfc-1compnt}) first, and then 
replacing $K_{i\nu}(\Delta z_h)$ by ``$K_{i\nu}(\Delta z_h)$'' in 
$\bar{A}(z_h)$, while replacing $K_{i\nu}(\Delta z_\gamma)$ by 
$I_{i\nu}(\Delta z_\gamma)$ in $\bar{\bar{A}}(z_\gamma)$.
Short distance (stringy) parameters such as AdS radius $R$ and 
string length $\sqrt{\alpha'}$ can be eliminated from this expression 
of the Compton tensor, so that it is written purely in terms of 
parameters of strongly coupled gauge theory / hadron physics; 
\begin{eqnarray}
 && (T^{\mu\nu} \epsilon^1_\mu \epsilon^{2*}_\nu)_{(j,j);(0,0,m)}
 \nonumber \\
& \simeq & \frac{1}{j!} \frac{t_\gamma \sqrt{\lambda}}{t_y \pi}
    \left(\frac{1}{2 \sqrt{\lambda}}\right)^j 
   \int_{-\infty}^{+\infty} \!\!\!\! d\nu \; 
               \frac{P^{(j-m)}_{ \rho_1 \cdots \rho_{j-m};
                         \sigma_1 \cdots \sigma_{j-m}}} 
                    { \frac{ {\cal E}_{0,0} + c_y}{\sqrt{\lambda}} +
		    N_{\rm eff.} - i \epsilon } 
      i\nu \\  
& &
 \int_0 \frac{dz_\gamma}{z_\gamma}
   \left[A_p (z_\gamma;q_2) (-i \overleftrightarrow{\nabla})^j_{k_1 \cdots k_j} 
         A_q (z_\gamma; q_1) \right] \delta^{\hat{p}\hat{q}} z^j 
  [ \delta^{\hat{k}_1 \hat{r}_1} \cdots   \delta^{\hat{k}_j \hat{r}_j} ]
  [e^{(2-j)A} \bar{\bar{A}}^{0,0,m}_{r_1 \cdots r_j}
     (z_\gamma; -\Delta, \nu)]^{\hat{\rho}'s}
  \nonumber \\
& &
 t_h \int_0 \frac{dz_h}{z^3_h}
   \left[\Phi (-i \overleftrightarrow{\nabla})^j_{l_1 \cdots l_j} 
          \Phi  \right] z^j 
  [ \delta^{\hat{l}_1 \hat{s}_1} \cdots   \delta^{\hat{l}_j \hat{s}_j} ]
  [e^{(2-j)A} \bar{A}^{0,0,m}_{s_1 \cdots s_j}
       (z_h; \Delta, \nu)]^{\hat{\sigma}'s} .
 \nonumber 
\label{eq:Compton-j-00m-full-physical}
\end{eqnarray}
Each line of this expression has zero mass-dimension, and hence 
$T^{\mu\nu}$ is also of zero mass-dimension, as expected from the 
Fourier transform of the matrix element (\ref{eq:current-current-ME}). 

The leading twist contribution to the Compton tensor $T^{\mu\nu}$ should 
be obtained by summing up the amplitudes of exchanging the spin-$j$ field
in the leading trajectory, with $m=0,\cdots,j$ also being summed up.
It is known in the literature that, for each spin-$j$, the second line 
of (\ref{eq:Compton-j-00m-full-physical}) becomes something close to 
the Wilson coefficient of the OPE, and the third line 
of (\ref{eq:Compton-j-00m-full-physical}) something close to the operator 
matrix element. We will elaborate more on it, with a particular emphasis 
on the role played by the summation over $m$. For now, we define 
\begin{eqnarray}
 C^{0,0,m} &:= & \int_0 \frac{dz}{z}
   \left[A_p (z;-q_2) (-i \overleftrightarrow{\nabla})^j_{k_1 \cdots k_j} 
         A_q (z; q_1) \right] \qquad 
 \times
    \left[ \frac{(2\Lambda)^{i\nu-j}}{\Delta^{i\nu}} \Gamma(i\nu+1) \right]
\nonumber \\
  & & \qquad 
  \delta^{\hat{p}\hat{q}} z^j 
  [ \delta^{\hat{k}_1 \hat{r}_1} \cdots   \delta^{\hat{k}_j \hat{r}_j} ]
  [e^{(2-j)A} \bar{\bar{A}}^{0,0,m}_{r_1 \cdots r_j}(z; -\Delta, \nu)
   ]^{\hat{\rho}_1\cdots \hat{\rho}_{j-m}} \epsilon_{\rho_1\cdots \rho_{j-m}}^{(0,0,m)}
\nonumber
\end{eqnarray}
and 
\begin{eqnarray}
\Gamma^{0,0,m} & := & t_h \int_0^{1/\Lambda} \frac{dz}{z^3}
   \left[\Phi (-i \overleftrightarrow{\nabla})^j_{l_1 \cdots l_j} 
          \Phi  \right] z^j
  \qquad   \times 
  \left[ \left(\frac{\Delta}{2\Lambda}\right)^{i\nu}
         \frac{\Lambda^j}{\Gamma(i\nu)} \right]   \nonumber \\
  & & \qquad 
  [ \delta^{\hat{l}_1 \hat{s}_1} \cdots   \delta^{\hat{l}_j \hat{s}_j} ]
  [e^{(2-j)A} \bar{A}^{0,0,m}_{s_1 \cdots s_j}(z; \Delta, \nu)
  ]^{\hat{\sigma}_1 \cdots \hat{\sigma}_{j-m}} 
  \epsilon_{\sigma_1 \cdots \sigma_{j-m}}^{(0,0,m)} 
\label{eq:Gamma-00m-def}
\end{eqnarray}
separately. The factor $[\Gamma(i\nu+1)(2\Lambda)^{i\nu-j}/\Delta^{i\nu}]$ 
in $C^{0,0,m}$ and a similar factor in $\Gamma^{0,0,m}$ are introduced 
so that $C^{0,0,m}$ and $\Gamma^{0,0,m}$ correspond 
to the OPE Wilson coefficients and hadron matrix elements, respectively, 
renormalized at $\mu_F \sim \Lambda$, as we will see later. 

We will focus on the spin-even contribution to a flavor-non-singlet 
component of the structure function $V_1$ 
in (\ref{eq:structure functions of Compton tensor}).
The $V_1$ structure function is picked up here, only because 
it is computed a little more easily than other structure functions. 
We will not touch flavor-singlet components in this article, 
apart from a brief discussion in section \ref{ssec:super-Pomeron}; this is 
because the cubic SFT with Chan--Paton factor in section \ref{sec:cubic-sft}
is not the adequate tool to study the singlet components.
The coefficient $C^{0,0,m}$ above is decomposed, just like 
$T^{\mu\nu} \epsilon_\mu^1 \epsilon^{2*}_\nu$ is; 
the spin-$j$ (with $j \in 2\Z$) contribution to the structure 
function $V_1^{+,\alpha}$---spin-even ($+$) and flavor 
non-singlet ($\alpha$)---is denoted by $C^{0,0,m}_{V_1; +,\alpha}$.

\subsubsection{Amplitude of the $(m=0)$-Mode Exchange}
\label{sssec:m=0-amplitude}

We first study $V_1^{+,\alpha}$ from the $m=0$ mode exchange.
With the Reggeon wavefunction given by 
$\Psi^{(j);0,0}_{i\nu;0,0,0}(t,z) = \Psi^{(j)}_{i\nu}(t,z)$ 
in (\ref{eq:Pomeron-wvfc-1compnt}), this $m=0$ contribution is 
expected to be the closest to what has been studied in the literatures
such as \cite{BPST-06, Hatta-07, BDST-10, NW-first}. Indeed, we reproduce 
the expression known in the literature, but with a little refinement 
in (\ref{eq:V1-m=0}).
 
Note first that the Reggeon wavefunctions 
$\bar{\bar{A}}^{0,0,m=0}_{r_1 \cdots r_j}$ and 
$\bar{A}^{0,0,m=0}_{s_1 \cdots s_j}$ are non-zero only when all the $r_i$'s
and $s_i$'s are in the 3+1 Minkowski directions,  
$(r_1 \cdots r_j)=(\rho_1 \cdots \rho_j)$ and 
$(s_1 \cdots s_j)=(\sigma_1 \cdots \sigma_j)$; furthermore, 
the wavefunction is 4D-transverse and 4D-traceless totally symmetric 
tensors of $\SO(3,1)$. 

This makes it much easier to evaluate the matrix element $\Gamma^{0,0,m=0}$.
Because  
\begin{equation}
 \left( \nabla^k \Phi \right)_{\sigma_1 \cdots \sigma_k} = 
  \partial_{\sigma_1} \cdots \partial_{\sigma_k} \Phi + 
   \left[ {\rm terms~proportional~to~} \eta_{\sigma_a \sigma_b} \right], 
\label{eq:4-hadron-current-m0k0}
\end{equation}
only 
\begin{eqnarray}
 & & \left[ \Phi(z;p_1) (-i \overleftrightarrow{\nabla})^j \Phi(z;-p_2) \right]_{\sigma_1 \cdots \sigma_j} 
     \nonumber \\
 &: = & \sum_{k=0}^j {}_j C_k 
     \left[ (i \nabla)^{j-k} \Phi(z; p_1) \right]_{\sigma_{k+1} \cdots \sigma_j}
     \left[ (-i \nabla)^k \Phi(z,-p_2) \right]_{\sigma_1 \cdots \sigma_k}
     \nonumber \\
 & \longrightarrow & (-1)^j (p_1+p_1)_{\sigma_1} \cdots (p_1+p_2)_{\sigma_j}
      \Phi(z;p_1) \Phi(z; -p_2)
\end{eqnarray}
contributes to $\Gamma^{0,0,m=0}$:
\begin{eqnarray}
 \Gamma^{0,0,m=0} & = & \left[ \epsilon^{(0,0,0)}_{\sigma_1 \cdots \sigma_j} 
    (-1)^j (p_1+p_2)^{\hat{\sigma}_1} \cdots (p_1+p_2)^{\hat{\sigma}_j} \right] \;
     \bar{g}^{0,0,0}(j,i\nu, \Delta), 
  \label{eq:Gamma-000} \\
 \bar{g}^{0,0,0}(j,i\nu,\Delta) & := &
   \int_0^{1/\Lambda} \frac{dz}{z^3} (\Lambda z)^j t_h (\Phi(z;m_h))^2 
     \frac{ \left\{ ``K_{i\nu}(\Delta z)'' \right\} }
          {[(\frac{\Delta}{2\Lambda})^{-i\nu} \Gamma(i\nu)]};  
   \label{eq:g000-def}
\end{eqnarray}
note here, that the confinement effect has been included in the form 
of i) introducing a cut in the holographic radius $z_h \leq 1/\Lambda$, 
and ii) $K_{i\nu}(\Delta z_h)$ modified to ``$K_{i\nu}(\Delta z_h)$'' in 
(\ref{eq:modified-K-inu}). The expression of $\bar{g}^{0,0,0}$ here, or 
that of $\Gamma^{0,0,m}$ in (\ref{eq:Gamma-00m-def}) implicitly ignores 
a possibility of ${\cal L}_{\rm bdry} \neq 0$. For practical purposes, 
though, this may not be a big deal, since Ref. \cite{BDST-10} reports 
that such confinement effects do not play a significant role quantitatively 
for most of kinematical region. 

Let us also evaluate the Wilson coefficient $C^{0,0,m=0}$. 
The expression 
\begin{eqnarray}
 & & \left[ A_p(z;-q_2) (-i \overleftrightarrow{\nabla})^j A_q(z; q_1)_q \right]_{\rho_1 \cdots \rho_j}
     \delta^{\hat{p}\hat{q}}     \nonumber \\
 & := & \sum_{k=0}^j {}_j C_k 
   \left[ (i\nabla)^{j-k} A(z;-q_2) \right]_{\rho_{k+1} \cdots \rho_j p}
   \left[ (-i \nabla)^k A(z; q_1) \right]_{\rho_1 \cdots \rho_k q} \; 
   \delta^{\hat{p}\hat{q}} 
\end{eqnarray}
appearing in $C^{0,0,m=0}$ can be evaluated by using the fact that 
\begin{eqnarray}
 (\nabla^k A)_{\rho_1 \cdots \rho_k \kappa} & \equiv &  
     (\partial_{\rho_1} \cdots \partial_{\rho_k} A_\kappa) 
   - \sum_{a=1}^k \frac{\eta_{\mu_a \kappa}}{z} 
             (\partial_{\rho_1} \cdots \check{\partial}_{\rho_a} 
                             \cdots \partial_{\rho_k} A_z) \nonumber \\ 
  & & \qquad  - \sum_{1 \leq a < b \leq k} \frac{\eta_{\rho_a \kappa}}{z^2} 
             ( \partial_{\rho_1} \cdots \check{\partial}_{\rho_a} \cdots 
                  \check{\partial}_{\rho_b} \cdots \partial_{\rho_k} A_{\rho_b}),  
 \label{eq:photon-derivative-a} \\
 (\nabla^k A)_{\rho_1 \cdots \rho_k z} & \equiv & 
     (\partial_{\rho_1} \cdots \partial_{\rho_k} A_z)
   + \frac{1}{z} \sum_{a=1}^k (\partial_{\rho_1} \cdots \check{\partial}_{\rho_a} 
       \cdots \partial_{\rho_k} A_{\rho_a}) 
 \label{eq:photon-derivative-b}
\end{eqnarray}
modulo terms proportional to $\eta_{\rho_c \rho_d}$.  
As we will focus only on the structure function $V_1^{+,\alpha}$,  
we can further drop the terms with $A_z$ 
in (\ref{eq:photon-derivative-a}, \ref{eq:photon-derivative-b}). 
Then the expression above becomes 
\begin{eqnarray}
  & &  [\eta^{\mu \nu} \epsilon^1_\mu \epsilon^{2*}_{\nu}] 
   (q_1+q_2)_{\rho_1} \cdots (q_1+q_2)_{\rho_j} 
      \nonumber \\
 & + & \frac{2}{z^2} \sum_{a \neq b} \epsilon(-q_2)_{\rho_a}\epsilon(q_1)_{\rho_b} 
    (q_1 + q_2)_{\rho_1} \cdots {}_{\check{\rho}_a}{}_{\check{\rho}_b} \cdots 
    (q_1+q_2)_{\rho_j}
\label{eq:photon-derivative-c}
\end{eqnarray}
multiplied by $[(q_1z)K_1(q_1z)][(q_2z)K_1(q_2z)]$.

There are two remaining tasks in evaluating the $(m=0)$-mode contribution to 
the $V_1^{+,\alpha}$ structure function; a) one is to to carry out 
the $z_\gamma$ integral, and b) the other is to sum up $C^{0,0,0} \Gamma^{0,0,0}$ 
for different polarizations of $\epsilon^{(0,0,0)}$.  
As for the $z_\gamma$ integral, the integrand sharply falls off 
at $z_\gamma \approx q^{-1}$ because of the photon wavefunctions of the form 
$[(q_iz)K_{i\nu}(q_i z)]$. The $z_\gamma$ integral in $C^{0,0,m}$ 
over the holographic radius $z_\gamma \in [0, \Lambda^{-1}]$ therefore 
comes mainly from a very small fraction of it, $\Lambda/q \ll 1$, 
in the regime of generalized Bjorken scaling (\ref{eq:Bjorken-general}).
It is then all right to make an approximation that 
\begin{equation}
 I_{i\nu}(\Delta z_\gamma) \approx \frac{1}{\Gamma(i\nu+1)} 
    \left( \frac{\Delta z_\gamma}{2} \right)^{i\nu}
  \left[ 1 + {\cal O}(\Delta/q) \right]
 \qquad {\rm when~} (\Delta z_\gamma) \lesssim \Delta/q \ll 1, 
\label{eq:approx-I-inu}
\end{equation}
and also to replace the range of integral $z_\gamma \in [0, \Lambda^{-1}]$ 
to $[0, +\infty)$, as in the literature; the error due to this approximation 
is only in the higher order in $(\Delta/q)$, and the twist-$(2+\gamma(j))$ 
contribution is still obtained properly.
The integral is then cast into the form of (\ref{eq:C1-def})  
with $\delta = j+i\nu$ for the first line of (\ref{eq:photon-derivative-c})
[resp. $\delta = j+i\nu-2$ for the second line 
of (\ref{eq:photon-derivative-c})], and $\vartheta = \eta/x$; thus we 
can use the analytic expression 
(\ref{eq:C1-from-integral-A}, \ref{eq:C1-from-integral-B}) in the appendix.

The other task, b) tensor computations, is carried out in 
the appendix \ref{ssec:tensor}. Using the results 
of (\ref{eq:q-proj-p-product}) and (\ref{eq:qee-proj-p-product}), 
one finds that the contribution to $(C^{0,0,0}_{V_1;+,\alpha})_{m=0}$ 
from the second line of (\ref{eq:photon-derivative-c}) is roughly 
\begin{equation}
 \frac{q^2 \; \Delta^2}{(q \cdot \Delta)(p \cdot q) } \ll 1
\label{eq:k-non0-suppressed-opecoeff}
\end{equation}
times smaller than the contribution from the first line of 
(\ref{eq:photon-derivative-c}) in the generalized Bjorken scaling regime 
(\ref{eq:Bjorken-general}), and is hence ignored, when 
only the twist-$(2+\gamma(j))$ contributions are retained. 

Combining all above, the spin-$j \in 2\Z$ contribution is 
\begin{eqnarray}
 (V_1^{+,\alpha})_{j,m=0} & \approx &  
    \frac{\sqrt{\lambda}}{\Gamma(j+1) \pi}
    \frac{t_\gamma}{t_y} 
    \int_{-\infty}^{+\infty} d\nu 
       \frac{1}{\frac{4+j+\nu^2 + c_j}{\sqrt{\lambda}}+j-1-i\epsilon} 
   \nonumber \\
   & & \quad 
    C_1(j+i\nu, \vartheta)
    \left(\frac{\Lambda}{q}\right)^{i\nu-j}
    \left(\frac{1}{\sqrt{\lambda}x}\right)^j 
    \bar{g}^{0,0,0}(j,i\nu,\Delta)  \hat{d}_j([\eta]),
\label{eq:V1-m=0-j}
\end{eqnarray}
where $C_1$ is given in (\ref{eq:C1-from-integral-B}), and 
$\hat{d}_j$ is a polynomial of degree $j$ in the argument 
\begin{equation}
 [\eta] := \eta \times \sqrt{\frac{-4p^2}{\Delta^2}} = 
  \eta \sqrt{\frac{4m_h^2+\Delta^2}{\Delta^2}},
\label{eq:[eta]-def}
\end{equation}
and is given in terms of Legendre polynomial, as in (\ref{eq:dhat-Legendre}).

Now that all the factors of the spin-$j$ contribution to 
$V_1^{+,\alpha}$ are given as analytic functions of $j$, it is possible
to convert the sum over the (spin-$j \in 2\N$) string states in the 
leading trajectory to a contour integral in the complex angular momentum 
plane; 
\begin{equation}
 (V_1^{+,\alpha})_{m=0} =  - \int \frac{dj}{4i} \frac{1+e^{\pi i j}}{\sin (\pi j)}
   (V_1^{+,\alpha})_{j,m=0},
\end{equation}
with the contour in the $j$ plane moving just below the real positive axis 
toward the left, and then just above the real positive axis toward the right. 
The integration contour in the $\nu$ plane is deformed so that it picks up 
the residue of the pole in the lower complex $\nu$ plane coming from the 
$t$-channel propagation of strings. Thus, 
\begin{eqnarray}
 (V_1^{+,\alpha})_{m=0} & \approx &
 - \int \frac{dj}{4i} \frac{1+e^{\pi i j}}{\sin (\pi j)}
  \frac{ t_\gamma/t_y }{\Gamma(j+1)}\frac{\lambda}{i\nu_j}
   \nonumber \\
&& C_1 \left( j+i\nu_j, \frac{\eta}{x} \right)  
   \left( \frac{\Lambda}{q} \right)^{\gamma(j)}
   \left( \frac{1}{\sqrt{\lambda} x} \right)^j     
   \bar{g}^{0,0,0}(j,i\nu_j,\Delta)  \hat{d}_j([\eta]),
\label{eq:V1-m=0}
\end{eqnarray}
where $\gamma(j) = i\nu_j-j$, and $i\nu_j \geq 0$ is a function of $j$ 
determined by the on-shell condition 
\begin{equation}
 j-1+\frac{4+j+\nu^2+c_j}{\sqrt{\lambda}} = 0.
\label{eq:on-shell-openstring}
\end{equation}

This is the result known in \cite{PS-02-DIS, BPST-06, Hatta-07, CC-DIS, 
BDST-10, NW-first} etc.; 
under an assumption that $\bar{g}^{0,0,0}(j,i\nu_j, \Delta)$ does not 
grow too rapidly for large ${\rm Re}(j)$ to cancel the large 
factor $\Gamma(j+1)$ in the denominator, the integration contour 
in the $j$ plane can be deformed toward the left in the $j$-plane, 
as in the classical Watson--Sommerfeld transformation;  
this is how the non-converging $j \in 2\N$ sum of the OPE 
is rendered well-defined for physical kinematics $x < 1$.
The integrand forms a saddle point due to the two factors 
$(1/x)^j$ and $(\Lambda/q)^{\gamma(j)}$;
let $j_*$ in the complex $j$-plane be where the saddle point 
is.\footnote{The saddle point value $j_*$ is determined by   
$\left. \frac{\partial \gamma(j)}{\partial j}\right|_{j=j_*}
= \frac{\ln(1/x)}{\ln(q/\Lambda)}$.} 
The integrand also has poles in the $j$-plane. 
The hadron matrix element $\bar{g}^{0,0,0}$ contains 
$c^{(j)}_{i\nu_j;0,0,0}$ in its definition, and $c^{(j)}_{i\nu;0,0,0}$ 
may have a pole in the $j$-plane \cite{BPST-06}.\footnote{ 
For example, imagine a case $(1-c^{(j)}_{i\nu;0,0,0})=  
[I_{-i\nu}(\Delta/\Lambda)/I_{i\nu}(\Delta/\Lambda)]$); the factor 
$c^{(j)}_{i\nu_j;0,0,0}$ has poles $j=\alpha_{\R,n}(t)$ ($n=1,2,\cdots$) 
in the $j$-plane given by the condition 
$j_{i\nu_j,n} = \sqrt{t}/\Lambda$; $j_{\mu,n}$'s are the $n$-the zero
of the Bessel function $J_\mu$.}
The saddle point value $j_*$ has larger real part than any one of the 
poles, when $\ln(q/\Lambda)$ is large relatively to $\ln(1/x)$;
the $j$-integral is well-approximated by the saddle point value 
of the integrand, and yields the DGLAP regime. 
When $\ln(1/x)$ is large relatively to $\ln(q/\Lambda)$, however, one 
of the poles may have a real part larger than ${\rm Re}(j_*)$. 
Then the integral is approximated by the residue at such leading pole.
In this way, the sting-theory result $(V_1^{+,\alpha})_{m=0}$ goes back and forth 
between the DGLAP phase and Regge phase, depending on the kinematical 
variables $x, (q^2/\Lambda^2)$ and $t= -\Delta^2$ \cite{BPST-06, NW-first}.

The derivation of (\ref{eq:V1-m=0}) was not just a review of preceding 
works, however. First, the integration over $z_\gamma$ yields a function 
$C_1(j+i\nu_j, \eta/x)$, which has precisely the same form as 
the one expected from the conformal OPE; 
comparing (\ref{eq:conf-OPE-2side}, \ref{eq:conf-OPE-center}) 
and (\ref{eq:C1-from-integral-A}, \ref{eq:C1-from-integral-B}), 
one finds that they agree, under the identification  
\begin{equation}
 \left[ (l_n+j_n-2) = 2j_n + (\tau_n-2) \right] 
    \Longleftrightarrow
 \left[ (j+i\nu_j) = 2j + \gamma(j) \right].
\end{equation}
The expression (\ref{eq:V1-m=0}) is indeed regarded as conformal OPE 
contributions from twist-$\tau_n = (2+\gamma(j))$ operators. 

Secondly, the $\eta$-dependence of the $m=0$ contribution is worked 
out, now. As we will see later in section \ref{sec:model}, it comes in a 
form that fits very well with what has been known as ``dual parametrization''
of GPD \cite{dual-para}. One will also notice that the argument of the 
degree-$j$ polynomial $\hat{d}_j([\eta])$ is $[\eta]$ 
in (\ref{eq:[eta]-def}), rather than $\eta$. This means that 
the coefficients of the $\eta^2$ term and higher diverge in the 
$t = -\Delta^2 \longrightarrow 0$ limit. This indicates that 
it is essential to sum up the $m \neq 0$ modes to obtain results 
that are physically sensible. We will address this issue 
in section \ref{ssec:D=0-cancellation}.

\subsubsection{Preparation}
\label{sssec:preparation}

Let us move on to the amplitudes of $m \geq 1$-mode exchange.
We begin with deriving a few general properties of those amplitudes, 
which make the subsequent computations less tedious. 

First, we observe that the hadron matrix element $\Gamma^{0,0,m}$ vanishes 
for any odd value of $m$. 
To see that this statement is true, we use a following property of 
$J^{hh}_{l_1 \cdots l_j}$: 
\begin{align}
\Phi(z,p_1) \overleftrightarrow{\nabla}_{\{ l_1}\dots
 \overleftrightarrow{\nabla}_{l_j\}}\Phi(z, -p_2) &=
(-1)^j
\Phi(z, -p_2) \overleftrightarrow{\nabla}_{\{l_1}\dots
 \overleftrightarrow{\nabla}_{l_j\}}\Phi(z, p_1); 
\end{align}
this is true in a process where the initial state hadron $h(p_1)$ 
remains to be the same hadron $h(p_2)$ in the final state, so that 
$-(p_1)^2 = - (p_2)^2 = m_h^2$. This property is used below, to study 
when $J^{hh}_{z^k \lambda_{k+1} \cdots \lambda_j} 
\bar{A}^{z^k \lambda_{k+1} \cdots \lambda_j}$ vanishes for various 
$k = 0,\cdots,m$.
 
For an even $j$, the $\SO(3,1)$-indices of 
$J^{hh}_{z^k \lambda_{k+1} \cdots \lambda_j}$ are provided by an even number 
of $(p_1+p_2)_\lambda$'s and even [resp. odd] number of $\Delta_\lambda$'s  
when $k$ is even [resp. odd]. The hadron matrix element $\Gamma^{0,0,m}$ 
receives non-vanishing contribution from 
$J^{hh}_{z^k \lambda_{k+1}\cdots \lambda_j} \bar{A}^{z^k \hat{\lambda}_{k+1}\cdots \hat{\lambda}_j}$
(no sum in $k$), only when the $D$ operator (\ref{eq:def-D}) 
is used for even [resp. odd] number of times in the Reggeon 
wavefunction (\ref{eq:Pomeron-wvfc-general-expans}).
This means that $s$ is even [resp. odd], and hence $\Gamma^{0,0,m}$ can be 
non-zero only when $m=k+s$ is even. 

For an odd $j$, the $\SO(3,1)$-indices of 
$J^{hh}_{z^k \lambda_{k+1} \cdots \lambda_j}$ are provided by an odd number 
of $(p_1+p_2)_\lambda$'s and an even [resp. odd] number of $\Delta_\lambda$'s 
when $k$ is even [resp. odd]. Thus, the matrix element $\Gamma^{0,0,m}$ 
receives non-zero contribution only when an even [resp. odd] number of 
the $D$ operator is used in (\ref{eq:Pomeron-wvfc-general-expans}).
This means, once again, that $s$ is even [resp. odd], and hence 
$\Gamma^{0,0,m}$ can be non-zero only when $m=k+s$ is even. 
This statement for an odd $j$ is not more than a side remark though, since 
we focus on the spin-even contribution 
$\propto [1+e^{-\pi i j}]/\sin(\pi j)$ in this article.

Secondly, $\Gamma^{0,0,m}$ can always be written in the form of 
\begin{equation}
 \Gamma^{0,0,m} = \left[ (-2)^{j-m} (p^{\hat{\sigma}_1} \cdots p^{\hat{\sigma}_{j-m}}) 
    \cdot \epsilon^{(0,0,m)}_{\sigma_1 \cdots \sigma_{j-m}} \right] \times 
   \bar{g}^{0,0,m}(j, i\nu, \Delta^2), 
\label{eq:Gamma-00m-factorize-tensor}
\end{equation}
and $\bar{g}^{0,0,m}$ is an $\SO(3,1)$-scalar of mass-dimension $m$; 
we have encountered a special case of this statement in 
(\ref{eq:Gamma-000}, \ref{eq:g000-def}). 
This statement itself is understood as follows. When we write down 
the covariant derivatives in $\bar{J}^{hh}_{z^k\lambda_1 \cdots \lambda_{j-k}}$ 
explicitly, the $\SO(3,1)$-indices---there are $(j-k)$ of them---are 
either one of $p_\lambda$, $\Delta_\lambda$ and $\eta_{\lambda \lambda'}$; 
$\eta_{\lambda \lambda'}$ can be further rewritten as 
$\eta_{\lambda \lambda'} - \Delta_\lambda \Delta_{\lambda'}/\Delta^2$ and 
$\Delta_\lambda \Delta_{\lambda'}$. Suppose that there are $N_p$ of the 
$\SO(3,1)$ indices from $\{p_\lambda\}$'s, $N_\Delta$ indices from 
$\{\Delta_\lambda\}$'s and $N_{\tilde{\eta}}$ from $\tilde{\eta}_{\lambda \lambda'}$'s
in a given term; $N_p + N_\Delta + 2 N_{\tilde{\eta}}=(j-k)$.
When such an $\SO(3,1)$ tensor is contracted with 
$\sum_N^{[(m-k)/2]} \tilde{E}^N D^{m-k-2N}[\epsilon^{(0,0,m)}]$ in the Reggeon 
wavefunction $\bar{A}^{0,0,m}_{z^k\lambda_1 \cdots \lambda_{j-k}}$, it remains non-zero 
only when $(m-k-2N)=N_\Delta$ and $N \geq N_{\tilde{\eta}}$, because 
of the relation (\ref{eq:extract-SO(3,1)-tensor}). 
It is not hard now to see that all the remaining terms are proportional 
to the prefactor of $\bar{g}^{0,0,m}$ in (\ref{eq:Gamma-00m-factorize-tensor}); 
the mass dimension of the remaining scalar factor (reduced matrix element) 
$\bar{g}^{0,0,m}$ follows from the fact that $\Gamma^{0,0,m}$ is defined to 
be of mass dimension $j$.

Finally, we note that the twist-$(2+\gamma(j))$ contribution to  
the coefficient $C^{0,0,m}$ arises only from the contraction 
$\bar{J}^{\gamma\gamma}_{z^k \kappa_{k+1} \cdots \kappa_{j}} 
 \bar{\bar{A}}^{\hat{z}^k \hat{\kappa}_{k+1} \cdots \hat{\kappa}_{j}}$ with $k=0$. 
We have already seen an example of this in the $m=0$ amplitude; 
the first term of (\ref{eq:photon-derivative-c}) contributes 
to (\ref{eq:V1-m=0}), while the second term does not because of 
(\ref{eq:k-non0-suppressed-opecoeff}), and the first term came from 
the $k=0$ contraction. 

In order to verify the claim above, note first that both an extra 
$\partial_z$ and an extra power of $1/z$ virtually change the integral of 
$C^{0,0,m}$ by about an extra power of 
$q \sim q_1 \sim q_2 \gg \Lambda, \Delta$.
Explicitly writing down covariant derivatives in 
$\bar{J}^{\gamma\gamma}_{z^k \kappa_1 \cdots \kappa_{j-k}}$, and evaluating 
the integrals only by the order of magnitudes, one can see that 
\begin{equation}
 (C^{0,0,m}_{V_1;+,\alpha})_{k}  \sim  
  \sum_M^{[\frac{j-k}{2}]} \left( \frac{\Lambda}{q}\right)^{i\nu-j}
   \frac{q^{k+2M} }{ (q^2)^j}
    \left[ \overbrace{(q_\kappa \cdots q_\kappa)}^{j-k-2M} (\eta_{\kappa \kappa} q^2)^M
    \right]  \cdot
   \sum_{N}^{[\frac{s}{2}]} \frac{1}{\Delta^{s-2N}}
       \tilde{E}^N D^{s-2N}[\epsilon^{(0,0,m)}]. 
\end{equation}
The $M=0$ contribution above is further evaluated by using the definition 
of $\tilde{E}$ and $D$ operators. Details of computation is found 
partially in (\ref{eq:q-[mmode]-product}); we find that 
\begin{equation}
(C^{0,0,m}_{V_1;+,\alpha})_{k,M=0} \sim \left( \frac{\Lambda}{q}\right)^{i\nu-j}
     \frac{\overbrace{(q_\kappa \cdots q_\kappa)}^{j-m} \epsilon^{(0,0,m)}}
          {(q^2)^j}
     \left(\frac{(q \cdot \Delta)}{\Delta}\right)^s q^k.
\end{equation}
Keeping the relation $m=k+s$ and also the 
result (\ref{eq:Gamma-00m-factorize-tensor}) in mind, we obtain 
\begin{eqnarray}
 C^{0,0,m}_{k, M=0} \cdot \Gamma^{0,0,m} & \sim & 
  \left( \frac{\Lambda}{q}\right)^{i\nu-j}
  \frac{(q \cdot p)^{j-m}}{(q^2)^j}
  \left(\frac{(q \cdot \Delta)}{\Delta}\right)^s q^k \times \bar{g}^{0,0,m} 
       \nonumber \\
 & \sim &  \left( \frac{\Lambda}{q}\right)^{i\nu-j}
   \left(\frac{1}{x}\right)^{j} \eta^{m-k} 
   \left( \frac{(q^2)(\Delta^2)}{(q \cdot p)^2} \right)^{\frac{k}{2}}
   \frac{\bar{g}^{0,0,m}}{\Delta^m}.
\end{eqnarray}
Therefore, this is regarded as a twist-$(2+\gamma(j)+k/2)$ contribution
in the generalized Bjorken scaling regime. 
Thus, only the $k=0$ term remains a twist-$(2+\gamma(j))$ contribution, 
and the terms with $k > 0$ are irrelevant to GPD.

The analysis becomes a little more complicated when $M > 0$ terms are also 
included, but not in an essential way. 
Contributions with some $(k,M)$ correspond to twist-$(2+\gamma+M+k/2)$, 
and only the $k=M=0$ terms contribute to GPD. 
This means that $C^{0,0,m}$ can be evaluated under the following approximation:
\begin{equation}
 [A_p(z;-q_2) (-i \overleftrightarrow{\nabla})^{j}A_q(z;q_1)
 ]^{\hat{m}_1 \cdots \hat{m}_j}   \delta^{\hat{p}\hat{q}} \bar{\bar{A}}_{m_1 \cdots m_j}
 \rightarrow 
 [A_\mu(z;-q_2) (-i \overleftrightarrow{\partial})^{j}A_\nu(z;q_1)
 ]^{\hat{\kappa}_1 \cdots \hat{\kappa}_j} \eta^{\mu\nu} 
  \bar{\bar{A}}_{\kappa_1 \cdots \kappa_j}. 
\end{equation}
%

\subsubsection{Wilson Coefficients, Conformal OPE and 
Hadron Matrix Elements}
\label{sssec:general-structure-amplitude}

The twist-$(2+\gamma(j))$ contribution to $C^{0,0,m}_{V_1;+,\alpha}$ can 
be determined completely, using the approximations above. 
\begin{eqnarray}
 C^{0,0,m}_{V_1;+,\alpha} & \simeq & \left(\frac{2\Lambda}{\Delta}\right)^{i\nu} 
   \frac{\Gamma(i\nu+1)}{(2\Lambda)^j}
   \int \frac{dz}{z} 
      [(q_1z)K_1(q_1z)][(q_2z)K_1(q_2z)] z^j  
    \label{eq:C00m-a} \\
 & & 
   \sum_{N=0}^{[\frac{m}{2}]}
   \left[ 2^j (q^{\hat{\rho}_1} \cdots q^{\hat{\rho}_j}) \cdot 
          \tilde{E}^N D^{m-2N}[\epsilon^{(0,0,m)}]_{\rho_1 \cdots \rho_j} \right]
   \frac{b^{(j-m)}_{m,N}}{\Delta^{m-2N}} 
    \left[e^{(2-j)A} \bar{\bar{\Psi}}^{(j);m,N}_{i\nu; 0,0,m} \right]. \nonumber
\end{eqnarray}
The product of rank-$j$ $\SO(3,1)$ tensors in the second line is 
reduced to a product of rank-$(j-m)$ tensors by the computation 
in (\ref{eq:q-[mmode]-product}). The Reggeon wavefunction $\bar{\bar{\Psi}}$
is also rewritten by using the small $(\Delta z_\gamma) \lesssim (\Delta/q)$ 
approximation (\ref{eq:approx-I-inu}): 
\begin{equation}
 [e^{(2-j)A} \bar{\bar{\Psi}}^{(j);m,N}_{i\nu;0,0,m}] \simeq  
   \sum_{a=0}^N (-1)^a{}_N C_a \left( 
  \zeta^{j+1} \partial_\zeta^{m-2a} [ \zeta^{-1-j+m} 
     \left(\zeta/2\right)^{i\nu} ]\right)_{\zeta \rightarrow (\Delta z)}
   \frac{N_{j,m}}{\Gamma(i\nu+1)}.
\end{equation}
The $a=0$ term in this expression has the lowest dimension in 
$\zeta = \Delta z_\gamma \lesssim (\Delta/q)$, and hence we only need to retain 
the $a=0$ term for a given $N$ for the twist-$(2+\gamma(j))$ contribution. 
Thus, 
\begin{equation}
 [e^{(2-j)A} \bar{\bar{\Psi}}^{(j);m,N}_{i\nu;0,0,m}] \simeq  
   2^{-i\nu} \frac{(-1)^m \Gamma(j+1-i\nu)}{\Gamma(j+1-i\nu-m)} 
   (\Delta z_\gamma)^{i\nu} 
   \frac{N_{j,m}}{\Gamma(i\nu+1)}.
\end{equation}
Using this expression and (\ref{eq:q-[mmode]-product}) 
in (\ref{eq:C00m-a}), we obtain 
\begin{eqnarray}
 C^{0,0,m}_{V_1;+,\alpha} & \simeq & \frac{\Lambda^{i\nu - j}}{q^{i\nu+j}}
\frac{j!}{(j-m)!} \sum_{N=0}^{[m/2]}
   \left[ \frac{(q\cdot \Delta)^2}{\Delta^2}\right]^N
   (-i)^m \frac{(q \cdot \Delta)^{m-2N}}{\Delta^{m-2N}} b^{(j-m)}_{m,N} \; 
   \left[ (q_{\mu_1} \cdots q_{\mu_{j-m}}) \cdot \epsilon^{(0,0,m)} \right] 
    \nonumber \\
 && 
  \int \frac{dz}{z} [(q_1z)K_1(q_1z)][(q_2z)K_1(q_2z)] (qz)^{j+i\nu}
    \frac{(-1)^m \Gamma(j+1-i\nu)}{\Gamma(j+1-i\nu-m)} 
    N_{j,m},  \\
 & = & \left( \frac{\Lambda}{q} \right)^{i\nu-j} 
   \frac{\left[ (q_{\mu_1} \cdots q_{\mu_{j-m}}) \cdot \epsilon^{(0,0,m)} \right] }
        {(q^2)^j} \frac{(q \cdot \Delta)^m}{\Delta^m} 
    C_1\left(j+i\nu, \frac{\eta}{x} \right) \nonumber \\
 & & i^m \frac{j!}{(j-m)!} N_{j,m}
     \frac{\Gamma(j+1-i\nu)}{\Gamma(j+1-i\nu-m)}    
     \left( \sum_{N=0}^{[m/2]} b^{(j-m)}_{m,N} \right),  \\
 & = & \left( \frac{\Lambda}{q} \right)^{i\nu-j} 
   \frac{\left[ (q_{\mu_1} \cdots q_{\mu_{j-m}}) \cdot \epsilon^{(0,0,m)} \right] }
        {(q^2)^j} \frac{(q \cdot \Delta)^m}{\Delta^m} 
    C_1\left(j+i\nu, \frac{\eta}{x} \right)  \nonumber \\
 & &  i^m 
   \frac{\Gamma(j+1+i\nu-m)}{N_{j,m}\Gamma(j+1+i\nu)},
\end{eqnarray}
where (\ref{eq:C1-def}, \ref{eq:C1-from-integral-B}) is used for the 
equality in the middle, while (\ref{eq:Njm-def-app}) is for the 
last one.

Repeating the same argument as in section \ref{sssec:m=0-amplitude},
we thus arrive at  
\begin{eqnarray}
 (V_1^{+,\alpha})_m & \simeq & - \int \frac{dj}{4i}
   \frac{1+e^{-\pi i j}}{\sin(\pi j)} 
   \frac{t_\gamma/t_y}{\Gamma(j+1)} \frac{\lambda}{i\nu_j} \; 
  C_1(j+i\nu_j) \left(\frac{\Lambda}{q}\right)^{\gamma(j)} 
  \label{eq:V1-m} \\
  & & 
   \left(\frac{1}{\sqrt{\lambda}x}\right)^j \; 
   \eta^m 
   \hat{d}_{j-m}([\eta]) \; \frac{\bar{g}^{0,0,m}}{\Delta^m} \; 
   \frac{i^m }{N_{j,m}}
   \frac{\Gamma(j+1+i\nu-m)}{\Gamma(j+1+ i\nu)}; \nonumber 
\end{eqnarray}
the computation in (\ref{eq:q-proj-p-product}, \ref{eq:dhat-Legendre}) 
for an even $j$ and $m$ was used once again.
Similarly to the case of $m=0$ amplitude, this expression is in the form 
of conformal OPE and inverse Mellin transformation 
in (\ref{eq:V1-inv-Mellin-conformal-QCD}). 
It should be noted that the integrand can be defined as a holomorphic 
function of $j$ (apart from poles and cuts), using the definition 
of $C_1$ in (\ref{eq:C1-def}), and that of $\hat{d}_{j-m}$ in 
(\ref{eq:dhat-Legendre}), not just for integer-valued $j$; 
at the same time, $\eta^m \hat{d}_{j-m}([\eta])$ becomes a polynomial of 
$\eta$ of degree $j$ for $j \in 2\N$, which is one of the important 
properties expected for the hadron matrix element \cite{GPD-review}.

The integration contour of (\ref{eq:V1-m}) is chosen so that it circles 
around the pole at $j=m$ after running just below the real positive axis 
in the $j$-plane and before running just above the real positive axis.
Only spin-$j$ stringy states with $m \leq j$ contribute then. 
It is not obvious whether the contour can be deformed so that it 
encircles $j=0,2, \cdots, m$ without changing $(V_1^{+,\alpha})_m$, and we 
leave it an open question. 
$\hat{d}_{j-m}$ in (\ref{eq:V1-m}) is given by a Legendre polynomial 
of degree $(j-m)$ when $j-m$ is an even positive integer, but 
otherwise it is defined by the hypergeometric function as in 
(\ref{eq:dhat-Legendre}), and it may or may not have a zero at negative 
even integer $(j-m)$ so that the pole from $\sin(\pi j)$ is canceled. 
Similarly, $\bar{g}^{0,0,m}(j,i\nu_j,\Delta)/N_{j,m}$ may or may not 
have a zero at negative integer $(j-m)$. The authors have not 
found a reason to believe that they have a zero, but we may be wrong. 

The twist-$(2+\gamma(j))$ contribution to the structure function 
$V_1^{+,\alpha}$ is obtained by summing up $(V_1^{+,\alpha})_m$ from 
the $(n,l,m)=(0,0,m)$ modes with $m=0,2,\cdots $:
\begin{equation}
 V_1^{+,\alpha} = \sum_{m = 0,2,\cdots}^\infty (V_1^{+,\alpha})_m \; .
\label{eq:V1-summing-V1-m}
\end{equation}
Combining (\ref{eq:V1-m=0}, \ref{eq:V1-m}) with (\ref{eq:V1-summing-V1-m}), 
a holographic version of (\ref{eq:V1-inv-Mellin-conformal-QCD}) is obtained. 
It is not obvious, though, whether or not the integration variable $j$ 
in (\ref{eq:V1-m}) for all the different $m$'s should be identified. If we 
are to define $j':=(j-m)$ and use it as a new variable of integration, then 
the integration contour of (\ref{eq:V1-m}) would be the same for all 
different $m$'s; the cost of doing so, however, is in this:
\begin{equation}
 C_1(j+i\nu_j,\vartheta)
 \left(\frac{\Lambda}{q}\right)^{\gamma(j)} \!\!\!\!
 \frac{1}{x^j} \eta^m \hat{d}_{j-m} 
 = \left[ C_1(j'+m+i\nu_{j'+m},\vartheta)\times \vartheta^m \right]
  \left(\frac{\Lambda}{q}\right)^{\gamma(j'+m)} \!\!\!
  \frac{1}{x^{j'}} \hat{d}_{j'}. 
\end{equation}
Certainly $\hat{d}_{j'}$ still remains to be a polynomial of degree
at most $j'$, but the expression no longer fits into the form 
of conformal OPE. For this reason, we identify the integration 
variable $j$ in (\ref{eq:V1-m}) for all $m=0,2,\cdots $ with 
that (complex angular momentum) of the inverse Mellin transformation
(\ref{eq:V1-inv-Mellin-conformal-QCD}).
This implies that the reduced hadron matrix element of the spin-$j$ 
primary operator is given a holographic expression 
\begin{equation}
 \overline{A}_j^{+,\alpha}(\eta,t) \propto \sum_{m=0}^j
   \frac{(-1)^{m/2}}{\sqrt{\lambda}^j \Gamma(j+1)}
   \frac{\bar{g}^{0,0,m}(j, i\nu_j, \Delta^2)}{N_{j,m} \Delta^m}
   \frac{\Gamma(j+1+i\nu_j-m)}{\Gamma(j+1+i\nu_j)} \times  
 \eta^m \hat{d}_{j-m}([\eta]).
\label{eq:Abar-eta-t-holographic}
\end{equation}
%

\subsubsection{The $(m=2)$-Mode Hadron Matrix Element}
\label{sssec:m=2-amplitude}

Most aspects of the expression (\ref{eq:V1-m}) are dictated by 
basic principles of field theory such as (conformal) OPE. 
Additional information from the holographic set-up is found 
primarily in the hadron matrix element $\bar{g}^{0,0,m}(j,i\nu,\Delta)$, 
apart from the anomalous dimension $\gamma(j) = i\nu_j - j$ of the 
twist-$(2+\gamma(j))$ operators. Now we have seen that 
$\bar{g}^{0,0,0}(j,i\nu_j,\Delta)$ is not the only hadron matrix element 
contributing to the non-perturbative information of 
$h+\gamma^{*} \longrightarrow h+\gamma^{(*)}$; let us take a moment here 
to have a closer look at one of the new hadron matrix elements we
encounter, $\bar{g}^{0,0,2}(j,i\nu,\Delta)$.

The hadron matrix element $\Gamma^{0,0,m}$ receives contributions 
from $\bar{J}^{hh}_{z^k \lambda_{k+1} \cdots \lambda_j} 
\bar{A}^{\hat{z}^k \hat{\lambda}_{k+1}\cdots \hat{\lambda}_j}$'s with 
$k=0,1,\cdots, m$. The contribution from each $k$ can be written in 
the form of (\ref{eq:Gamma-00m-factorize-tensor}), and hence 
$(\bar{g}^{0,0,m}(j,i\nu,\Delta))_{k}$ is defined ($k\leq m$).
We compute $(\bar{g}^{0,0,2})_k$ explicitly for $k=0,1,2$.

For this purpose, we need the following technical results: 
\begin{equation}
 \left( \nabla^l \Phi \right)_{\lambda_1 \cdots \lambda_l}
  \equiv \left( \partial_{\lambda_1} \cdots \partial_{\lambda_l} \Phi \right)
  - \sum_{1 \leq a<b \leq l} \frac{\eta_{\lambda_a \lambda_b}}{z}
      \left( \partial_{\lambda_1} \cdots {}_{\check{\lambda_a}}{}_{\check{\lambda}_b} 
        \cdots \partial_{\lambda_l} 
      \left( \partial_z + \frac{l-a-1}{z} \right) \Phi \right)
\end{equation}
modulo terms proportional to $\eta_{\lambda_a \lambda_b} \eta_{\lambda_c \lambda_d}$ 
instead of (\ref{eq:4-hadron-current-m0k0}), and 
\begin{eqnarray}
 \left(  \nabla^l \Phi \right)_{\lambda_1 \cdots z \cdots \lambda_l}
 & \equiv & \left( \partial_z + \frac{l-1}{z} \right) 
        \partial_{\lambda_1} \cdots {}_{\check{\lambda}_a} \cdots \partial_{\lambda_l} \Phi, \\
 \left( \nabla^l \Phi \right)_{\lambda_1 \cdots z \cdots z \cdots \lambda_l}
  & \equiv & 
     \left[ \left(\partial_z + \frac{l-1}{z}\right) 
            \left(\partial_z + \frac{l-2}{z}\right) + \frac{a-1}{z^2}
     \right] 
     \partial_{\lambda_1} \cdots {}_{\check{\lambda}_a} {}_{\check{\lambda}_b} \cdots 
     \partial_{\lambda_l} \Phi, \nonumber 
\end{eqnarray}
modulo terms proportional to $\eta_{\lambda_c \lambda_d}$.

It is now a straightforward computation to use the relations above 
as well as the explicit Reggeon wavefunctions $\bar{A}$ determined 
in section \ref{sec:mode-fcn}, to derive the following:
\begin{eqnarray}
 \frac{\bar{g}^{0,0,2}_{k=2}}{N_{j,2} \Delta^2} & = & 
   \frac{j(j-1)}{2} 
     \int_0^{1/\Lambda} \frac{dz}{z^3} (\Lambda z)^j 
    \frac{\left\{ z^2 ``K_{i\nu}(\Delta z)'' \right\} }
         {[\left(\frac{\Delta}{2\Lambda}\right)^{-i\nu} \Gamma(i\nu)]}
      \nonumber \\
 & & \qquad \qquad 
   t_h \left[ 2\left\{- \Phi (\partial^2_z \Phi) + (\partial_z \Phi)^2 \right\}
     -\frac{2}{z} \Phi (\partial_z \Phi)
     - \frac{4(j-2)}{3z^2} \Phi^2 \right], \nonumber \\
  \frac{\bar{g}^{0,0,2}_{k=1}}{N_{j,2} \Delta^2} & = & 
   \frac{j(j-1)}{2} 
     \int_0^{1/\Lambda} \frac{dz}{z^3} (\Lambda z)^j 
     \frac{ \left\{ z^{j+1} \partial_z \left(z^{1-j} ``K_{i\nu}(\Delta z)'' \right)
      \right\} }{[\left(\frac{\Delta}{2\Lambda}\right)^{-i\nu} \Gamma(i\nu)]}
   \left[\frac{-2 t_h}{z} \Phi^2 \right], \nonumber \\
 \frac{\bar{g}^{0,0,2}_{k=0}}{N_{j,2} \Delta^2} & = & j(j-1)
     \int_0^{1/\Lambda} \frac{dz}{z^3} (\Lambda z)^j \times \nonumber \\
 & & \left( 
    \frac{\left\{ [z^{j+1} \partial_z^2 z^{1-j}-(z\Delta)^2] ``K_{i\nu}(\Delta z)'' 
     \right\}}{[\left(\frac{\Delta}{2\Lambda}\right)^{-i\nu} \Gamma(i\nu)]}
   \times \left[\frac{p^2}{\left(j-\frac{1}{2}\right)\Delta^2}t_h \Phi^2 \right] 
     \right. \nonumber \\
 & & \left. + \frac{\left\{ -z^2 ``K_{i\nu}(\Delta z)'' \right\}}
            {[\left(\frac{\Delta}{2\Lambda}\right)^{-i\nu} \Gamma(i\nu)]}
    \times t_h 
     \left[ \frac{1}{z} \Phi (\partial_z \Phi)+ \frac{j-2}{3z^2} \Phi^2 
     \right]
    \right).
\end{eqnarray}
These results are used in the study below.

\section{A Holographic Model of GPD}
\label{sec:model}

The differential cross section of DVCS process involves integral of GPD; 
GPD needs to be parametrized first, and then the parameters are determined 
by fitting the data \cite{DVCS-GPD-dispersion}. The idea of dual 
parametrization of GPD \cite{dual-para}---also known as collinear factorization 
approach \cite{Mueller-S-05, KK-Mueller-07}---is to expand the reduced 
hadron matrix element $\overline{A}^{+,\alpha}_j(\eta,t)$ as 
\begin{equation}
 \overline{A}^{+,\alpha}_j(\eta,t) = \sum_{m=0}^j
    \overline{\Gamma}^{+,\alpha}_{m}(j,t) \; \eta^{m} \times 
     \left[\eta^{j-m} d_{j-m}(1/\eta) \right],
\label{eq:dual-parametrize}
\end{equation}
where $d_\ell(\cos \theta)$'s are polynomials of degree $\ell$ in the 
argument $(\cos\theta)$; Legendre polynomials, Gegenbauer polynomials or 
Jacobi polynomials are used depending on the helicity change of the target 
hadron $h$ in the scattering process \cite{GPD-review}. When the target 
hadron is a scalar, as in the study of this article, Legendre polynomial 
is chosen for $d_\ell$ \cite{dual-para}. 
With no ambiguity introduced in the polynomials $d_{j-m}(x)$, 
$\overline{\Gamma}^{+,\alpha}_{m}(j,t)$'s are the fully general, 
yet non-redundant parametrization for the reduced hadron matrix element 
for GPD.

At the end of the study in the preceding sections, we arrived at a holographic 
model of GPD, with the reduced hadron matrix element given by 
(\ref{eq:Abar-eta-t-holographic}) for the flavor-non-singlet sector. 
String theory---the descendant of the dual resonance model---yields 
a result that fits straightforwardly with the format of the dual 
parametrization (\ref{eq:dual-parametrize}); this should not be 
a surprise, but must be something the authors of \cite{dual-para} have 
anticipated. With the string-theory implementation provided, one can 
now move forward; now 
\begin{equation}
\overline{\Gamma}^{+,\alpha}_m(j,t) \sim  
   (-1)^{m/2} \frac{\bar{g}^{0,0,m}(j,i\nu_j,\Delta)}{N_{j,m} \Delta^m} 
\end{equation}
can be computed using holographic backgrounds, independently from 
experimental data. Certainly the matrix elements $[\bar{g}^{0,0,m}/\Delta^m]$
will depend on holographic backgrounds to be used for computation, and 
predictions from individual holographic backgrounds should not be taken 
seriously at the quantitative level. But it is still worth looking 
closely into qualitative features of the holographic hadron matrix elements 
$\bar{g}^{0,0,m}/\Delta^m$ to learn non-perturbative aspects of 
$\overline{\Gamma}^{+,\alpha}_m(j,t)$.

\subsection{$\Delta^2 \longrightarrow 0$ Limit}
\label{ssec:D=0-cancellation}

As we have already remarked earlier in this article, the holographic 
result (\ref{eq:Abar-eta-t-holographic}) is not precisely in the same form 
of parametrization as in (\ref{eq:dual-parametrize}); the argument 
of the polynomial $\hat{d}_{j-m}$ is $[\eta]$ defined in (\ref{eq:[eta]-def}), 
rather than $\eta$. This difference itself does not raise an issue immediately; 
$[\eta]$ is the same as $\eta$ in the hard scattering regime, 
$\Delta^2 \gg m_h^2$. 

Let us study how the hadron matrix element behaves in the 
$t = -\Delta^2 \longrightarrow 0$ limit, however. The matrix 
element $\bar{g}^{0,0,0}(j,i\nu_j,\Delta)$ has already been studied 
in the literature, and is known not to diverge or vanish in the 
$\Delta^2 \longrightarrow 0$ limit. The polynomial $\hat{d}_j([\eta])$ 
to be multiplied with this $\bar{g}^{0,0,0}(j,i\nu_j,\Delta)$, however, 
has diverging coefficients in all of the terms $\eta^2$, $\eta^4, \cdots$ 
except the $\eta^0$ term. Therefore, the $m=0$ contribution (\ref{eq:V1-m=0}) 
alone does not have a physically reasonable behavior in the 
$\Delta^2 \longrightarrow 0$ limit. A natural expectation will be 
that the hadron matrix element $\overline{A}^{+,\alpha}_j(\eta,t)$ still 
has a reasonable behavior, after summing up $m=0,2,\cdots, j$.

To get started, we focus on the $\eta^2$ term. It is generated from 
the $m=0$ mode exchange, and also from the $m=2$ mode exchange. 
There is a $(p^2)/\Delta^2$ factor both in 
$\bar{g}^{0,0,0} \times \hat{d}_j([\eta])|_{\eta^2}$ 
and $\bar{g}^{0,0,2}/\Delta^2 \times \eta^2$, and hence both diverge in the 
$\Delta^2 \longrightarrow 0$ limit. When they are summed up, however, 
the divergence may cancel, as we see in the following.
Let us study the coefficient of the $\eta^2$ term 
\begin{equation}
 - \int \frac{dj}{4i} \frac{1+e^{-\pi i j}}{\sin(\pi j)}
 \left(\frac{\Lambda}{q}\right)^{i\nu-j} 
 \left(\frac{1}{\sqrt{\lambda}x}\right)^j
 C_1\left(j+i\nu, \frac{\eta}{x} \right)
 \frac{\lambda}{i\nu_j} \frac{t_\gamma/t_y}{\Gamma(j+1)} \times \eta^2
\label{eq:common-prefactor-eta2}
\end{equation}
in the $\Delta^2 \longrightarrow 0$ limit, picking up 
contribution to the integral $\bar{g}^{0,0,0}$ and $\bar{g}^{0,0,2}$ from 
the $I_{-i\nu}(\Delta z_h)$ component in ``$K_{i\nu}(\Delta z_h)$'' 
first.\footnote{\label{fn:I-minu-LO}
The leading divergence in the $\Delta^2 \longrightarrow 0$ 
limit comes from 
\begin{equation}
K_{i\nu}(\Delta z) \sim \left(\frac{\pi}{2}\right)
  \frac{I_{-i\nu}(\Delta z)}{\sin(\pi i \nu)} \simeq  
 \left(\frac{\pi}{2}\right)
  \frac{(\Delta z/2)^{-i\nu}}{\sin(\pi i \nu)\Gamma(-i\nu+1)} 
 = \frac{\Gamma(i\nu)}{2}(\Delta z/2)^{-i\nu}.
\end{equation}
}
Then in that limit, the coefficient of the 
expression (\ref{eq:common-prefactor-eta2}) becomes 
\begin{eqnarray}
  \frac{p^2}{\Delta^2} \lim_{\Delta^2 \longrightarrow 0}
   \left[ \bar{g}^{0,0,0}(j,i\nu_j,\Delta)
          \frac{j(j-1)}{(j-\frac{1}{2})} - 
          \frac{\bar{g}^{0,0,2}(j,i\nu_j,\Delta)/(p^2)}
               {(j-1+i\nu_j)(j+i\nu_j)N_{j,2}} \right]
 + {\cal O}(\Delta^0).
\end{eqnarray}
The two terms in $\lim_{\Delta^2 \longrightarrow 0}[\cdots ]$ cancel each other, 
as one can see by using the approximation in footnote \ref{fn:I-minu-LO}.
Thus, the $\eta^2$ term in $\overline{A}^{+,\alpha}_j(\eta,t)$ also has a 
finite limit value in the $\Delta^2 \longrightarrow 0$ limit. 

It is quite likely, however, that the $I_{i\nu}(\Delta z)$ component 
in ``$K_{i\nu}(\Delta z)$'' has just as important contribution 
as the $I_{-i\nu}(\Delta z)$ component does in the 
$\Delta^2 \longrightarrow 0$ limit to the hadron matrix elements 
$\bar{g}^{0,0,0}$ and $\bar{g}^{0,0,2}$;
the coefficient $(1-c^{(j)}_{i\nu;0,0,m})$ may behave as $(\Delta/\Lambda)^{-2i\nu}$
in the $\Delta^2 \longrightarrow 0$ limit. 
Because we have seen above that the divergence $(p^2/\Delta^2)$ cancels 
when only the $I_{-i\nu}(\Delta z)$ component is taken into account, 
the contributions from the $I_{i\nu}(\Delta z)$ should also have some 
cancellation mechanism. Using an approximation for the $I_{i\nu}(\Delta z)$ 
components in ``$K_{i\nu}(\Delta z)$'' similar to the one in 
footnote \ref{fn:I-minu-LO}, one finds that the $(p^2/\Delta^2)$ divergence 
cancels in the $\eta^2$ coefficient, if and only if 
\begin{equation}
 \lim_{\Delta^2/\Lambda^2 \longrightarrow 0} \left[
   \left(\frac{\Delta}{2\Lambda}\right)^{2i\nu}
   \left\{ (1-c^{(j)}_{i\nu_j;0,0,0})-(1-c^{(j)}_{i\nu_j;0,0,2})
        \frac{(j-1-i\nu_j)(j-i\nu_j)}{(j-1+i\nu_j)(j+i\nu_j)}
   \right\} \right] = 0.
\label{eq:cond-no-div-eta2}
\end{equation}
The coefficients $c^{(j)}_{i\nu;0,0,m}$ are functions of $\Delta/\Lambda$, 
rather than complex numbers. The discussion above shows that 
physically sensible implementations of the confining effect require 
one condition above between the two functions $c^{(j)}_{i\nu;0,0,0}$ 
and $c^{(j)}_{i\nu;0,0,2}$.

The $\eta^{2M}$ term with $M=2,\cdots $, instead of the $\eta^2$ term 
in (\ref{eq:common-prefactor-eta2}), also receives divergent contributions 
from amplitudes of the $m=0,2,\cdots,2M$ mode exchange. There will be 
apparent divergence of order $(p^2/\Delta^2)^M$, $(p^2/\Delta^2)^{M-1}, \cdots, 
(p^2/\Delta^2)$. The cancellation of divergence in the 
$\Delta^2 \longrightarrow 0$ limit will set $M$ conditions on 
the $\Delta^2/\Lambda^2 \longrightarrow 0$ limit of $(1-c^{(j)}_{i\nu_j;0,0,2M})$.

In a phenomenological approach of implementing the confining effect, 
that is all we can say for now. With a little more model-building mind set, 
however, we can find some solutions to the conditions above. 
It is not hard to verify that the combination of 
\begin{equation}
 \left. 
 \left[\partial_z \left(\Psi^{(j);0,0}_{i\nu;0,0,0}(t,z) \right)
  \right]\right|_{z \Lambda = 1} = 0, \qquad 
  \left.
  \left[\partial_z \left(\Psi^{(j);2,0}_{i\nu;0,0,2}(t,z) \right) 
  \right]\right|_{z \Lambda = 1} = 0
\label{eq:confine-0and2-Neuman}
\end{equation}
results in $c^{(j)}_{i\nu;0,0,0}$ and $c^{(j)}_{i\nu;0,0,2}$ satisfying 
the condition (\ref{eq:cond-no-div-eta2}). It is tempting to 
generalize this and impose the boundary condition 
$\partial_z[\Psi^{(j);2M,0}_{i\nu;0,0,2M}]=0$ to determine $c^{(j)}_{i\nu;0,0,2M}$, 
though we do not know whether all the $m_h^2/\Delta^2$ divergences above 
are removed under this boundary condition. 
The top-down approach is much more authentic and well-motivated than 
such a hand-waving and wishful approach, and we do not try to speculate 
beyond that; we use this implementation of the confining effect, 
(\ref{eq:confine-0and2-Neuman}), only to ``get the feeling'' 
in the numerical presentation in section \ref{ssec:numerical}.

\subsection{Large $\Delta^2$ Behavior}

Certainly the holographic model of GPD yields a result of 
the reduced hadron matrix element that fits perfectly with the 
dual parametrization. The holographic result, however, turns out to be 
a little more complicated than the models that have often been explored 
for the purpose of phenomenological fit of the DVCS data. An example of 
model for phenomenological fit (see e.g., \cite{KK-Mueller-07}) was to 
introduce an ansatz that 
\begin{equation}
 \overline{\Gamma}^{+,\alpha}_{m}(j,t) = f_{j,m} \Sigma_{j-m}(t),
\label{eq:pheno-fit-form}
\end{equation}
where only one $(t= -\Delta^2)$-dependent {\em function} is involved 
in the form of a ``form factor'' $\Sigma_{j-m}(t)$ for some ``spin $(j-m)$'',
and all the remaining non-perturbative information is reduced to 
some {\it numbers} $f_{j,m} \in \R$. 
The function $\Sigma_J(t)$ may also be parametrized by an ansatz like 
\begin{equation}
  \Sigma_J(t) = \frac{1}{J-\alpha_0 - \alpha'_{\rm eff} t}
             \frac{1}{\left[ 1- \frac{t}{m^2(J)} \right]^{p}}, 
\label{eq:form-factor-KKM-model}
\end{equation}
in order to implement both the Regge behavior and the power-law form factor 
in the hard regime  $1 \ll -t/\Lambda^2$.
To fit the data in practice, it is certainly unavoidable to reduce the unknown 
information into a finite set of real {\it numbers}. 

A theoretical picture based on the holographic model, on the other hand, 
suggests that the $t=-\Delta^2$ dependence is more complicated than this. 
If we strictly stick to the expansion (\ref{eq:dual-parametrize}), 
then individual $\overline{\Gamma}^{+,\alpha}_m(j,t)$'s may diverge 
at $t = -\Delta^2 =0$, as we have seen above, and are not like form factors. 
The $\overline{\Gamma}^{+,\alpha}_m(j,t)$ would not depend only on the 
difference $(j-m)$ as in (\ref{eq:pheno-fit-form}) either; 
we have already seen that $\overline{\Gamma}^{+,\alpha}_{m=2}(j,t) \propto 
\bar{g}^{0,0,m=2}/\Delta^2$ diverges at $t = - \Delta^2 \longrightarrow 0$ 
for arbitrary $j$, but there is no such divergence in 
$\overline{\Gamma}^{+,\alpha}_{m=0}(j,t) \propto \bar{g}^{0,0,0}$, for example. 
Therefore, holographic models of GPD might be used as a theoretical guide 
to think of parametrization (for fitting) that is different from 
(\ref{eq:pheno-fit-form}).

The holographic model provided by the calculation in the previous section 
involves infinitely many spin-dependent form factors, 
$\bar{g}^{0,0,m}(j,i\nu_j,\Delta)/\Delta^m$. We can still find that they 
share a common behavior at large $\Delta^2 = -t$. To see this, 
note that ``$K_{i\nu}(\Delta z_h)$'' in the Reggeon wavefunction 
effectively cuts off the integral over the holographic radius $z_h$ 
at $z_h \lesssim 1/\Delta$ in the regime 
\begin{equation}
 \Lambda^2, \; m_h^2 \quad \ll \quad \Delta^2 
  \quad \ll \quad |q^2|, \; (p \cdot q), \; |(q \cdot \Delta)|.
\label{eq:hard-scatter-regime}
\end{equation}
The explicit form of $\Psi^{(j);s,N}_{i\nu; 0,0,m}(z;\Delta)$ 
in (\ref{eq:00m-mode-fcn}) is not more than modification ``$K_{i\nu}(\Delta z)$''
by a function of $\Delta z_h$, and hence they still play just the role of 
cutting-off the integral at $z_h \Delta \lesssim 1$.
The ``current'' $\bar{J}^{hh}_{z^k \lambda_{k+1} \cdots \lambda_j}$ 
provides extra $m$-th powers of either $1/z$ or $\partial_z$ and 
$(j-m)$ momenta $p_\lambda$, in addition to $[\Phi]^2$, which behaves like 
\begin{equation}
 [\Phi] \sim z (\Lambda z)^{\Delta_\phi-1}
\end{equation}
in the region $z \lesssim 1/\Delta \ll 1/\Lambda$; $\Delta_\phi$ is the 
conformal dimension of an operator in a strongly coupled gauge theory
dual to the holographic model, which is a property of the target hadron $h$.
The $\tilde{E}^ND^{s-2N}[\epsilon]/\Delta^{2-2N}$ operation on the $\SO(3,1)$ 
tensor in (\ref{eq:Pomeron-wvfc-general-expans}) does not introduce any 
power of $(\Delta/\Lambda)$ or $(\Lambda z)$. Therefore, we find 
in the hard scattering regime (\ref{eq:hard-scatter-regime}) that 
\begin{equation}
\frac{\bar{g}^{0,0,m}}{\Delta^m} \sim 
\left(\frac{\Delta}{\Lambda}\right)^{i\nu} \times 
  (\Lambda/\Delta)^{j + 2 (\Delta_\phi-1)} \times \Delta^m/\Delta^m
 \sim \frac{1}{(\Delta/\Lambda)^{2 \Delta_\phi-2 -\gamma(j)}}. 
\label{eq:largeD2-g00m}
\end{equation}
Interestingly, the reduced hadron matrix elements $\bar{g}^{0,0,m}/\Delta^m$ 
for $(j,m)$ have the large $\Delta^2$ power-law behavior that is 
independent of $m$; $2\Delta_\phi$ reflects a property of the target hadron 
$h$, and $-(2+\gamma(j)) = -\tau_n$ is $j$-dependent, but the power 
does not depend on $m$.\footnote{This scaling was known already for 
$\bar{g}^{0,0,0}$ \cite{NW-first}.} Holographic models suggest this 
$j$-dependent $p = {\rm const}. - \gamma(j)/2$ scaling behavior as an 
alternative to the fixed-power $p = {\rm const}.$ scaling 
of (\ref{eq:form-factor-KKM-model}).

We have chosen a factorization into the Wilson coefficient and 
the matrix element that corresponds to renormalization at $\mu = \Lambda$; 
this choice was made implicitly when we chose a factor 
$[\Delta^{i\nu}/\Lambda^{i\nu-j}]^{\pm 1}$ at the time the amplitude was 
factorized into $C^{0,0,m}$ and $\Gamma^{0,0,m}$ in (\ref{eq:Gamma-00m-def}).
When we keep the renormalization scale $\mu$ to be arbitrary
(e.g., taking $\mu$ higher than $\Delta$ when $\Delta  \gg \Lambda$), the 
Wilson coefficient contains a factor $(\mu/q)^{\gamma(j)}$ instead of 
$(\Lambda/q)^{\gamma(j)}$, and the reduced matrix element also has the following 
large $\Delta^2$ behavior, 
\begin{equation}
\frac{\bar{g}^{0,0,m}}{\Delta^m} \sim 
  \frac{1}{(\Delta/\Lambda)^{2 \Delta_\phi-2}} \times
  \frac{1}{(\mu/\Delta)^{\gamma(j)}}. 
\end{equation}
%

\subsection{Pomeron and Superstring}
\label{ssec:super-Pomeron}

We have so far talked about Reggeon and the flavor-non-singlet sector in  
sections \ref{sec:organize}--\ref{sec:model}, instead of Pomeron. 
Since flavor-singlet sector ($\approx$ gluon) dominates in the 
small-$x$ physics, that was not a desired choice.  

This is due to technical limitation in string theory at this moment. 
In order to deal with propagation of string states on a curved spacetime, 
vertex operators and $L_0$ (Virasoro generator) need to be defined 
properly as composite operators; the non-linear sigma model for 
AdS$_5 \times W_5$ on the world-sheet becomes conformal and the 
renormalization of the composite operators well-defined, however, only 
after the Ramond--Ramond background is also implemented 
(e.g., \cite{Tseytlin-RR}). 
Presumably an option in the future will be to implement 
the Klebanov--Strassler model and its variations 
in the Green--Schwarz formalism. 
One then computes the spectrum of stringy excited states, and further 
works out the world-sheet OPE, in the form of 
\begin{equation}
V^{(q_1)}(z)V^{(-q_2)}(-z) \sim \sum_I C_I(z) {\cal O}_I(0) 
\label{eq:OPE-mid-point}
\end{equation}
using operators ${\cal O}_I(0)$ at the middle point, where $V^{(q_1)}$ and 
$V^{(-q_2)}$ are the 
vertex operators corresponding to the incoming and outgoing photons 
(\ref{eq:vx-op-Killing-vector}). 
In this way, we would not have to use string field theory. 

It may also be possible to use the bosonic string field theory 
for closed string, instead of the bosonic open string field theory 
we used in section \ref{sec:cubic-sft} of this article. 
Bosonic closed string field theory is also well-understood 
already \cite{ClosedSFT-Zwiebach92}. 
Certainly the bosonic closed string field theory is not for 
Type IIB superstring, but it will still allow us to get the feeling of how much 
open string (flavor non-singlet sector) and closed string (flavor singlet 
sector) are different, from theoretical perspectives, as well as in 
phenomenological consequences.  At least it is known that the 
Virasoro--Shapiro amplitude is generated, not just by the 1-string 
exchange in the $t$-channel, $s$-channel and $u$-channel, but also 
a four-point contact interaction vertex in the string field 
theory \cite{ClosedSFT-4pt}.
The Virasoro--Shapiro amplitude does not have a simple $s$--$t$ duality 
of the Veneziano amplitude, either. Certainly it is possible to write it 
down in the form of ``$t$-channel'' expansion only (cf \cite{BPST-06} and 
\cite{NW-first}), but we also need to be aware that the discussion 
in these two references did not use the OPE at the middle point 
as in (\ref{eq:OPE-mid-point}), but used an OPE of the form 
$V(z)V(0) \sim \sum_IC_I(z){\cal O}_I(0)$. To get the skewness-dependence 
right, this difference really matters. Thus, an analogue of the prescription 
(\ref{eq:SFT-Veneziano-prescr}) needs to be worked out separately for 
the closed string amplitude. 

Orthodox approaches such as those above are way beyond the scope of 
this article. One can hardly overestimate importance of such a solid 
approach, but at the same time, very few would find that the following 
guess would not be terribly off the mark. For practical purposes, therefore, 
one can live with that for the time being. First of all, the on-shell 
relation for the bosonic open string in (\ref{eq:on-shell-openstring}) 
will be replaced by 
\begin{equation}
 \frac{j}{2}-1 + \frac{4+j+\nu^2 + c_j}{4\sqrt{\lambda}} = 0, 
\label{eq:on-shell-closedstring}
\end{equation}
with the constraint $c_{j=1}=-4$ for bosonic open string replaced 
by $c_{j=2} = -2$.
Interaction vertices should also be different; looking at the 
the difference between the Veneziano amplitude and the Virasoro--Shapiro 
amplitude, one finds that the following replacements should be made:
\begin{equation}
  \frac{t_\gamma/t_y}{\Gamma(j+1)} \longrightarrow 
  \frac{t_\gamma/t_y}{[\Gamma(j/2)]^2}, \qquad 
  \left(\frac{1}{\sqrt{\lambda}x}\right)^j \longrightarrow 
  \left(\frac{1}{4\sqrt{\lambda}x}\right)^j.
\end{equation}
The overall normalization $t_\gamma/t_y$ is like $N_c / N_c^{-2} \sim N_c^{-1}$ 
now, when the Pomeron (closed string) contribution is used in the 
$t$-channel, and the source field for the ``QED current'' is implemented 
in the form of D7-brane gauge field; the $1/N_c$ scaling 
(see footnotes \ref{fn:closed-kin-normalization} and \ref{fn:open-kin-normalization}) 
is also the natural expectation in the large $N_c$ argument. 

\subsection{Numerical Results}
\label{ssec:numerical}

At the end of this article, we leave a few plots of numerical evaluation 
of various results that have been obtained. We do not intend to 
provide a quantitative (precise) prediction from holography, as we have 
repeatedly emphasized our perspective on this issue in this article;
the holographic approach to GPD will provide at best a qualitatively 
new way to think of how to parametrize the matrix elements for GPD.
Having said that, it is still desirable to grasp various expressions 
in a more intuitive form and bring them down to more practical situations. 
This section \ref{ssec:numerical} serves for this purpose. 

There are a couple of parameters that need to be specified, in order 
to obtain numerical outputs in a few summary plots. We used the on-shell 
relation (\ref{eq:on-shell-openstring}), which means that we should 
understand the numerical results to be that of Reggeon contribution. 
We adopted $c_j +4 = 0$ for all $j$, although there is no rationale to 
specify the $j$-dependence in this way (see \cite{Cornalba-cj, Costa-1209} 
and literatures therein for how to work out the $j$-dependence of $c_j$).
The confining effect was implemented in the form of the boundary condition 
 (\ref{eq:confine-0and2-Neuman}) for the Reggeon wavefunction.
As for the target hadron, we set the mass term of the scalar field 
to be $5/R^2$ (i.e., $c_y = 5$), just like the lowest non-trivial 
spherical harmonics on $W_5 = S_5$ for the Type IIB dilaton 
field \cite{KimNeuwen}. The operator dimension in the dual CFT becomes 
$\Delta_\phi = 2 + \sqrt{4 + c_y} = 5$.

\begin{figure}[tbp]
\begin{center}
\begin{tabular}{cc}
 \includegraphics[scale=0.55]{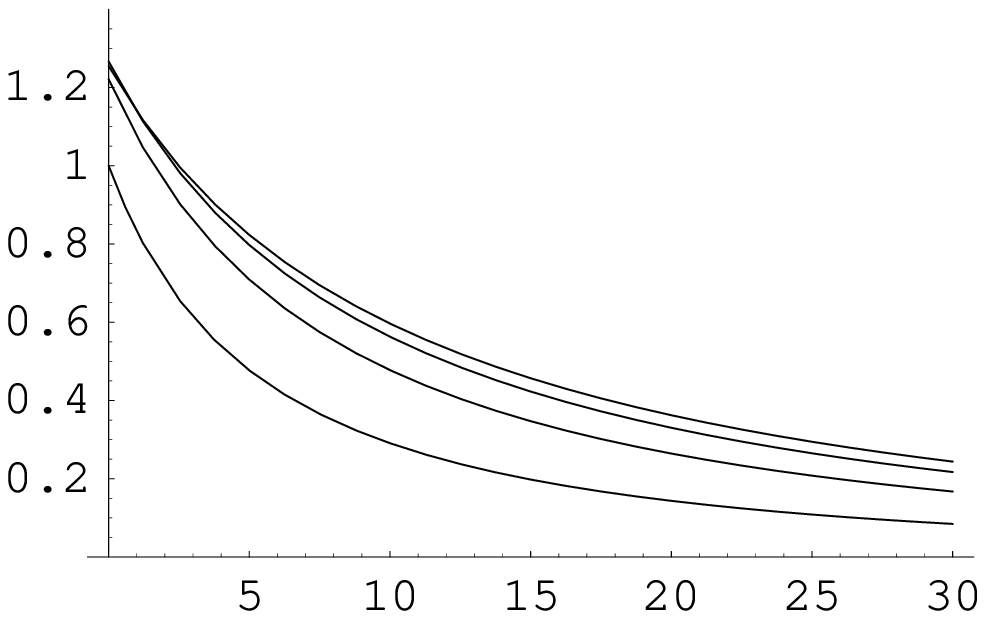} &
 \includegraphics[scale=0.55]{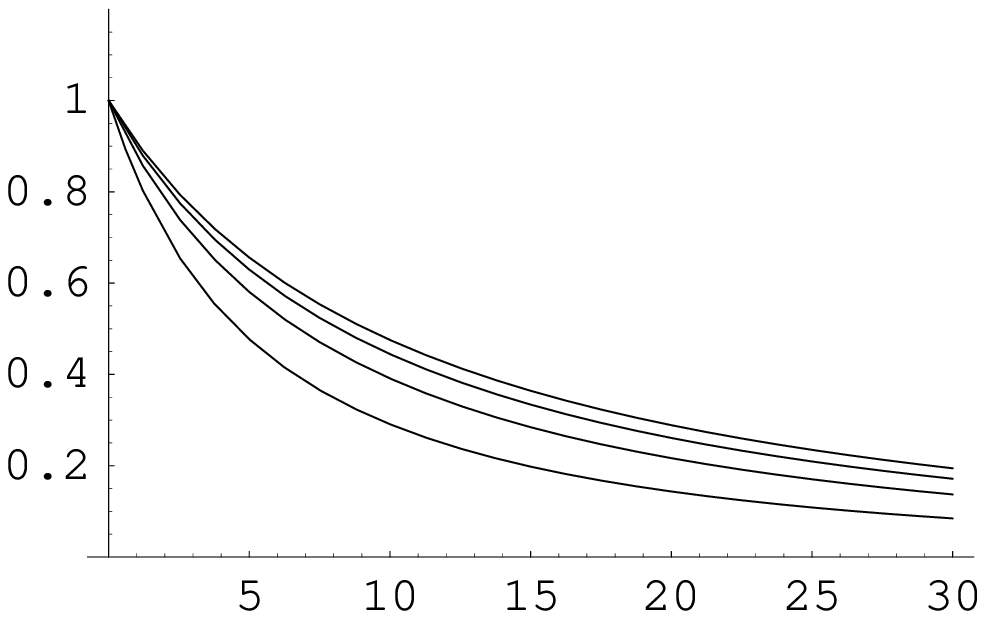} \\
(a) & (b) 
\end{tabular}
\caption{
\label{fig:g000}
The panel (a) shows $\bar{g}^{0,0,0}(j,i\nu_j, \Delta)$ as 
a function of $\Delta^2/\Lambda^2$.
The curve at the bottom is for $j=1$, while the one at the top is 
for $j=2.5$; two in the middle correspond to $j=1.5$ and $j=2$.
The panel (b) shows $\bar{g}^{0,0,0}(\Delta)/\bar{g}^{0,0,0}(\Delta = 0)$, i.e., 
$\bar{g}^{0,0,0}(j,i\nu_j, \Delta)$ normalized at the value of $\Delta^2 = 0$.
The curve at the bottom is for $j=1$, and the curve goes up for $j=1.5$, 2 and 
$2.5$; this softer behavior for larger $j$ is consistent 
with (\ref{eq:largeD2-g00m}). }
\end{center}
\end{figure}

Figure \ref{fig:g000} shows the reduced matrix element 
$\bar{g}^{0,0,0}(j,i\nu_j,\Delta)$ for the $m=0$-mode exchange; 
the results for different values of spin $j=1, 1.5, 2, 2.5$ are shown in 
the figure. Lattice computation can be used to determine matrix elements 
at integer valued spins, but the analytic expression (\ref{eq:g000-def}) 
allows us to determine the matrix element even for non-integer spin, so that 
the inverse Mellin transformation is possible, and we can also talk of 
the matrix elements evaluated at the saddle point value of spin $j=j^*$.
The panel (b) in Figure~\ref{fig:g000} is essentially the same 
as that of Fig. 5 in \cite{NW-first}, while the panel (a) shows 
$\bar{g}^{0,0,0}$ without normalizing the matrix element by its value at 
$t = -\Delta^2 = 0$. Since they are not the matrix element 
of a ``conserved current'' for $j \neq 1$, the matrix element 
does not necessarily approach $1$ in the $\Delta^2 \longrightarrow 0$ limit.
The panel (b) has a property that $\bar{g}^{0,0,0}$ is soft 
($\bar{g}^{0,0,0}$ gets smaller slowly in $\Delta^2$) for larger $j$; 
this is consistent with the observation in (\ref{eq:largeD2-g00m}), 
because $\partial \gamma(j)/\partial j > 0$.

\begin{figure}[tbp]
\begin{center}
\begin{tabular}{cc}
(a) & (b) \\
 \includegraphics[scale=0.65]{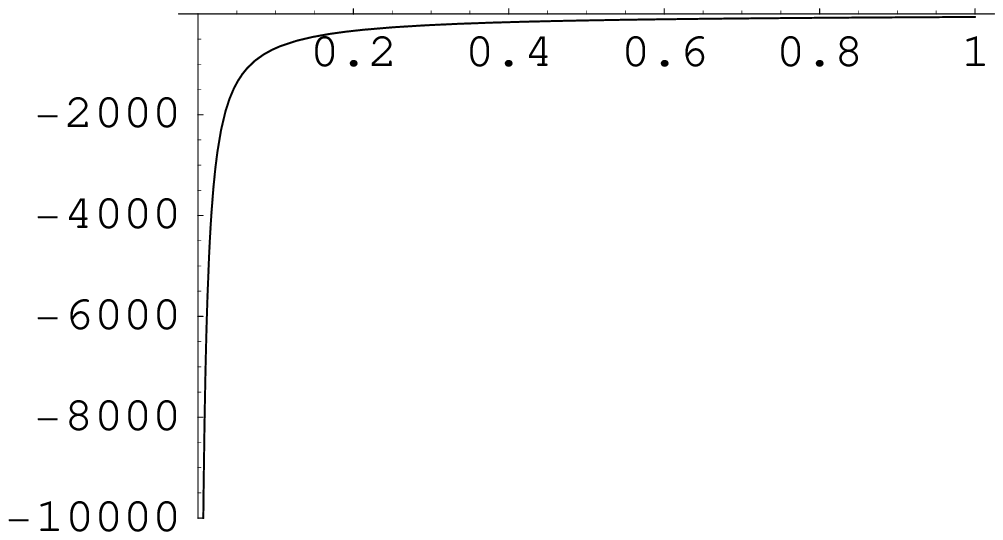} &
 \includegraphics[scale=0.65]{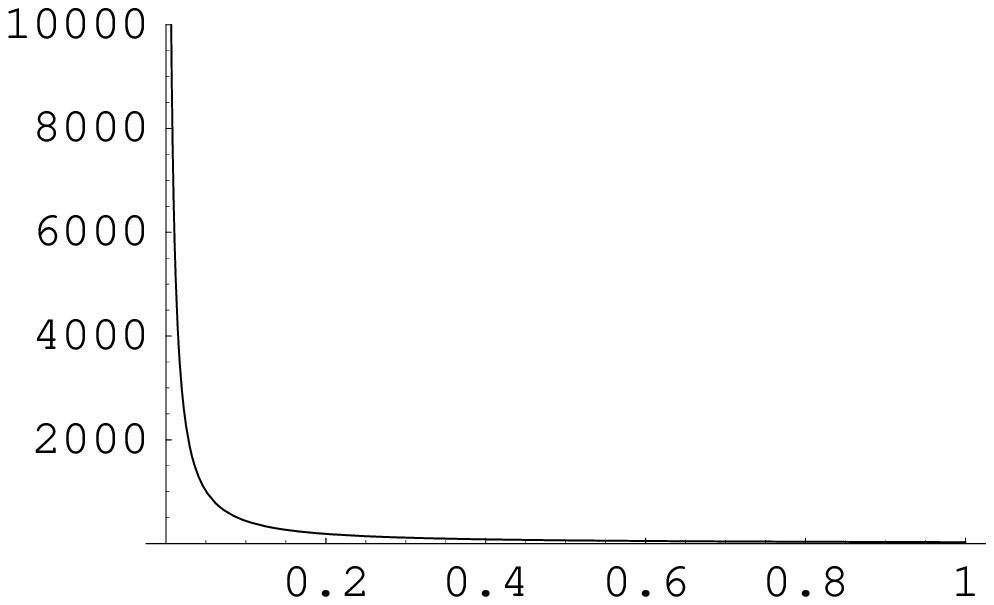} \\
 \includegraphics[scale=0.65]{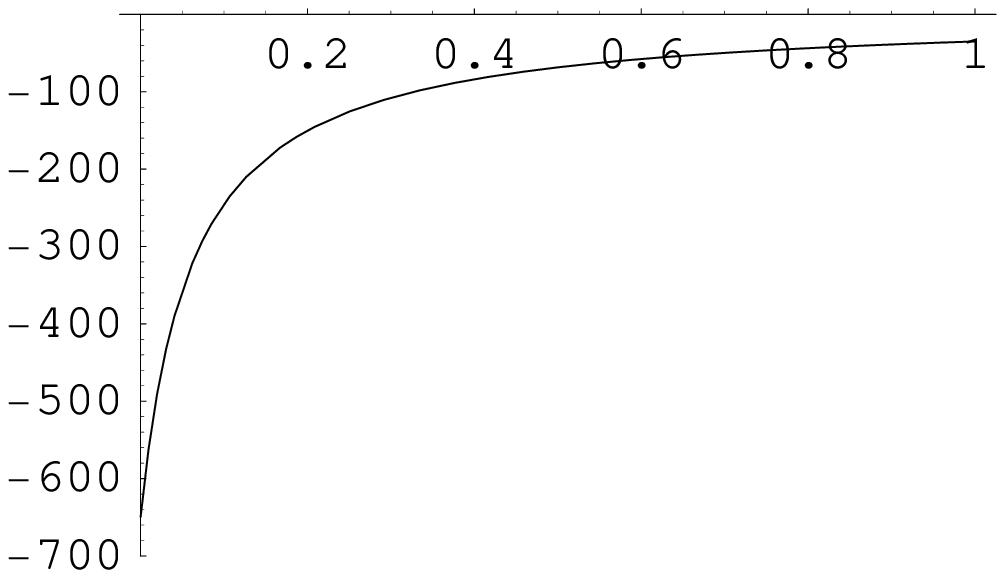} & 
 \includegraphics[scale=0.65]{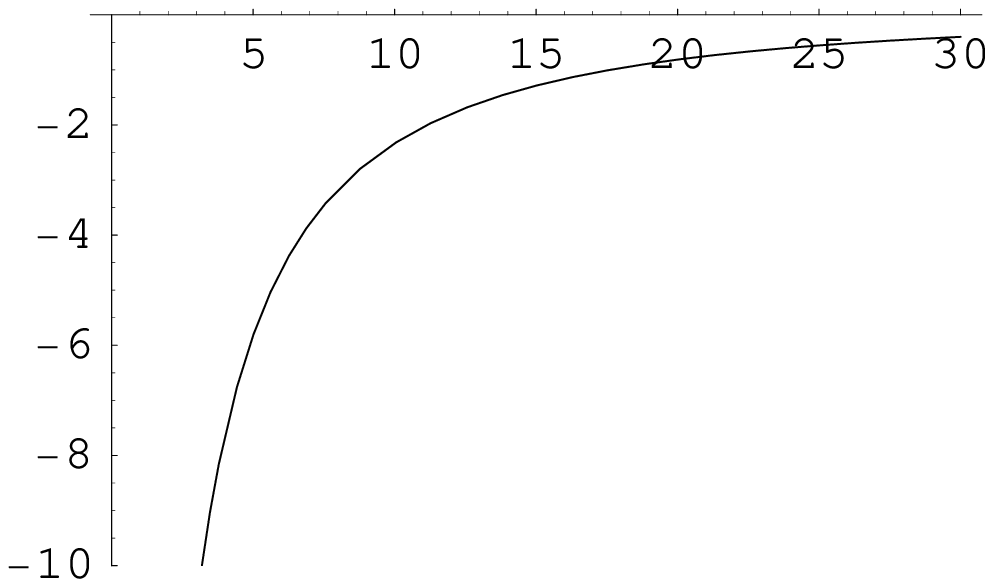} \\
(c) & (d) 
\end{tabular}
\caption{\label{fig:eta-2}
The first and second term of (\ref{eq:eta2-terms}) are plotted 
in (a) and (b), respectively, as functions of $\Delta^2/\Lambda^2$; 
parameters are set to the values described in the text, and we used 
$j=2$ in these figures. Although both (a) and (b) diverge
at $\Delta^2 \longrightarrow 0$, they add up to be (c), where 
the $\Delta \longrightarrow 0$ limit is finite. 
The large $\Delta^2$ behavior is seen better in the panel (d).}
\end{center}
\end{figure}

A numerical result for the $\eta^2$ term in $\overline{A}^{+,\alpha}_j$, 
which is proportional to 
\begin{equation}
 \bar{g}^{0,0,0}(j,i\nu_j, \Delta) \times \left[ \frac{p^2}{\Delta^2} 
 \frac{j(j-1)}{j-\frac{1}{2}} \right] \quad + \quad  
\frac{\bar{g}^{0,0,2}(j,i\nu_j,\Delta)}{N_{j,2} \Delta^2} \times
 \frac{-1}{(j+i\nu_j)(j-1+i\nu_j)}, 
\label{eq:eta2-terms}
\end{equation}
is shown in Figure~\ref{fig:eta-2}, using $j=2$.
The first and second terms of (\ref{eq:eta2-terms}) both diverge at 
the $\Delta^2 \longrightarrow 0$ limit, as we have seen in 
section \ref{ssec:D=0-cancellation}, but their sum has a finite value at 
$\Delta^2 = 0$, as one can see in the figure. It is worth 
mentioning that this finite limit value $\approx - 700$ is much 
larger than that of $\bar{g}^{0,0,0}$. This is likely to be due, 
at least partially, to the hadron mass $m_h$ value in this case; for 
the value of parameters we chose, $m_h = j_{\Delta_h-2,1}\Lambda$, 
$j_{3,1}\simeq 6.4$, and $m_h^2/\Lambda^2 \approx 40$. An extra derivative 
$\partial_z$ in the matrix elements $\bar{g}^{0,0,2}_k$ is more like 
$m_h$ than $\Lambda$, and hence the second term can be larger 
than the first term by about $(m_h/\Lambda)^2$.
The factor $(m_h/\Lambda)^2 \approx 40$ does not explain all of the 
moderately large value $-700$, however. 
The $t = -\Lambda^2$-dependence of the $\eta^0$ term (i.e., 
$\bar{g}^{0,0,0}(j, i\nu_j, \Delta)$) is quite different from 
that of the coefficient of the $\eta^2$, at least at small $\Delta^2$.

\begin{figure}[tbp]
\begin{center}
 \includegraphics[scale=0.65]{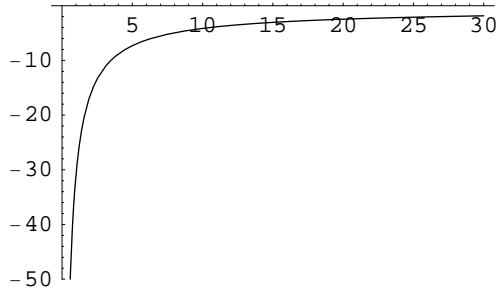} 
\caption{\label{fig:eta-2-vs-0}
The coefficient of the $\eta^2$ term of $\overline{A}^{+,\alpha}_j(\eta,t)$ 
to that of the $\eta^0$ term, as a function of 
$-t/\Lambda^2 = \Delta^2/\Lambda^2$. We used $j=2$ and other parameters 
described in the text. This is the ratio of Figure~\ref{fig:eta-2}~(d) 
to Figure~\ref{fig:g000}~(a).}
\end{center}
\end{figure}

In the DGLAP phase, a crude approximation to the GPD is given by 
\begin{equation}
\overline{H}^{+,\alpha}(x,\eta,t; q^2) \approx 
  \left(\frac{1}{x}\right)^{j_*} \left(\frac{\Lambda}{q}\right)^{\gamma(j_*)}
  \overline{A}^{+,\alpha}_{j*}(\eta,t),
\end{equation}
where $j^*$ is the saddle point value of $j$ depending primarily 
on $\ln(1/x)$, $\ln(q/\Lambda)$ and $t=-\Lambda^2$. 
Apart from applications to the time-like Compton scattering 
with very large (positive) lepton invariant mass-square, relevant 
range of $|\eta|$ is not much more than $x$ in such processes as TCS, DVCS
and VMP. Suppose, in the power series expansion of $\overline{A}^{+,\alpha}_{j*}$ 
in $\eta$, that all the terms with different power of $\eta$ have a 
($t$-dependent) coefficient at most of ${\cal O}(1)$. Then the GPD 
(or $\overline{A}^{+,\alpha}_{j^*}(\eta,t)$) in the small-$x$ regime would 
not have skewness dependence very much in the range of interest, 
$|\eta| \lesssim x$, because $\eta^2$ and higher-order terms are small 
relatively to the $\eta^0$ term.
The coefficient of the $\eta^2$ term, however, turns out to be 
of ${\cal O}(-700)$ for $\Delta^2 \approx 0$, which at least contains a 
factor $m_h^2/\Lambda^2$.
Thus, for the range of moderately small $x$, such as $x \sim 10^{-1}$ and 
$|\eta| \lesssim x$, the $\eta^2$ term in $\overline{A}^{+,\alpha}_{j^*}(\eta, t)$
can be just as important as the $\eta^0$ term for small $\Delta^2$. 
Consequently the prediction/fit of the slope parameter ($t$-dependence) 
may also be affected, since the $\eta^2$ term with a steeper $t$-dependence
is involved. 
Toward higher $\Delta^2$, however, the ratio of the coefficient of the 
$\eta^2$ term to that of the $\eta^0$ term changes as in 
a numerical computation shown in Figure~\ref{fig:eta-2-vs-0}.
Since the $\eta^2$-term coefficient becomes not more than 10 times 
the $\eta^0$ term for $5 \Lambda^2 \lesssim (\Delta^2 = -t)$ at $j=2$ in this 
numerical computation, the $\eta^0$-term alone will become a good enough 
approximation in this range of $t$ even for moderately small 
$|\eta| \lesssim x \approx {\cal O}(10^{-1})$; for an even smaller $x$, 
the $\eta^2$-term can be negligible for a broader range of $t = - \Lambda^2$. 
We have nothing more to say about the $\eta^4$ term and higher at this 
moment, or whether this moderately large value $\approx 700$ is an artifact 
of a specific implementation of confining effects we adopted for the numerical 
presentation in this section. If this relatively large coefficient of the 
$\eta^2$ term (and also higher order terms) turns out to be a robust 
consequence of holographic models, that may be regarded as an unexpected 
lesson from holography to phenomenology. 

\section*{Acknowledgment}

We thank Wen Yin for discussion, with whom we worked together at earlier 
stage of this project, and Simeon Hellerman and Teruhiko Kawano for useful 
comments. This work is supported in part by JSPS Research Fellowships for 
Young Scientists (RN), WPI Initiative and a Grant-in-Aid for 
Scientific Research on Innovative Areas 2303 from MEXT, Japan (RN, TW). 

\appendix 

\section{More on the Mode Decomposition on AdS$_5$}
\label{sec:appendix-mode-fcn}

For convenience, let us copy here the eigenmode equation
(\ref{eq:Eigenequation}) for a totally symmetric rank-$j$ tensor field 
on AdS$_5$; the equation consists of the following equations labeled by 
$k=0, \cdots, j$:
\begin{eqnarray}
&& \left( (R^2 \Delta_j) 
- \left[(2k+1)j-2k^2+3k\right]\right) A_{z^k \mu_1 \cdots \mu_{j-k} } 
   \nonumber  \\ 
&& + 2zk \partial^{\hat{\rho}} A_{z^{k-1}\rho\mu_1\cdots \mu_{j-k}}
+ k(k-1) A^{\hat{\rho}}_{z^{k-2}\rho \mu_1 \cdots \mu_{j-k} } 
   \nonumber \\
&& - 2z (D[A_{z^{k+1} \cdots}])_{\mu_1 \cdots \mu_{j-k}}
+ (E[A_{z^{k+2} \cdots}])_{\mu_1 \cdots \mu_{j-k}} = -{\cal E} A_{z^k \mu_1
\cdots \mu_{j-k}}.  
 \label{eq:Eigenequation-4+1-app}
\end{eqnarray}
%

\subsection{Eigenvalues and Eigenmodes for $\Delta^\mu = 0$}
\label{ssec:appendix-eigen-q=0}

{\bf block diagonal decomposition}

In the main text, we considered a decomposition of the rank-$j$ totally
symmetric tensor field with $(-i \partial_\mu) = \Delta_\mu = 0$ in the
form of 
\begin{equation}
 A_{z^k \mu_1 \cdots \mu_{j-k}}(z; \Delta^\mu = 0) =
 \sum_{N=0}^{[(j-k)/2]} \left(E^N[ a^{(k,N)}] \right)_{\mu_1 \cdots \mu_{j-k}}, 
 \nonumber 
\end{equation}
where $a^{(k,N)}$'s are $z$-dependent rank-$(j-k-2N)$ totally symmetric 
tensor field of $\SO(3,1)$, satisfying the 4D-traceless condition 
(\ref{eq:cond-4D-traceless}).  
This is indeed a decomposition, in that all the degrees of freedom in 
$A_{z^k \mu_1 \cdots \mu_j-k}(z; \Delta^\mu = 0)$ are described by 
$a^{(k,N)}(z)_{\mu_1 \cdots \mu_{j-k-2N}}$ with $0 \leq N \leq [(j-k)/2]$ 
without redundancy. To see this, 
one only needs to note that there is a relation\footnote{
This relation can be verified recursively in $N$.} that, for 
a totally symmetric 4D-traceless rank-$r$ $\SO(3,1)$-tensor $a$, 
\begin{equation}
 \eta^{\hat{\rho} \hat{\sigma}} 
 E^N[a]_{\rho \sigma \mu_1 \cdots \mu_{r+2N-2}} = 
4N(r+N+1) E^{N-1}[a]_{\mu_1 \cdots \mu_{r+2N-2}}. 
\label{eq:recursion-E-by-eta}
\end{equation}
Using this relation, $a^{(k,N)}_{\mu_1 \cdots \mu_{j-k-2N}}$ can be 
retrieved from $A_{z^k\mu_1 \cdots \mu_{j-k}}$, starting from ones with 
larger $N$ to ones with smaller $N$.

Let us now see that the eigenmode equation 
(\ref{eq:Eigenequation-4+1}=\ref{eq:Eigenequation-4+1-app}) can be made 
block diagonal by using this decomposition.
The eigenmode equation (\ref{eq:Eigenequation-4+1-app}) with the label $k$ 
for $\Delta^\mu = 0$ can be rewritten by using this 
relation (\ref{eq:recursion-E-by-eta}) as follows:
\begin{eqnarray}
\sum_N &&  \left(R^2 \Delta_j - \left[(2k+1)j-2k^2+3k\right] + {\cal E} \right) 
   E^N[a^{(k,N)}]
  \nonumber   \\ 
&+& k(k-1) \left[4 (N+1) (j-k-N+2) \right] E^N[a^{(k-2,N+1)}]
+ E[E^{N-1}[a^{(k+2, N-1)}]] = 0. \nonumber
\end{eqnarray}
Although this equation has to hold only after the summation in $N$, 
it actually has to be satisfied separately for different $N$'s.
To see this, let us first multiply $\eta^{\hat{\rho}\hat{\sigma}}$
for $[(j-k)/2]$ times and contract indices just like in 
(\ref{eq:recursion-E-by-eta}); we obtain an equation that involves 
only $a^{(k,[(j-k)/2])}$, $a^{(k-2,[(j-k)/2]+1)}$ and
$a^{(k+2,[(j-k)/2]-1)}$. Next, multiply $\eta^{\hat{\rho}\hat{\sigma}}$
for $[(j-k)/2] - 1$ times, to obtain another equation involving 
$a^{(k,[(j-k)/2]-1)}$, $a^{(k-2,[(j-k)/2])}$ and
$a^{(k+2,[(j-k)/2]-2)}$. In this way, we obtain 
\begin{eqnarray}
&& \left(R^2 \Delta_j - \left[(2k+1)j-2k^2+3k\right] + {\cal E} \right) 
   a^{(k,N)}    \label{eq:Eigen-eq-q=0-app} \\ 
&+& k(k-1) \left[4 (N+1) (j-k-N+2) \right] a^{(k-2,N+1)}
+ a^{(k+2, N-1)} = 0.   \qquad \qquad ({\rm for~} {}^{\forall} k,N)  \nonumber 
\end{eqnarray}
Fields $a^{(k,N)}$'s with the same $k+2N = n$ form a system of coupled 
equations, but those with different $n = k + 2N$ do not mix.   
Thus, the eigenmode equation for $\Delta^\mu=0$ is decomposed into 
sectors labeled by $n$. The $n$-th sector consists of 
$z$-dependent fields that are all in the rank-$(j-n) = (j-k-2N)$ 
totally symmetric tensor of $\SO(3,1)$.

{\bf classification of eigenmodes for $\Delta^\mu=0$}

Let us now study the eigenmode equations more in detail for the 
separate diagonal blocks we have seen. 
Simultaneous treatment is possible for all the $n$-th sectors with even 
$n$, and for all the sectors with odd $n$.

Let us first look at the $n$-th sector of the eigenmode problem for 
an $n=2\bar{n} \leq j$.
In the eigenmode equation of $\Delta^\mu=0$, we can assume\footnote{
This is because, in the absence of $z^2 \partial^2$ term, the operator 
$\Delta_j$ becomes a constant multiplication when it acts on a simple power of
$z$. Upon $z^{2-j-i\nu}$, for example, $R^2 \Delta_j$ returns $-(4+\nu^2)$.} 
the same $z$-dependence for all the fields in this diagonal block:
\begin{equation}
 a^{(k,N)}(z)_{\mu_1\cdots \mu_{j-n}} = 
   \bar{a}^{(k,N)}_{\mu_1 \cdots \mu_{j-n}} z^{2-j-i\nu}, \qquad
   \qquad  k+2N = n, 
\end{equation}
where $\bar{a}^{(k,N)}$'s are $(x,z)$-independent 4D-traceless 
rank-$(j-n)$ tensor of $\SO(3,1)$. The eigenmode equations with the 
label $(k,N) = (2\bar{k},\bar{n}-\bar{k})$ with 
$\bar{k}=0, \cdots, \bar{n}$ are relevant to the $n = 2\bar{n}$ sector, 
and are now written in a matrix form:
\begin{equation}
 \sum_{\bar{k}' = 0}^{\bar{n}} {\cal D}_{2\bar{k},2\bar{k}'}  \; 
   \bar{a}^{(2\bar{k}',\bar{n}-\bar{k}')} = 
   ((4+\nu^2)-{\cal E}) \; \bar{a}^{(2\bar{k}, \bar{n}-\bar{k})}, 
\label{eq:EGeq-q=0-matrix-n-even}
\end{equation}
where 
\begin{itemize}
 \item diagonal $(k,k')=(2\bar{k}, 2\bar{k})$ entry: 
   ${\cal D}_{2\bar{k},2\bar{k}} = -[(2k+1)j-2k^2+3k]$,
 \item diagonal$^+$; $(k,k') = (2\bar{k}, 2\bar{k}+2)$ entry: 
   ${\cal D}_{2\bar{k}, 2\bar{k}+2} = 1$, 
 \item diagonal$^-$; $(k,k') = (2\bar{k}, 2\bar{k}-2)$ entry: 
   ${\cal D}_{2\bar{k},2\bar{k}-2} = k(k-1) \times 
  4(\bar{n}-\bar{k}+1)(j-\bar{n}-\bar{k}+2)$. 
\end{itemize}
There must be $(\bar{n}+1)$ independent eigenmodes 
in this $(\bar{n}+1) \times (\bar{n}+1)$ matrix equation. 
Let ${\cal E}_{n,l}$ denote the collection of eigenvalues in this 
$n=2\bar{n}$-th diagonal block, and 
$l=0,\cdots, \bar{n} = n/2 $ label distinct eigenmodes.  
Corresponding eigenmode wavefunction for the $(n=2\bar{n}, l)$ 
mode is in the form of 
\begin{equation}
 a^{(k,N)}(z; \Delta^\mu=0) = a^{(2\bar{k},\bar{n}-\bar{k})} 
= c_{2\bar{k},l,n}  \epsilon^{(n,l)} z^{2-j-i\nu},  
\end{equation}
where $\epsilon^{(n,l)}$ is an $(x,z)$-independent 4D-traceless 
totally symmetric rank-$(j-n)=(j-2\bar{n})$ tensor of $\SO(3,1)$, 
and $c_{2\bar{k},l,n}$ are $(x,z)$-independent constants determined 
as the eigenvector corresponding to the eigenvalue ${\cal E}_{n,l}$.

Similarly, in the $n=2\bar{n}+1 \leq j$-th sector of the 
eigenmode problem, with an odd $n$, we can assume a simple power law 
for all the component fields involved in this sector;  
\begin{equation}
 a^{(k,N)}(z)_{\mu_1 \cdots \mu_{j-n}} =
  \bar{a}^{(k,N)}_{\mu_1 \cdots \mu_{j-n}} z^{2-j-i\nu}, \qquad \qquad 
   k+2N = n,
\end{equation}
where $\bar{a}^{(k,N)}$ are $(x,z)$-independent 4D-traceless totally
symmetric tensor of $\SO(3,1)$. The eigenmode equation with the label 
$(k,N) = (2\bar{k}+1,\bar{n}-\bar{k})$ with $\bar{k}=0,\cdots,\bar{n}$ 
are relevant to this sector, and in the matrix form, the eigenmode
equation now looks 
\begin{equation}
 \sum_{\bar{k}' = 0}^{\bar{n}} {\cal D}_{2\bar{k}+1, 2\bar{k}'+1} 
  \bar{a}^{(2\bar{k}'+1,\bar{n}-\bar{k})} = 
  ((4+\nu^2)-{\cal E}) \; \bar{a}^{(2\bar{k}+1, \bar{n}-\bar{k})}, 
\label{eq:EGeq-q=0-matrix-n-odd}
\end{equation}
where 
\begin{itemize}
 \item diagonal $(k,k') = (2\bar{k}+1, 2\bar{k}+1)$ entry:
   ${\cal D}_{2\bar{k}+1,2\bar{k}+1} = -[(2k+1)j+(-2k^2+3k)]$,  
 \item diagonal$^+$ $(k,k') = (2\bar{k}+1, 2\bar{k}+3)$ entry: 
   ${\cal D}_{2\bar{k}+1, 2\bar{k}+3} = 1$, 
 \item diagonal$^-$ $(k,k') = (2\bar{k}+1, 2\bar{k}-1)$ entry:
   ${\cal D}_{2\bar{k}+1, 2\bar{k}-1} = k(k-1) \times
    4(\bar{n}-\bar{k}+1)(j - \bar{n} - \bar{k} + 1)$.
\end{itemize}
From here, $\bar{n}+1$ independent modes arise; 
their eigenvalues are denoted by ${\cal E}_{n,l}$, and 
$l = \{ 0, \cdots, \bar{n}\}$ is the label distinguishing different modes.
The eigenmode labeled by $(n=2\bar{n}+1, l)$ has a wavefunction 
\begin{equation}
 a^{(k,N)}(z; \Delta^\mu=0) = 
 a^{(2\bar{k}+1, \bar{n}-\bar{k})} = 
 c_{2\bar{k}+1,l,n} \epsilon^{(n,l)} z^{2-j-i\nu},
\end{equation}
where $\epsilon^{(n,l)}$ is an $(x,z)$-independent 4D-traceless 
rank-$(j-n)$ totally symmetric tensor of $\SO(3,1)$, and 
$c_{2\bar{k}+1,l,n}$ is the eigenvector for the $(n,l)$ eigenmode
determined in the matrix equation above. 

{\bf explicit examples}

Let us take a moment to see how the general theory above works out 
in practice. 

{\bf The easiest of all is the $n = 0$ sector}, which contains 
only one rank-$j$ 4D-traceless field, $a^{(0,0)}$. The eigenmode equation is 
\begin{equation}
 \left[\Delta_j - \frac{\left[(2k+1)j-2k^2+3k\right]_{k=0}}{R^2} \right]
  a^{(0,0)}  =  \left[\Delta_j - \frac{j}{R^2} \right]  a^{(0,0)}  = 
 - \frac{ {\cal E}_{0,0}}{R^2} a^{(0,0)}.
\label{eq:EGeq-q=0-n=0-app}
\end{equation}
The eigenmode wavefunction has the form of 
\begin{equation}
 a^{(0,0)}(z)_{\mu_1 \cdots \mu_j} =
  \epsilon^{(0,0)}_{\mu_1 \cdots \mu_j} \; z^{2-j-i\nu}, 
\end{equation}
and the eigenvalue ${\cal E}_{n,l}$ is 
\begin{equation}
 {\cal E}_{0,0} = (j + 4 + \nu^2).
\label{eq:bgn-of-empirical-egval}
\end{equation}

{\bf Also to the $n=1$ sector}, only one rank-$(j-1)$
4D-traceless tensor field contributes. That is $a^{(1,0)}$.
The eigenmode equation becomes 
\begin{equation}
 \left[R^2 \Delta_j - [(2k+1)j-2k^2+3k]|_{k=1} \right] \; a^{(1,0)} =
 \left[R^2 \Delta_j - (3j+1) \right] \; a^{(1,0)}
 = - {\cal E}_{1,0} \; a^{(1,0)}.
\end{equation}
The solution is 
\begin{equation}
a^{(1,0)}(z)_{\mu_1 \cdots \mu_{j-1}} = 
 \epsilon^{(1,0)}_{\mu_1 \cdots \mu_{j-1}} z^{2-j-i\nu}, \qquad 
{\cal E}_{1,0} = (3j + 5 + \nu^2). 
\end{equation}

{\bf In the $n=2$ sector}, two rank-$(j-2)$ 4D-traceless fields are
involved. They are $a^{(0,1)}$ and $a^{(2,0)}$. After introducing the 
$z$-dependence $\propto z^{2-j-i\nu}$, the eigenmode equation 
(\ref{eq:EGeq-q=0-matrix-n-even}) in the $n=2$ sector becomes 
\begin{equation}
 \left[ \begin{array}{cc}
  -j & 1 \\ 8j & - (5j-2)
	\end{array}\right]
\left(\begin{array}{c}
 \bar{a}^{(0,1)} \\ \bar{a}^{(2,0)}
      \end{array}\right)
 = ((4+\nu^2)-{\cal E}) 
\left(\begin{array}{c}
 \bar{a}^{(0,1)} \\ \bar{a}^{(2,0)}
      \end{array}\right).
\end{equation}

One of the two eigenmodes is 
\begin{equation}
 {\cal E}_{2,0} =(4+5j+\nu^2), \qquad 
\left( \begin{array}{c}
 a^{(0,1)}(z)_{\mu_1 \cdots \mu_{j-2}} \\
 a^{(2,0)}(z)_{\mu_1 \cdots \mu_{j-2}}
       \end{array}\right) =
 \left( \begin{array}{c}
  1 \\ -4j
	\end{array}\right) \; 
 \epsilon^{(2,0)}_{\mu_1 \cdots \mu_{j-2}} \; z^{2-j-i\nu},
\end{equation}
and the other 
\begin{equation}
 {\cal E}_{2,1} = (2+j + \nu^2), \qquad 
\left( \begin{array}{c}
 a^{(0,1)}(z)_{\mu_1 \cdots \mu_{j-2}} \\
 a^{(2,0)}(z)_{\mu_1 \cdots \mu_{j-2}}
       \end{array}\right) =
 \left( \begin{array}{c}
  1 \\ 2
	\end{array}\right) \; 
 \epsilon^{(2,1)}_{\mu_1 \cdots \mu_{j-2}} \; z^{2-j-i\nu}.
\label{eq:graviton-wavefnc}
\end{equation}

{\bf In the $n=3$ sector}, two rank-$(j-3)$ 4D-traceless tensor fields
are involved: $a^{(1,1)}$ and $a^{(3,0)}$.
The eigenmode equations (\ref{eq:EGeq-q=0-matrix-n-odd}) become
\begin{equation}
 \left[ \begin{array}{cc}
  -(3j+1) & 1 \\ 24 (j-1) & - (7j-9)
	\end{array}\right]
\left(\begin{array}{c}
 \bar{a}^{(1,1)} \\ \bar{a}^{(3,0)}
      \end{array}\right)
 = ((4+\nu^2)-{\cal E}) 
\left(\begin{array}{c}
 \bar{a}^{(1,1)} \\ \bar{a}^{(3,0)}
      \end{array}\right). 
\end{equation}
So, one of the two eigenmodes is
\begin{equation}
{\cal E}_{3,0} = (7j + 1 + \nu^2), \qquad 
 \left( \begin{array}{c}
  a^{(1,1)}(z)_{\mu_1 \cdots \mu_{j-3}} \\
  a^{(3,0)}(z)_{\mu_1 \cdots \mu_{j-3}} 
	\end{array}\right) = 
 \left( \begin{array}{c}
  1 \\ -4(j-1)
	\end{array}\right)
 \epsilon^{(3,0)}_{\mu_1 \cdots \mu_{j-3}} \; z^{2-j-i\nu}, 
\end{equation}
and the other one  
\begin{equation}
 {\cal E}_{3,1} = (3j -1 + \nu^2), \qquad 
 \left( \begin{array}{c}
  a^{(1,1)}(z)_{\mu_1 \cdots \mu_{j-3}} \\
  a^{(3,0)}(z)_{\mu_1 \cdots \mu_{j-3}} 
	\end{array}\right) = 
 \left( \begin{array}{c}
  1 \\ 6
	\end{array}\right)
 \epsilon^{(3,1)}_{\mu_1 \cdots \mu_{j-3}} \; z^{2-j-i\nu}.
\end{equation}

{\bf Finally, in the $n=4$ sector}, 
the eigenmode equation (\ref{eq:EGeq-q=0-matrix-n-even}) is given by 
\begin{equation}
 \left[ \begin{array}{ccc}
  -j & 1 & 0 \\ 16(j-1) & -(5j-2) & 1 \\ 0 & 48(j-2) & -(9j-20)
	\end{array}\right]
 \left( \begin{array}{c}
  \bar{a}^{(0,2)} \\ \bar{a}^{(2,1)} \\ \bar{a}^{(4,0)}
	\end{array}\right) = 
 ((4+\nu^2)-{\cal E}) 
 \left( \begin{array}{c}
  \bar{a}^{(0,2)} \\ \bar{a}^{(2,1)} \\ \bar{a}^{(4,0)}
	\end{array}\right).
\end{equation}
There are three solutions. First, 
\begin{eqnarray}
 {\cal E}_{4,0} & = & (9j-4+\nu^2),  \\
 (a^{(0,2)}, a^{(2,1)}, a^{(4,0)}) & = & 
 (1, -8(j-1),32(j-1)(j-2)) \; \epsilon^{(4,0)} \; z^{2-j-i\nu}, 
\end{eqnarray}
second, 
\begin{eqnarray}
 {\cal E}_{4,1} & = & (5j-6+\nu^2),  \\
(a^{(0,2)}, a^{(2,1)}, a^{(4,0)}) & = & 
 (1, -(4j-10), -48(j-2)) \; \epsilon^{(4,1)} \; z^{2-j-i\nu}, 
\end{eqnarray}
and finally, 
\begin{eqnarray}
 {\cal E}_{4,2} & = & (j + \nu^2),  \\
(a^{(0,2)}, a^{(2,1)}, a^{(4,0)}) & = & 
 (1, 4, 24) \; \epsilon^{(4,2)} \; z^{2-j-i\nu}.  
\label{eq:end-of-empirical-egval}
\end{eqnarray}

An empirical relation is observed in the $j$-dependence of the 
eigenvalues we have worked out so far. The eigenvalues 
in the $n$-the sector are in the form of 
${\cal E}_{n,l} = \nu^2 + (2n+1-4l)j + {\cal O}(1)$ for 
$0 \leq l \leq [n/2]$.

{\bf 5D-traceless modes: the $l=0$ modes}

Although the precise expressions for the eigenvalues ${\cal E}_{n,l}$ 
and eigenvectors $c_{k,l,n}$ are not given for all the eigenmodes, 
there is a class of eigenmodes whose eigenvalues and eigenvectors 
(wavefunctions) are fully understood. 

As we discussed in p.~\pageref{page:5D-trl-trv}, it is possible to 
require both a field is an eigenmode and satisfies the 5D-traceless
condition (\ref{eq:cond-5D-traceless}) at the same time.
In the $n=(k+2N)$-th sector, the 5D-traceless condition becomes 
\begin{eqnarray}
0 & = & 
 \left( E^N[a^{(k,N)}] \right)^{\hat{\rho}}_{\; \rho \mu_3 \cdots \mu_{j-n}}
+ \left( E^{N-1}[a^{(k+2,N-1)}] \right)_{\mu_3 \cdots \mu_{j-n}}, 
 \label{eq:5D-traceless-q=0-explicit} \\
 & = & E^{N-1} \left[ 4N(j-n+N+1) a^{(k,N)} + a^{(k+2,N-1)} \right] 
 \quad 
 \left\{ 
\begin{array}{l}
 k=0,2,\cdots, 2(\bar{n}-1)   \quad ({\rm even~}n), \\
 k=1,3,\cdots, 2\bar{n}-1  \quad ({\rm odd~}n).
\end{array}
 \right. \nonumber
\end{eqnarray}
Thus, the 5D-traceless condition uniquely determines one eigenmode 
in each one of the $n$-th sector. 
\begin{equation}
  {\cal E}_{n,0} = (2n+1)j + 2n - n^2 + 4+\nu^2, 
\end{equation}
and 
\begin{equation}
 c_{2\bar{k},0,2\bar{n}} = (-)^{\bar{k}} 4^{\bar{k}}
   \frac{ \bar{n}! }{ (\bar{n}-\bar{k})! }
   \frac{ (j-\bar{n}+1)! }{ (j-\bar{n}-\bar{k}+1)! }, \quad 
 c_{2\bar{k}+1,0,2\bar{n}+1} = (-)^{\bar{k}} 4^{\bar{k}}
   \frac{ \bar{n}! }{ (\bar{n}-\bar{k})! }
   \frac{ (j-\bar{n})! }{ (j-\bar{n}-\bar{k})! }. 
\end{equation}
%

\subsection{Mode Decomposition for non-zero $\Delta_\mu$}
\label{ssec:appendix-eigen-q=not0}

\subsubsection{Diagonal Block Decomposition for the $\Delta^\mu \neq 0$ Case}

Let us now turn our attention to the eigenmode 
equation (\ref{eq:Eigenequation}, \ref{eq:Eigenequation-4+1}) 
with $\Delta^\mu \neq 0$. Because of the 2nd and 4th terms in 
(\ref{eq:Eigenequation-4+1}), the eigenmode problem becomes much more 
complicated. We begin by finding diagonal block decomposition 
suitable for the case with $\Delta^\mu \neq 0$.

In the main text, we introduced a decomposition of a totally symmetric 
rank-$j$ tensor field $A_{m_1 \cdots m_j}$ of $\SO(4,1)$ into a collection 
of totally symmetric 4D-traceless 4D-transverse tensor fields of
$\SO(3,1)$. Instead of (\ref{eq:block-dcmp-q=0}), a new
decomposition is given by (\ref{eq:block-dcmp-q=not0-tilde}=\ref{eq:block-dcmp-q=not0-tilde-app}):  
\begin{equation}
 A_{z^k \mu_1 \cdots \mu_{j-k}}(z; \Delta^\mu)  = 
 \sum_{s=0}^{j-k} \sum_{N=0}^{[s/2]} 
   \left( \tilde{E}^N D^{s-2N} [ a^{(k,s,N)} ] \right)_{\mu_1 \cdots \mu_{j-k}},
 \label{eq:block-dcmp-q=not0-tilde-app}  
\end{equation}
where $a^{(k,s,N)}$ are totally symmetric 4D-traceless 4D-transverse 
rank-$(j-k-s)$ tensor fields of $\SO(3,1)$. 
An operation $a \mapsto \tilde{E}[a]$ on a totally symmetric 
$\SO(3,1)$ tensor $a$ is given by (\ref{eq:def-E-tilde}).

In order to see that the parameterization of 
$A_{z^k \mu_1 \cdots \mu_{j-k}}$ by 
$(a^{(k,s,N)})_{\mu_1 \cdots \mu_{j-k-s}}$'s above is indeed a
decomposition, one needs to see that $a^{(k,s,N)}$'s can be retrieved 
from $A_{z^k \mu_1 \cdots \mu_{j-k}}$, so that the degrees of freedom 
$a^{(k,s,N)}$ are independent.
For this purpose, it is convenient to derive some relations analogous to 
(\ref{eq:recursion-E-by-eta}). First of all, note that 
$E [D[a]] = D[E[a]]$ and\footnote{%
\begin{equation}
 E^t D^{s-2t}[ a ] = \sum
  \eta_{\mu_{p_1} \mu_{p_2}} \cdots \eta_{\mu_{p_{2t-1}} \mu_{p_{2t}} }
  \partial_{\mu_{p_{2t+1}} } \cdots \partial_{\mu_{p_s}} 
  \left[ a \right]_{\mu_1 \cdots \check{} \check{} \cdots
  \mu_{r+s}},
\label{eq:def-DandE}
\end{equation}
where the sum is taken over all possible ordered choices of 
$p_1, p_2, \cdots, p_s \in \{ 1, \cdots, j \}$ such that 
$p_i \neq p_j$ for $i \neq j$. 
} $\tilde{E}[ D[a]] = D[\tilde{E}[a]]$ for a totally symmetric 
$\SO(3,1)$ tensor $a$. 
If the rank-$r$ tensor $a$ is also 4D-transverse and 4D-traceless, then 
one can derive the following relations:
\begin{eqnarray}
 \partial^{\hat{\rho}} 
  \left( E^t D^{s-2t} \left[ a \right] \right)_{\rho \mu_2 \cdots
  \mu_{r+s}} & = & 
  -\Delta^2 (s-2t) E^t D^{s-2t-1}\left[a \right]
  + (2t) E^{t-1} D^{s-2t+1} \left[ a \right], 
    \label{eq:recursion-ED-by-partial}\\
 \eta^{\hat{\rho} \hat{\sigma}}
  \left(E^t D^{s-2t} \left[ a \right] \right)_{\rho \sigma \mu_3
  \cdots \mu_{r+s}} & = &
   - \Delta^2 (s-2t)(s-2t-1) E^t D^{s-2t-2} \left[ a \right] 
   \nonumber \\
  & & + 4t(r+s-t+1) E^{t-1} D^{s-2t} \left[ a \right]. 
    \label{eq:recursion-ED-by-eta}
\end{eqnarray}
\begin{eqnarray}
\partial^{\hat{\rho}}
 \left( \tilde{E}^N D^{s-2N} [ a ]\right)_{\rho \mu_2 \cdots \mu_{r+s}} & = &  
-(s-2N) \Delta^2 \tilde{E}^N D^{s-2N-1} [ a ],  \\
 \left( \eta^{\hat{\rho}\hat{\sigma}} -
        \frac{ \partial^{\hat{\rho}} \partial^{\hat{\sigma}} }
             {\partial^2} 
 \right)
 \left(\tilde{E}^N D^{s-2N}[ a ]
 \right)_{\rho \sigma \mu_3 \cdots \mu_{r+s}} & = & 
 4N(r+N+1/2) \tilde{E}^{N-1} D^{s-2N}[ a ]. 
\end{eqnarray}
With the relations above, it is now possible to compute 
\begin{eqnarray}
&&
 \left(\eta^{\hat{\mu}_1 \hat{\mu}_2}
      - \frac{\partial^{\hat{\mu}_1} \partial^{\hat{\mu}_2}}
             {\partial^2} \right)
   \cdots 
 \left(\eta^{\hat{\mu}_{2p-1} \hat{\mu}_{2p}}
    - \frac{\partial^{\hat{\mu}_{2p-1}} \partial^{\hat{\mu}_{2p}}}
           {\partial^2} \right)
 \; 
 \frac{\partial^{\hat{\mu}_{2p+1}}}{\partial^2} \cdots
 \frac{\partial^{\hat{\mu}_{2p+q}}}{\partial^2}  
 \; 
 \left( \tilde{E}^N D^{s-2N}[a] \right)_{\mu_1 \cdots \mu_{r+s}}
   \nonumber  \\
& = & \left\{ \begin{array}{ll}
   \frac{b^{(r)}_{s-2p-q, N-p}}{b^{(r)}_{s,N}}
   \left(\tilde{E}^{N-p} D^{s-2N-q}[a] \right)_{\mu_{2p+q+1} \cdots \mu_{r+s}}
      & {\rm if~} p \leq N {\rm ~and~} q \leq s-2N,\\
0 & {\rm otherwise}, 
      \end{array}\right. 
\label{eq:extract-SO(3,1)-tensor}
\end{eqnarray}
where we assume that $a$ is a totally symmetric 4D-traceless
4D-transverse rank-$r$ tensor of $\SO(3,1)$.
In the last line, 
\begin{equation}
 b^{(r)}_{s,N} := \frac{1}{4^N N! (s-2N)!}
  \frac{\Gamma\left(r+3/2\right)}{\Gamma \left( r+N+3/2 \right)}.
\label{eq:def-of-b}
\end{equation}

It is now clear how to retrieve $a^{(k,s,N)}$ 
from $A_{z^k \mu_1 \cdots \mu_{j-k}}$ given by
(\ref{eq:block-dcmp-q=not0-tilde}=\ref{eq:block-dcmp-q=not0-tilde-app}). 
First, one has to multiply 
$\eta^{\hat{\rho}\hat{\sigma}} - \partial^{\hat{\rho}} 
\partial^{\hat{\sigma}}/\partial^2$ and
$\partial^{\hat{\sigma}}/\partial^2$ to $A_{z^k \mu_1 \cdots \mu_{j-k}}$
as many times as possible in order to obtain $a^{(k,s,N)}$ with larger 
$N$ and $(s-2N)$. Then $a^{(k,s,N)}$'s with smaller $N$ or $(s-2N)$ can 
be determined by multiplying 
$\eta^{\hat{\rho}\hat{\sigma}} - \partial^{\hat{\rho}} 
\partial^{\hat{\sigma}}/\partial^2$ and
$\partial^{\hat{\sigma}}/\partial^2$ fewer times. 

Let us now return to the eigenmode equation for the cases with 
$\Delta^\mu \neq 0$. Following precisely the same argument as in 
section \ref{ssec:appendix-eigen-q=0}, one can see that the eigenmode 
equation can be separated into the following independent equations 
labeled by $k, s$ and $N$: 
\begin{eqnarray}
& &
 \left[R^2 \Delta_j - \left[(2k+1)j-2k^2+3k \right]+ {\cal E}\right] \; 
      a^{(k,s,N)}     \nonumber \\
& + &  2z k(s+1-2N) (\partial^2) \; a^{(k-1,s+1,N)}  \nonumber \\
& + & k(k-1)(s+2-2N)(s+1-2N) (\partial^2) \; a^{(k-2,s+2,N)}  
   \label{eq:EGeq-q=not0-matrix-m}  \\
&& + 4k(k-1)(N+1) (j-m+N+3/2) \; a^{(k-2,s+2,N+1)}  \nonumber \\
& - & 2z \; a^{(k+1,s-1,N)}
+ a^{(k+2,s-2,N-1)}
+ (\partial^2)^{-1} \; a^{(k+2,s-2,N)} = 0 
\qquad {\rm for~}{}^\forall k,s,N.   \nonumber 
\end{eqnarray}
The relations (\ref{eq:recursion-ED-by-partial},
\ref{eq:recursion-ED-by-eta}) were used to evaluate the 2nd--4th terms 
of (\ref{eq:Eigenequation-4+1-app}).
One can see that $a^{(k,s,N)}$'s with a common value of $m := k+s$
form a coupled eigenmode equations, but those with different $m$'s 
do not. Thus, $a^{(k,s,N)}(z; \Delta^\mu)$'s with $k+s = m$ form 
the $m$-th subspace of $A_{m_1 \cdots m_j}(z; \Delta^\mu)$, and the 
eigenmode equation becomes block diagonal in the decomposition 
into the subspaces labeled by $m =0, \cdots, j$.

The eigenmode equation on the $m$-th subspace is given by 
the equation above with $0 \leq k =(m-s) \leq m$, and 
$0 \leq N \leq [s/2]$. Thus, the total number of equations is 
\begin{equation}
 \sum_{s=0}^m \left( [s/2] + 1 \right), 
\label{eq:nmb-eqn-m-sector-app}
\end{equation} 
and the same number of eigenvalues should be obtained from the $m$-th sector.

\subsubsection{Examples}

{\bf The sector} $m=0$: There is only one field $a^{(0,0,0)}$ in this
sector, and the eigenmode equation is 
\begin{equation}
 \left[\Delta_j - \frac{j}{R^2} \right] a^{(0,0,0)}(z; \Delta^\mu) 
 = - \frac{{\cal E}}{R^2} a^{(0,0,0)}(z; \Delta^\mu).
\end{equation}
Assuming a power series expansion for the solution to this equation, 
beginning with some power $z^{2-j-i\nu}$, the eigenvalue is 
determined as a function of $(i\nu)$:
\begin{equation}
 {\cal E}_{0,0} = (j+4+\nu^2), \nonumber 
\end{equation}
and the wavefunction can be chosen as  
\begin{eqnarray}
 a^{(0,0,0)}(z; \Delta^\mu)_{\mu_1 \cdots \mu_j} & = &  
 \epsilon^{(0,0,0)}_{\mu_1 \cdots \mu_j} \;
 \Psi^{(j)}_{i\nu}(- \Delta^2, z), \\
\Psi^{(j)}_{i\nu}(\Delta^2,z) & := & 
  \frac{2}{\pi}\sqrt{\frac{\nu \sinh (\pi \nu)}{2R}} \; 
  e^{(j-2)A} K_{i\nu}(\Delta z).
\end{eqnarray}

{\bf The sector} $m=1$: The eigenmode equation in this sector becomes 
\begin{equation}
 \left[ \begin{array}{cc}
  R^2 \Delta_j - j & -2z \\ -2z \Delta^2 & R^2 \Delta_j - (3j+1) 
	\end{array}\right]
\left(  \begin{array}{c}
 a^{(0,1,0)} \\ a^{(1,0,0)}
	\end{array}\right) = 
 - {\cal E}
\left(  \begin{array}{c}
 a^{(0,1,0)} \\ a^{(1,0,0)}
	\end{array}\right). 
\label{eq:EGeq-q=not0-m=1}
\end{equation}
Assuming the power series expansion in $z$, beginning with 
$z^{2-j-i\nu}$ terms, we obtain two eigenvalues depending on $i\nu$. 
They are given by evaluating $R^2 \Delta_j - j$ and 
$R^2 \Delta_j - (3j+1)$ on $z^{2-j-i\nu}$: 
\begin{equation}
 {\cal E}_{0,0} = (j+4+\nu^2), \quad {\rm and} \quad 
 {\cal E}_{1,0} = (3j+5+\nu^2).
\end{equation}

{\bf The sector} $m=2$: The eigenmode equation becomes 
\begin{equation}
 \left(
(R^2 \Delta_j + {\cal E}) {\bf 1}_{4\times 4} + 
 \left[ \begin{array}{cccc}
  -j & & -2z & 1/\partial^2 \\
  & -j & & 1 \\
  4z \partial^2 &  & -(3j+1) & -2z \\
  4\partial^2 & 8j-4 & 4z \partial^2 & -(5j-2)
	\end{array}\right]
\right)
\left( \begin{array}{c}
 a^{(0,2,0)} \\
 a^{(0,2,1)}  \\
 a^{(1,1,0)}  \\
 a^{(2,0,0)}
       \end{array}\right) = 0.
\end{equation}
The indicial equation relating the exponent $(2-j-i\nu)$ at $z = 0$ and 
the eigenvalues split into two parts; three eigenvalues of this matrix 
\begin{equation}
 \left(\begin{array}{ccc}
  -j & & 1 \\ & -j & 1 \\ 4 & (8j-4) & -(5j-2)
       \end{array}\right),
\end{equation}
determine $-{\cal E}-(4+\nu^2)$ for the three eigenmodes, 
and $-({\cal E}-(4+\nu^2)) = - (3j+1)$ for the last eigenmode.
Therefore, the four eigenvalues in the $m=2$ sector are 
\begin{equation}
 {\cal E}_{0,0}=(j+4+\nu^2), \quad 
 {\cal E}_{1,0}=(3j+5+\nu^2), \quad 
 {\cal E}_{2,0}=(5j+4+\nu^2), \quad 
 {\cal E}_{2,1}=(j+2+\nu^2).
\end{equation}

In all the examples above, the $m$-th sector consists of 
eigenmodes with eigenvalues ${\cal E}_{n,l}$ for 
$0 \leq n \leq m$, $0 \leq l \leq [n/2]$. The number of eigenmodes 
is, of course, the same as (\ref{eq:nmb-eqn-m-sector-app}).

\subsection{Wavefunctions of 5D-Traceless 5D-Transverse Modes}
\label{ssec:appendix-5D-trltrv}

As we discussed toward the end of section \ref{ssec:eigen-q=not0}, 
it is possible to require for a rank-$j$ totally symmetric tensor 
field configuration $A_{m_1 \cdots m_j}(z; \Delta^\mu)$ to be 
an eigenmode and to be 5D-traceless 5D-transverse
(\ref{eq:cond-5D-traceless}, \ref{eq:cond-5D-transverse}) at the same
time. We will see in the following that these two extra conditions 
(\ref{eq:cond-5D-traceless}, \ref{eq:cond-5D-transverse}) leave
precisely one eigenmode in each one of the block-diagonal sectors labeled
by $m = 0, \cdots, j$. We will further determine the wavefunction
profile of such eigenmodes. 

Let us first rewrite the 5D-traceless condition 
(\ref{eq:cond-5D-traceless}) in a more convenient form. 
\begin{equation}
  \eta^{\hat{\rho} \hat{\sigma}} 
  A_{z^{k-2}\rho\sigma \mu_1 \cdots \mu_{j-k}}
   + A_{z^k \mu_1 \cdots \mu_{j-k}} = 0,   
\label{eq:5D-trac-cond}
\end{equation}
which, in the $m$-th sector, means 
\begin{equation}
 a^{(k,s,N)} = (s+2-2N)(s+1-2N) \Delta^2 a^{(k-2,s+2,N)}
    + 4 (N+1)(j-m+N+3/2) a^{(k-2,s+2,N+1)}  
\label{eq:5D-trac-cond-EZ}
\end{equation}
for $N=0, \cdots , [s/2]$; $k+s = m$ is understood. 
Under the 5D-traceless condition, the 5D-transverse condition 
\begin{equation}
 (k-1) \eta^{\hat{\rho} \hat{\sigma}}
     A_{z^{k-2}\rho\sigma \mu_1 \cdots \mu_{j-k}} + 
  z \partial^{\hat{\rho}} 
     A_{z^{k-1}\rho \mu_1 \cdots \mu_{j-k}} + 
  \left(z \partial_z + (k-4) \right)
     A_{z^k \mu_1 \cdots \mu_{j-k}} = 0, 
\end{equation}
becomes
\begin{equation}
   z \partial^{\hat{\rho}} 
       A_{z^{k-1}\rho \mu_1 \cdots \mu_{j-k}} + 
   \left(z \partial_z -3 \right) A_{z^k \mu_1 \cdots \mu_{j-k}} = 0.
  \label{eq:5D-transv-cond-splfd}
\end{equation}
In the $m$-th sector ($k+s = m$), therefore, 
\begin{equation}
 (s+1-2N) \Delta^2 a^{(k-1,s+1,N)} = z^3 \partial_z z^{-3} a^{(k,s,N)}
\label{eq:5D-transv-cond-splfd-EZ}
\end{equation}
for $N=0,\cdots, [s/2]$. Hereafter, we use a simplified notation 
$\mathscr{D} := z^3 \partial_z z^{-3}$.
One can see that all of $a^{(k,s,N)}$'s with $k+s = m$ and $N \leq
[s/2]$ can be determined from $a^{(m,0,0)}$ by using the 
relations (\ref{eq:5D-trac-cond-EZ}, \ref{eq:5D-transv-cond-splfd-EZ}).
This observation already implies that there can be at most one 
eigenmode in a given $m$-th sector that satisfies both the 
5D-traceless and 5D-transverse conditions. 

For now, let us assume that there is one, and proceed to determine 
the wavefunction. 
The wavefunction---$z$-dependence---of $a^{(m.0.0)}(z; \Delta^\mu)$ 
can be determined from the eigenmode equation
(\ref{eq:EGeq-q=not0-matrix-m}) with $k = m$, $s=N=0$.
Using (\ref{eq:5D-trac-cond-EZ}) and (\ref{eq:5D-transv-cond-splfd-EZ}), 
we can rewrite the equation as 
\begin{equation}
 \left[ R^2 \Delta_j - \{ (2m+1)j -m^2 + 2m \} 
       - 2 m \left(z \partial_z - 3 \right)
       +{\cal E} \right] 
   a^{(m,0,0)}(z;\Delta)
 = 0. 
\end{equation}
For this equation, 
\begin{equation}
 \left(a^{(m,0,0)}(z;\Delta)\right)_{\mu_1 \cdots \mu_{j-m}} = 
 \epsilon_{\mu_1 \cdots \mu_{j-m}} 
  \left(\frac{z}{R}\right)^{2-j} (\Delta z)^m K_{i\nu}(\Delta z), 
    \qquad 
{\cal E} = (j+4+\nu^2), 
\end{equation}
is a solution, where $\epsilon_{\mu_1 \cdots \mu_{j-m}}$ is a
$z$-independent 4D-traceless 4D-transverse totally symmetric 
rank-$(j-m)$ tensor of $\SO(3,1)$. From the value of the eigenvalue, 
it turns out that the 5D-traceless 5D-transverse mode in the $m$-th
sector corresponds to the $(n,l,m) = (0,0,m)$ mode. 
The $z$-dependence we determined above implies that 
\begin{equation}
 \Psi^{(j);0,0}_{i\nu; 0,0,m} (-\Delta^2,z) 
  \propto   (\Delta z)^{m} \Psi^{(j);0,0}_{i\nu; 0,0,0} (-\Delta^2, z).
\label{eq:result-Psi-00}
\end{equation}
This result corresponds to the $(s,N) = (0,0)$ case of (\ref{eq:00m-mode-fcn}).
The normalization constant $N_{j,m}$ is determined later in this 
section.

Let us now proceed to determine other $\Psi^{(j);s,N}_{i\nu;0,0,m}$, 
not just for $(s,N)=(0,0)$. 
Using the 5D-transverse condition, (\ref{eq:5D-transv-cond-splfd-EZ}), 
$a^{(m-1,1,0)}(z; \Delta)$ can be determined from $a^{(m,0,0)}(z;\Delta)$. 
\begin{equation}
 a^{(m-1,1,0)} = \frac{\mathscr{D}}{\Delta^2} a^{(m,0,0)}, \qquad 
 \Psi^{(j);1,0}_{i\nu;0,0,m} = 
   \frac{\mathscr{D}}{\Delta} \Psi^{(j);0,0}_{i\nu;0,0,m}.
\end{equation}

In order to determine the $s=2$ components $a^{(m-2,2,N)}$ 
($N=0,1$) of the $(n,l)=(0,0)$ mode in the $m$-th sector, 
one has to use both the 5D-transverse condition and 5D-traceless condition: 
\begin{eqnarray}
 2 \Delta^2 a^{(m-2,2,0)} & = & \mathscr{D} \; a^{(m-1,1,0)}, \\ 
 2 \Delta^2 a^{(m-2,2,0)}
 -  4(j-m+3/2) a^{(m-2,2,1)} & = & a^{(m,0,0)} .
\end{eqnarray}
Therefore, 
\begin{equation}
 a^{(m-2,2,0)} = \frac{1}{2\Delta^2} 
   \left(\frac{\mathscr{D}}{\Delta}\right)^2 a^{(m,0,0)},  \qquad 
 a^{(m-2,2,1)} = \frac{1}{4(j-m+3/2)}
   \left\{ \left(\frac{\mathscr{D}}{\Delta}\right)^2 - 1 \right\}
   a^{(m,0,0)}.
\end{equation}
After factoring out the normalization factor $(b^{(j-m)}_{s,N}/\Delta^{s-2N})$ 
and the common 4D-tensor $\epsilon^{(0,0,m)}$, we obtain 
\begin{equation}
 \Psi^{(j);2,0}_{i\nu;0,0,m} = \left(\frac{\mathscr{D}}{\Delta}\right)^2
    \Psi^{(j);0,0}_{i\nu;0,0,m}, \qquad 
 \Psi^{(j);2,1}_{i\nu;0,0,m} = 
   \left\{ \left(\frac{\mathscr{D}}{\Delta}\right)^2 - 1 \right\}
   \Psi^{(j);0,0}_{i\nu;0,0,m}. 
\end{equation}

The 5D-transverse conditions (\ref{eq:5D-transv-cond-splfd-EZ}) 
determine the $s=3$ components $a^{(m-3,3,N)}(z;\Delta)$ ($N=0,1$) 
from the $s=2$ components.
\begin{equation}
 a^{(m-3,3,0)} = \frac{1}{6\Delta^3} 
   \left(\frac{\mathscr{D}}{\Delta}\right)^3 a^{(m,0,0)},  \quad 
 a^{(m-3,3,1)} = \frac{1}{4(j-m+3/2)\Delta}
   \left\{ \left(\frac{\mathscr{D}}{\Delta}\right)^3
         -  \left(\frac{\mathscr{D}}{\Delta}\right) \right\}
   a^{(m,0,0)}, 
\end{equation}
and after factoring out the normalization factor 
$(b^{(j-m)}_{s,N}/\Delta^{s-2N})$ and $\epsilon^{(0,0,m)}$ as before, 
we obtain 
\begin{equation}
 \Psi^{(j);3,0}_{i\nu;0,0,m} =
    \left(\frac{\mathscr{D}}{\Delta}\right)^3 \Psi^{(j);0,0}_{i\nu;0,0,m},
   \qquad 
\Psi^{(j);3,1}_{i\nu;0,0,m} = 
   \left\{  \left(\frac{\mathscr{D}}{\Delta}\right)^3 
          -  \left(\frac{\mathscr{D}}{\Delta}\right) \right\}
     \Psi^{(j);0,0}_{i\nu;0,0,m}.
\end{equation}
The $s=3$ components determined purely by the conditions
(\ref{eq:5D-transv-cond-splfd-EZ}) satisfy the 5D-traceless condition 
(\ref{eq:5D-trac-cond-EZ}) with the $s=1$ component:
\begin{equation}
 6\Delta^2 a^{(m-3,3,0)} - 4(j-m+3/2)a^{(m-3,3,1)} =
   \frac{\mathscr{D}}{\Delta^2} a^{(m,0,0)} = a^{(m-1,1,0)}.
\end{equation}

In this way, the wavefunctions $\Psi^{(j);s,N}_{i\nu;0,0,m}(-\Delta^2, z)$ 
for all $(s,N)$ are determined, and
the result is 
\begin{equation}
  \Psi^{(j);s,N}_{i\nu;0,0,m}(-\Delta^2,z) =
    \sum_{a=0}^N (-)^a {}_N C_a 
      \left( \frac{\mathscr{D}}{\Delta}\right)^{s-2a} 
        \left[(z \Delta)^m \Psi^{(j);0,0}_{i\nu;0,0,0}(-\Delta^2, z)
	\right] \times N_{j,m}.
\label{eq:00m-mode-fcn-app}
\end{equation}
The only remaining concern was that there are more conditions 
from (\ref{eq:5D-trac-cond-EZ}, \ref{eq:5D-transv-cond-splfd-EZ}) 
than the number of components $a^{(k,s,N)}$ in the $m$-th sector; 
there can be at most one eigenmodes satisfying these 5D-traceless 
5D-transverse conditions, as we stated earlier, but there may be no 
eigenmode left, if the conditions are overdetermining. We have confirmed, 
however, that the wavefunctions 
(\ref{eq:00m-mode-fcn}=\ref{eq:00m-mode-fcn-app}) satisfy all of the 
relations given by (\ref{eq:5D-trac-cond-EZ}, \ref{eq:5D-transv-cond-splfd-EZ}).

\subsubsection{Normalization}
\label{sssec:normalization}

We have yet to determine the normalization factor $N_{j,m}$; 
as in the main text, we choose (\ref{eq:wvfc-normalization-cond}) 
to be the normalization condition. Orthogonal nature among the
eigenmodes is guaranteed because of the Hermitian nature of the 
operator $\alpha' \left( \nabla^2 - M^2 \right)$.
It is thus sufficient to focus only on the divergent part of the 
integral in the normalization condition in order to determine $N_{j,m}$.

The divergent part of the integral in (\ref{eq:wvfc-normalization-cond})
comes only from terms with $s = m$, $k=0$, $(0 \leq N \leq [m/2])$ 
and $a = 0$. For a given $m$, 
\begin{eqnarray}
&& [ \epsilon \cdot \epsilon'] \; \delta(\nu-\nu') 
   \label{eq:normalization-temp-app}\\
& \sim & 
 N_{j,m}^2 \int_0 dz \sqrt{-g(z)} e^{-2jA} \nonumber \\
& & \quad 
  \left(\sum_{N=0}^{[m/2]} \tilde{E}^N D^{m-2N}[\epsilon^{(0,0,m)}]
     \frac{b^{(j-m)}_{m,N}}{\Delta^{m-2N}} 
     \frac{z^3 \partial_z^{m} z^{-3}}{\Delta^m}
     (z\Delta)^m \Psi^{(j);0,0}_{i\nu;0,0,m}(- \Delta^2,z)
  \right)_{\mu_1 \cdots \mu_j}
   \nonumber  \\
 & & \quad 
  \left(\sum_{M=0}^{[m/2]} \tilde{E}^M D^{m-2M}[\epsilon^{(0,0,m)}]
     \frac{b^{(j-m)}_{m,M}}{\Delta^{m-2M}} 
     \frac{z^3 \partial_z^{m} z^{-3}}{\Delta^m}
     (z\Delta)^m \Psi^{(j);0,0}_{i\nu;0,0,m}(- \Delta^2,z)  
  \right)^{\hat{\mu}_1 \cdots \hat{\mu}_j}.
   \nonumber 
\end{eqnarray}
Divergent part of the integral in this expression comes from 
\begin{eqnarray}
 & &
 \left(\frac{2}{\pi}\right)^2 \frac{\nu \sinh(\pi\nu)}{2}
  \int dx x^{2j-5}
    \left[x^3 \partial_x^m x^{-1-j+m} K_{i\nu}(x) \right]
    \left[x^3 \partial_x^m x^{-1-j+m} K_{i\nu'}(x) \right] \nonumber \\
& \simeq & 
   \prod_{p=1}^m \left[ (j-p+1)^2 + \nu^2 \right] \delta(\nu-\nu') 
 = \frac{ \Gamma(j+1-i\nu) \Gamma(j+1+i\nu) }
        { \Gamma(j+1-m-i\nu) \Gamma(j+1-m+i\nu) } \delta(\nu- \nu').
   \nonumber 
\end{eqnarray}
Noting that 
\begin{eqnarray}
&&
   \left(\sum_{N=0}^{[m/2]} \tilde{E}^N D^{m-2N}[\epsilon^{(0,0,m)}]
     \frac{b^{(j-m)}_{m,N}}{\Delta^{m-2N}} \right)
  \left(\sum_{M=0}^{[m/2]} \tilde{E}^M D^{m-2M}[\epsilon^{'(0,0,m)}]
     \frac{b^{(j-m)}_{m,M}}{\Delta^{m-2M}} \right), \nonumber  \\
&=& 
 \frac{j!}{(j-m)!} \left( \sum_{N=0}^{[m/2]} b^{(j-m)}_{m,N} \right) 
 \; \epsilon^{(0,0,m)}_{\mu_1 \cdots \mu_{j-m}} \cdot 
    \epsilon^{'(0,0,m) \; \hat{\mu}_1 \cdots \hat{\mu}_{j-m}}, \nonumber  
\end{eqnarray}
we find that (\ref{eq:normalization-temp-app}) implies 
\begin{eqnarray}
 N_{j,m}^{-2} & = &
 \frac{ \Gamma(j+1-i\nu) \Gamma(j+1+i\nu) }
        { \Gamma(j+1-m-i\nu) \Gamma(j+1-m+i\nu) } 
 \frac{j!}{(j-m)!} \left( \sum_{N=0}^{[m/2]} b^{(j-m)}_{m,N} \right), 
    \label{eq:Njm-def-app} \\ 
& = &  
  \frac{\Gamma(j+1-i\nu)}{\Gamma(j+1-m-i\nu)}
  \frac{\Gamma(j+1+i\nu)}{\Gamma(j+1-m+i\nu)} \; 
  {}_jC_m 
  \frac{\Gamma(3/2+j-m)}{2^m\Gamma(3/2+j)}
  \frac{\Gamma(2+2j)}{\Gamma(2+2j-m)}. \nonumber 
\end{eqnarray}
%

\subsection{A Note on Wavefunction of Massless Vector Field}
\label{ssec:vector}

For a rank-1 tensor (vector) field on AdS$_5$, we can determine 
the wavefunction of the $(n,l,m) = (1,0,1)$ eigenmode, 
not just for the $(n,l,m) = (0,0,m)$ modes with $m = 0,1$.
With the eigenvalue ${\cal E}_{1,0} = (3j+5+\nu^2)|_{j=1}$, 
\begin{equation}
 a^{(0,1,0)} = \epsilon^{(1,0,1)} z^2 K_{i\nu}(\Delta z), \qquad 
 a^{(1,0,0)} = \epsilon^{(1,0,1)}
  \partial_z \left(z^2 K_{i\nu}(\Delta z) \right)  
\end{equation}
is the eigenvector solution to (\ref{eq:EGeq-q=not0-m=1}).

The $(n,l,m) = (0,0,1)$ mode and $(n,l,m)=(1,0,1)$ mode are
independent, even after the mass-shell condition (\ref{eq:mass-shell}) 
for generic vector fields in the bosonic string theory. However, 
for the massless vector field $A_m$ obtained by simple dimensional 
reduction of the massless vector field $A^{(Y)}_M$ with 
$Y = \left\{1,0,0 \right\}$, those two modes become degenerate. 
To see this, note that $c_y = -4$ for this mode, so that the 
mass-shell condition (\ref{eq:mass-shell}) implies, 
\begin{equation}
 (j+4+\nu^2 + c_y)|_{j=1} = 0 \quad (0,0,1){\rm ~mode}, \qquad 
 (3j+5+\nu^2+c_y)|_{j=1} = 0 \quad (1,0,1){~\rm mode},
\end{equation}
or equivalently, $i\nu = 1$ and $i\nu = 2$, respectively, for these two
modes. It is now obvious that the terms proportional to 
$(\epsilon \cdot q)$ in (\ref{eq:photon-wvfc}) are in the 
form of this $(n,l,m) = (1,0,1)$ mode. With the relations 
$x^3 \partial_x \left[x^{-3+2} K_1(x) \right] =
  - x^3 \left[x^{-1} K_2(x)\right]$ and  $
\partial_x \left[x^2 K_2(x) \right] = - x^2 K_1(x)$, one can also see 
that the wavefunction for the $(n,l,m)=(0,0,1)$ mode is also
proportional to the form given in (\ref{eq:photon-wvfc}) when the 
on-shell condition is imposed. 

\subsection{Projection operator of $\SO(3,1)$ tensors}
\label{ssec:appendix-projector}

Note first that 
\begin{equation}
 a = \sum_{s=0}^r \sum_{N=0}^{[s/2]} \tilde{E}^N D^{s-2N}_\Delta
  [a^{(s,N)}]
\label{eq:4D-tensor-dcmp-app}
\end{equation}
is an \underline{orthogonal} decomposition of a totally symmetric $\SO(3,1)$ 
tensor $a$ of rank-$r$ into totally symmetric 4D-traceless 4D-transverse 
$\SO(3,1)$ tensors $a^{(s,N)}$ of rank-$(r-s)$. Here, the metric is 
given by 
\begin{equation}
[b_{(-\Delta)}] \cdot [a_{(\Delta)}] := 
 \left[b_{(-\Delta)}\right]_{\rho_1 \cdots \rho_r} 
 \left[a_{(+\Delta)}\right]_{\sigma_1 \cdots \sigma_r} 
 \eta^{\hat{\rho}_1 \hat{\sigma}_1 } \cdots 
 \eta^{\hat{\rho}_r \hat{\sigma_r}}
\end{equation}
as in the main text. To see that the decomposition is orthogonal under
this metric, one only needs to use (\ref{eq:extract-SO(3,1)-tensor}) 
to verify that 
\begin{equation}
 \left[ \tilde{E}^M D^{t-2M}_{-\Delta} [b^{(t,M)}] \right] \cdot 
 \left[ \tilde{E}^N D^{s-2N}_{\Delta} [a^{(s,N)}] \right] =
  \delta_{M,N} \delta_{t-2M, s-2N} \; 
   \frac{\Delta^{2(s-2N)}}{b^{(r-s)}_{s,N}}  [b^{(t,M)}] \cdot [a^{(s,N)}].
\end{equation} 

Using the fact that (\ref{eq:4D-tensor-dcmp-app}) is an orthogonal
decomposition, let us construct projection operators $\bar{P}^{(r;s,N)}$ 
that extract various components $a^{(s,N)}$ from a totally symmetric 
$\SO(3,1)$ tensor $a$ of rank-$r$.
We introduced an operator $P^{(r)}$ in (\ref{eq:projection-000-r}), 
which acts on rank-$r$ $\SO(3,1)$ tensors. From what we have seen above, 
it can be used to extract the $a^{(s,N)=(0,0)}$ component from a 
rank-$r$ tensor $a$. That is, $\bar{P}^{(r;0,0)} = P^{(r)}$.
It is straightforward to see that the projection operator for 
other components $a^{(s,N)}$ with general $(s,N)$ is given by 
\begin{equation}
 \bar{P}^{(r;s,N)} := \sum_{a}
   \frac{b^{(r-s)}_{s,N}}{\Delta^{2(s-2N)}}
   \frac{1}{D_a}
   \left(\tilde{E}^N D^{s-2N}_{\Delta}[\epsilon_a ]
   \right)_{\rho_1 \cdots \rho_r}
   \left(\tilde{E}^N D^{s-2N}_{-\Delta}[\epsilon_a ]
   \right)_{\sigma_1 \cdots \sigma_r},
\end{equation}
where $\epsilon_a$'s are an orthogonal basis of totally symmetric 
4D-traceless 4D-transverse $\SO(3,1)$ tensors of rank-$(r-s)$.

It is also useful to have a concrete form of the projection operator 
$P^{(r)}$, not just its abstract definition in (\ref{eq:projection-000-r}).
We find that it is given by 
\begin{equation}
 P^{(r)} \cdot a = \sum_{M=0}^{\left[ \frac{j}{2} \right]} 
   \frac{(-1)^M \; \Gamma\left(r+\frac{1}{2}-M\right)}
              {4^M M! \; \Gamma\left(r+\frac{1}{2}\right)}
    \sum_{k=0}^{r-2M} \frac{(-1)^k}{k!} 
     [\tilde{E}^M D^k \; OP_{(p,q)=(M,k)}] \cdot a,
\label{eq:4D-tensor-projection-op}
\end{equation}
where $OP_{(p,q)}$ is the operator given 
in (\ref{eq:extract-SO(3,1)-tensor}). A totally symmetric rank-$r$ tensor 
$a$ is converted once into rank-$(r-2M-k)$ tensors, and then they are 
converted back to a rank-$r$ tensor under the operator $P^{(r)}$.
To see that all the $\tilde{E}^ND^{s-2N}[a^{(s,N)}]$ components are projected 
out by $P^{(r)}$, one only needs to use the following formula \cite{Mathematica}
\begin{equation}
 \sum_{M=0}^N (-1)^M {}_N C_M 
   \frac{\Gamma\left(r-N+\frac{3}{2}\right)}
        {\Gamma\left(r-N+\frac{3}{2}-M\right)}
   \frac{\Gamma\left(r+\frac{1}{2}-M \right)}
        {\Gamma\left(r+\frac{1}{2}\right)} 
  = \frac{\Gamma\left(\frac{1}{2}-r\right)}
         {\Gamma\left(\frac{1}{2}-r+N\right)\Gamma(1-N)},
\end{equation}
which vanishes for an integer $N\geq 0$.

\subsection{Some Tensor Computations}
\label{ssec:tensor}

Let us derive a more concrete expression for the product 
$\left(q^{\mu_1}\dots q^{\mu_r}\right) \cdot 
[P^{(r)}]_{\mu_1\dots \mu_r}^{\nu_1\dots \nu_r} \cdot 
\left(p_{\nu_1}\dots p_{\nu_r}\right)$, by using the explicit 
expression for the projection operator $P^{(r)}$ to the 
$\SO(3,1)$-transverse $\SO(3,1)$-traceless rank-$r$ tensor. 
\begin{align}
& \left(q^{\mu_1}\dots q^{\mu_r}\right) \cdot 
[P^{(r)}]_{\mu_1\dots \mu_r}^{\nu_1\dots \nu_r} \cdot 
\left( p_{\nu_1}\dots p_{\nu_r} \right) \notag
\\
&= \sum_{M=0}^{[r/2]} 
 \frac{ (-1)^M \; \Gamma \left( j+\frac{1}{2} -M \right)}
      { 4^M M! \; \Gamma \left( j+\frac{1}{2} \right) }
 \frac{r!}{(r-2M)! } 
     \left[ q^2 - \frac{(q \cdot \Delta)^2}{\Delta^2}\right]^M
 (p^2)^M (q \cdot p)^{r-2M}, 
\end{align}
where we used that $p \cdot \Delta = 0$. 
Within the regime of $q^2, (p \cdot q), (q \cdot \Delta) \gg 
\Lambda^2, \Delta^2, p^2$ we have been interested in in this article, 
$(q \cdot \Delta)^2/\Delta^2 \gg q^2$. Thus, after ignoring $q^2$, 
\begin{align}
& \left(q^{\mu_1}\dots q^{\mu_r}\right) \cdot 
[P^{(r)}]_{\mu_1\dots \mu_r}^{\nu_1\dots \nu_r} \cdot 
\left( p_{\nu_1}\dots p_{\nu_r} \right) \notag
\\
\approx & 
(p\cdot q)^r
\sum_{M=0}^{[r/2]}
 \frac{ \; \Gamma \left( r+\frac{1}{2} -M \right)}
      { 4^M M! \; \Gamma \left( r+\frac{1}{2} \right) }
 \frac{r!}{(r-2M)! } 
\left[\sfrac{q \cdot \Delta}{q\cdot p}^2 \frac{p^2}{\Delta^2}
\right]^M =: (p \cdot q)^r \times \hat{d}_{r}(\eta,\Delta^2).
\label{eq:q-proj-p-product}
\end{align}
This introduces $\hat{d}_{r}$, which is a polynomial of skewness 
$(q \cdot \Delta)/(p \cdot q)= -2\eta$ of degree $2[r/2]$.

When $r$ is even, this polynomial of $\eta$ can also be rewritten 
by using Legendre polynomial, $P_\ell(x)$, which is defined 
by (p.82, \cite{Iwanami-formula-3})
\begin{equation}
 P_{\ell}(x) = {}_2F_1 \left(-\ell, \ell+1,1;\frac{1-x}{2} \right)
  = \frac{(2\ell-1)!!}{\ell !} x^{\ell}
   {}_2F_1 \left(-\frac{\ell}{2},\frac{1-\ell}{2},\frac{1}{2}-\ell; 
     \frac{1}{x^2}\right).
\end{equation}
For an even $r$, 
\begin{align}
\hat{d}_r(\eta, \Delta^2) = &  \sum_{M=0}^{r/2} \frac{
     \left(- \frac{r}{2}\right)_M 
     \left(\frac{1-r}{2}\right)_M } 
  {M! \left(\frac{1}{2}-r\right)_M}
  \left( -\frac{4p^2}{\Delta^2} \eta^2 \right)^M
= {}_2F_1\left( - \frac{r}{2}, \frac{1-r}{2}, \frac{1}{2}-r; 
   \frac{(4m_h^2+\Delta^2)\eta^2}{\Delta^2} \right) \notag \\
= & \frac{r!}{(2r-1)!!} \; 
  \left[ \sqrt{\frac{4m_h^2+\Delta^2}{\Delta^2}} \eta \right]^r
  P_r \left(\sqrt{\frac{\Delta^2}{4m_h^2+\Delta^2}}\frac{1}{\eta}\right) 
  =: \hat{d}_r([\eta]), 
\label{eq:dhat-Legendre}
\end{align}
where we used the kinematical relation $4p^2 = - (4m_h^2+\Delta^2)$.

Similarly, it is also necessary to compute the following expression 
in order to study the $m=0$ exchange amplitude in 
section \ref{sssec:m=0-amplitude}:
\begin{eqnarray}
  \left[  \sum_{a \neq b} \epsilon^{2*}_{\rho_a} \epsilon^1_{\rho_b} 
      q_{\rho_1} \cdots {}_{\check{\rho_a}}{}_{\check{\rho_b}} \cdots q_{\rho_j} \right]
   \cdot [P^{(j)}]^{\hat{\rho}_1 \cdots \hat{\rho}_j}_{\sigma_1 \cdots \sigma_j} \cdot 
   \left[ p^{\hat{\sigma}_1} \cdots p^{\hat{\sigma}_j} \right], 
\end{eqnarray}
which is also evaluated as above. The term proportional to 
$\eta^{\hat{\mu}\hat{\nu}} \epsilon^{2*}_{\nu} \epsilon^1_{\mu}$ 
(contribution to the structure function $V_1$) is 
\begin{eqnarray}
& & 2^2  \sum_{M=1}^{[j/2]} 
   \frac{(-1)^M \; \Gamma \left( j+\frac{1}{2}-M \right)}
        { 4^M M! \; \Gamma \left( j + \frac{1}{2} \right) }
   \frac{j!}{(j-2M)! 2!} 
   \left[ q^2 - \frac{(q \cdot \Delta)^2}{\Delta^2} \right]^{M-1}
   (p^2)^M (q \cdot p)^{j-2M} \nonumber \\
& \approx & - 2 \frac{\Delta^2}{(q \cdot \Delta)^2} \times 
  (q \cdot p)^j \sum_{M=1}^{[j/2]} 
   \frac{ \; \Gamma \left( j+\frac{1}{2}-M \right)}
        { 4^M M! \; \Gamma \left( j + \frac{1}{2} \right) }
   \frac{j!}{(j-2M)!} 
   \left[ \left(\frac{q \cdot \Delta}{q \cdot p} \right)^2 
          \frac{p^2}{\Delta^2} \right]^M.
  \label{eq:qee-proj-p-product} 
\end{eqnarray}
This expression is once again a polynomial of $\eta$ of degree 
$2[j/2]-2$, and is roughly of order $\Delta^2/(q \cdot p)^2$ times 
the expression (\ref{eq:q-proj-p-product}). 

We will also need the following computation in 
sections \ref{sssec:preparation} and \ref{sssec:general-structure-amplitude}:
\begin{align}
 & (q_{\mu_1} \cdots q_{\mu_{j-k}}) \cdot 
 \left( \tilde{E}^ND^{s-2N}_{-\Delta} [\epsilon^{(0,0,m)}] 
 \right)^{\hat{\mu}_1 \cdots \hat{\mu}_{j-k}}   \notag \\
 = & \frac{(j-k)!}{(j-m)!}
   \left[ q^2-\frac{(q\cdot \Delta)^2}{\Delta^2}\right]^N
   (-i q \cdot \Delta)^{s-2N} \; 
   \left[ (q_{\mu_1} \cdots q_{\mu_{j-m}}) \cdot \epsilon^{(0,0,m)} \right].
\label{eq:q-[mmode]-product}
\end{align}
%

\section{Conformal OPE Coefficients from AdS Integrals}
\label{sec:conformal-coeff-AdS}

Let us introduce an integral 
\begin{align}
C_1(\delta,\vartheta)&:= (1-\vartheta^2)^{1/2}\int_0^\infty dy \; y^{1+\delta}
   \; K_1(y\sqrt{1+\vartheta}) \; K_1(y\sqrt{1-\vartheta}),   
  \label{eq:C1-def}
\end{align}
which we encounter as the photon--photon--Pomeron/Reggeon vertex on $AdS_5$.
$\vartheta = \eta/x$ and $\delta = j+i\nu$ in that context.

It is known (p.101, \cite{Magnus:1966}), if 
${\rm Re} (\alpha+\beta) >0$ and ${\rm Re}(1\pm \nu\pm \mu-\rho) > 0$, 
that 
\begin{align}
 &\int_0^\infty dt t^{-\rho}K_\mu(\alpha t)K_\nu(\beta t) =
2^{-\rho-2}\alpha^{\rho-\nu-1}\beta^\nu[\Gamma(1-\rho)]^{-1} \nonumber \\
&\times \Gamma\left(\frac{1+\nu+\mu-\rho}{2} \right)
        \Gamma\left(\frac{1+\nu-\mu-\rho}{2} \right)  
        \Gamma\left(\frac{1-\nu+\mu-\rho}{2} \right)
        \Gamma\left(\frac{1-\nu-\mu-\rho}{2} \right)  \nonumber \\
&\times{}_2F_1\left( \frac{1+\nu+\mu-\rho}{2}, \frac{1+\nu-\mu-\rho}{2};
     1-\rho ;1-\frac{\beta^2}{\alpha^2} \right). 
\end{align}
Substituting $\rho=-1-\delta$, $\mu=1$, $\nu=-1$, 
$\alpha=\sqrt{1-\vartheta}$, and $\beta=\sqrt{1+\vartheta}$, we obtain
\begin{align}
 C_1(\delta,\vartheta) = 
  \frac{ \Gamma(\frac{\delta}{2}) ( \Gamma(\frac{\delta}{2}+1) )^2 
         \Gamma(\frac{\delta}{2}+2) }
       { \Gamma(\delta+2) } 
  2^{\delta -1} (1-\vartheta )^{-\frac{\delta }{2}} 
  {}_2{F}_1\left(\frac{\delta }{2},\frac{\delta +2}{2};\delta +2;
     \frac{2 \vartheta }{\vartheta -1} \right)
\label{eq:C1-from-integral-A}
\end{align}
An equivalent, but a little different expression is also obtained by 
using the following relation (p.60, \cite{Iwanami-formula-3})
\begin{align}
 {}_2F_1(\alpha,\beta,2\beta;2z)=(1-z)^{-\alpha}{}_2F_1\left(\frac{\alpha}{2},\frac{\alpha+1}{2},\beta+\frac{1}{2};\sfrac{z}{1-z}^2\right); 
\end{align}
namely, 
\begin{align}
 C_1(\delta,\vartheta)&=2^{\delta-1} \frac{\delta+2}{\delta}
 \frac{(\Gamma(\frac{\delta}{2}+1))^4}{\Gamma(\delta+2)} {}_2F_1 \left(\frac{\delta}{4},\frac{\delta}{4}+\frac{1}{2};\frac{\delta}{2}+\frac{3}{2};\vartheta^2\right).
\label{eq:C1-from-integral-B}
\end{align}
As a function of $\vartheta = \eta/x$, (\ref{eq:C1-from-integral-A}) and 
(\ref{eq:C1-from-integral-B}) are precisely of the 
form (\ref{eq:conf-OPE-2side}) and (\ref{eq:conf-OPE-center}), respectively, 
required in the conformal OPE coefficients.

%

\end{document}